\documentclass[final,3p,times]{elsarticle}

\usepackage{amssymb}
\usepackage{amsmath}
\usepackage{subfigure}
\usepackage{graphicx}
\usepackage{amsmath,amssymb,boldmath}
\usepackage[latin1]{inputenc}
\usepackage{colortbl}
\usepackage[english]{babel}
\usepackage{psboxit}
\usepackage{rotating}
\usepackage{times}
\usepackage{color}
\usepackage{pifont}
\usepackage{pstricks}
\usepackage{picinpar}
\usepackage{psboxit}
\usepackage{xspace}
\usepackage{latexsym}
\usepackage{moreverb}
\usepackage{floatfig}
\usepackage{wrapfig}
\usepackage{appendix}

\journal{Computer Physics Communications}

\begin{document}

\begin{frontmatter}


\title{Regularization with Numerical Extrapolation
for Finite and \\ UV-Divergent Multi-loop Integrals
}
\author{E de Doncker\fnref{label1}}
\ead{elise.dedoncker@wmich.edu}
\ead[url]{http://www.cs.wmich.edu/elise}
\author{F Yuasa\fnref{label2}}
\ead{fukuko.yuasa@kek.jp}
\author{K Kato\fnref{label3}}
\ead{kato@cc.kogakuin.ac.jp}
\author{T Ishikawa\fnref{label2}}
\ead{tadashi.ishikawa@kek.jp}
\author{J Kapenga\fnref{label1}}
\ead{john.kapenga@wmich.edu}
\author{O Olagbemi\fnref{label1}}
\ead{omofolakunmiel.olagbemi@wmich.edu}
\fntext[label1]{Department of Computer Science,
Western Michigan University, 1903~West Michigan Avenue,
Kalamazoo, MI 49008, United States}
\fntext[label2]{High Energy Accelerator Research Organization (KEK),
1-1 OHO Tsukuba, Ibaraki 305-0801, Japan}
\fntext[label3]{Department of Physics, Kogakuin University,
Shinjuku, Tokyo 163-8677, Japan
}


%
\begin{abstract}
We give numerical integration results for Feynman loop diagrams 
such as those covered by Laporta~\cite{laporta01} 
and by Baikov and Chetyrkin~\cite{baikov10}, and which may give rise to 
loop integrals with UV singularities. We explore automatic
adaptive integration using multivariate techniques from the 
{\sc ParInt} package for multivariate integration, as well as iterated
integration with programs from the {\sc Quadpack} package, and a trapezoidal
method based on a double exponential transformation. {\sc ParInt} is 
layered over {\sc MPI} (Message Passing Interface), and incorporates
advanced parallel/distributed techniques including load balancing among 
processes that may be distributed over a cluster or a network/grid of nodes. 
Results are included for 2-loop vertex and box diagrams 
and for sets of 2-, 3- and 4-loop self-energy diagrams with or without UV terms.
Numerical regularization of integrals with singular terms is achieved by 
linear and non-linear extrapolation methods. 
\end{abstract}

\end{frontmatter}


\section{Introduction}
\label{intro}
High energy physics collider experiments target the precise measurement of parameters in the standard model and beyond, and detection of any deviations of the experimental data from the theoretical predictions, leading to the study of new phenomena. In modern physics, there are three basic interactions acting on particles: weak, electromagnetic and strong interactions. When we consider a scattering process of elementary particles, the cross section reflects the dynamics that govern the motion of the particles, caused by the interaction.

All information on a particle interaction is contained in the amplitude according to the (Feynman) rules of Quantum Field Theory. Generally, with a given particle interaction, a large number of configurations (represented by Feynman diagrams) is associated. Each diagram represents one of the possible configurations of virtual processes, and it describes a part of the total amplitude.
The square sum of the amplitudes delivers the probability or cross section of the process. Based on the Feynman rules, the amplitude can be obtained in an automatic manner:
\emph{(i)} {determine the physics process (external momenta and order of perturbation);}
\emph{(ii)} {draw all Feynman diagrams relevant to the process;}
\emph{(iii)} {describe the contributions to the amplitude.}

Feynman diagrams are constructed in such a way that the initial state particles are connected to the final state particles by propagators and vertices. Particles meet at vertices according to a coupling constant $g,$ which indicates the strength of the interaction. The amplitude is expanded as a perturbation series in $g,$ where the leading (lowest) order of approximation corresponds to the tree level of the Feynman diagrams. Higher orders require the evaluation of loop diagrams,
so that the computation of loop integrals is very important for the present and future high-energy experiments.

When few masses occur in the computation of loop integrals, analytic approaches are generally feasible.
However, in the presence of a wide range of masses, analytic evaluation becomes very complicated or impossible. 
For one-loop integrals, explicit analytic methods have been established by many authors,
but alternative approaches are compulsory  
for multi-loop integrals with a variety of masses and momenta.
We propose a fully numerical approach based on multi-dimensional integration and extrapolation, 
and demonstrate results of the technique for multi-loop integrals with and without masses.
 
In the computation of loop integrals we have to handle singularities. 
Depending on the value of internal masses and external momenta, the integrand denominator 
may vanish in the interior of the integration domain.
The term $i\rho$ (subtracted from $V$) in the denominator of the loop integral
representation of Eq~\eqref{Lloop} is intended to prevent the integral from diverging
if $V$ vanishes in the domain.
The idea of our numerical extrapolation approach is to consider $\rho$ not as an 
infinitesimal small number for the analytic continuation but
as a finite number, to make the integral non-singular.
We choose a sequence of $\rho$ values, $\rho_\ell\rightarrow 0$ 
(e.g., a geometric sequence), so that multi-dimensional integration yields consecutive
$I(\rho_{\ell})$ corresponding to $\rho_\ell.$
The sequence of $I(\rho_{\ell})$ is extrapolated numerically to approximate the value
of the loop integral in the limit as $\rho_\ell\rightarrow 0.$ 
For physical kinematics where an imaginary part is present, it can be treated numerically 
as well as the real part,
since the integrand is not singular for finite $\rho_\ell.$
In previous work we have demonstrated various loop integral computations 
using this type of method not only in the Euclidean but also in the
physical region~\cite{edcpp03,eddacat03,acat07,acat08,ddacat10,cpc11y,jocs11}.

For the infrared divergent case we have two prescriptions. One is to introduce a 
small fictitious mass for the massless particles and the other is to use 
dimensional regularization. 
We have shown results for several problem classes in~\cite{iccs05b,acat07,cpp10,acat11,jocs11}.

In this paper we concentrate on loop integrals with UV singularities, which satisfy asymptotic 
expansions in the dimensional regularization parameter $\varepsilon$
(see Eq~\eqref{Lloop}, where the space-time dimension $n$ will be set to $n = 4-2\varepsilon$
to account for UV singularity).
Based on multi-dimensional integration and numerical extrapolation, 
we present a novel numerical regularization method for integrals with UV singularities, 
applied to 1-, 2-, 3- and 4-loop diagrams.
We compare with results in the literature, including those of Laporta~\cite{laporta01} $-$ whose method is based on 
the numerical solution of systems of difference equations, 
the sector decomposition approach by Smirnov and Tentyukov~\cite{smirnov10}, 
and the analytic results by Baikov and Chetyrkin~\cite{baikov10}.

The integration strategies in this paper adhere to \emph{automatic integration}, which is
a black-box approach for generating an approximation ${\mathcal Q}f$ to an integral
\begin{equation}\label{blackbox}
{\mathcal I\hspace*{-0.4mm}}f = \int_{\mathcal D} f(\vec{x}) ~d\vec{x},
\end{equation}
as well as an absolute error estimate $E_af,$ 
in order to satisfy a specified accuracy requirement of the form
\begin{equation}
\label{accuracy}
|\,{\mathcal Q}f-{\mathcal I\hspace*{-0.4mm}}f\,| ~\le ~ E_af ~\le~ \max\,\{\,t_a\,,\,t_r\,|\,{\mathcal I\hspace*{-0.4mm}}f\,|~\} 
\end{equation}
for a given integrand function 
$f: {\mathcal D}\subset {\mathbb R}^d \rightarrow {\mathbb R},$
a $d$-dimensional domain $\mathcal D,$ and (absolute/relative)
error tolerances $t_a$ and $t_r.$ 
If it is found that Eq~\eqref{accuracy} cannot be achieved, an error indicator
should be returned.
In order to achieve the accuracy requirement, the actual error should not exceed the error estimate
$E_af,$ and the error estimate should not exceed the weaker of the
absolute and relative error tolerances (indicated by the maximum taken on the right
of`\eqref{accuracy}). 
When a relative or an absolute accuracy (only) needs to be satisfied we set $t_a = 0$ or $t_r = 0,$ respectively.
If both $t_a \ne 0$ and $t_r \ne 0,$ the weaker of the two error tolerances is imposed; if
$t_a = t_r = 0$ then the program will reach an abnormal termination.
This type of accuracy requirement is based on~\cite{deboor71}
and used extensively in {\sc Quadpack}~\cite{pi83}.

Known methods for parallelization of these procedures include:\\
\emph{(i)} Parallelization on the {\color{black}rule} or {\color{black}points} level: typically in \emph{non-adaptive}
algorithms, e.g., for \emph{Monte-Carlo (MC)} algorithms and composite rules
using \emph{grid} or \emph{lattice} points. Then in $If = \int_{\mathcal D}f
\approx \sum_k w_k {\color{black}f(\vec{x}_k)},$ the function evaluations
{\color{black}$f(\vec{x}_k)$} are performed in parallel.\\
\noindent
\emph{(ii)} Parallelization on the {\color{black}region} level: in \emph{adaptive}
(region-partitioning) methods. These lead to {\color{black}task pool strategies},
which may benefit from load balancing on distributed memory systems; 
or maintain a shared priority queue on shared memory systems.\\
\noindent
\emph{(iii)} We added multi-threading to {\color{black}iterated integration}~\cite{icmsq12,ccp12,ddacat13}:
the {\color{black}inner integrals} are independent and computed in parallel.
For example, over a subregion $\mathcal S = {\mathcal D}_1\times {\mathcal D}_2$ ~(with inner region ${\mathcal D}_2$) consider
$\int_{\mathcal S} F({\vec x}) \,d{\vec x} \approx \sum_k w_k {\color{black}\vec F({\vec x_k})},$ ~with~
${\color{black}\vec F({\vec x_k})} = \int_{{\mathcal D}_2}f(\vec{x_k},{\vec y}) \,d{\vec y}.$
The integrations in the different coordinate directions can be performed {\color{black}adaptively}, which we
achieved with iterated versions of the 1D programs {\sc Dqags} or {\sc Dqage} from
{\sc Quadpack}~\cite{pi83,jocs11}.

We further apply numerical extrapolation techniques for convergence acceleration 
of a sequence of integrals with respect to a parameter $\gamma.$ For linear
extrapolation, an asymptotic expansion of the form
\begin{equation}
{\color{black}{\mathcal I}(\gamma) \sim \sum_{k\ge \kappa} C_k \,\varphi_k(\gamma), ~~~~~~\mbox{as~~} \gamma\rightarrow 0}
\label{asymp}
\end{equation}
is assumed, where ${\cal I}(\gamma)$ represents the integral and the
sequence of $\varphi_k(\gamma)$ is known.
If the structure of the expansion is unknown we resort to a non-linear extrapolation with the 
$\epsilon$-algorithm~\cite{shanks55,wynn56,sidi96,sidi03,sidi11}.

This paper gives an overview of our recent work. 
Section~\ref{Feynman} provides background and notations for multi-loop Feynman integrals
and diagrams, and discusses the use of extrapolation or convergence acceleration.
Section~\ref{integration} describes iterated integration, 
the {\sc ParInt} adaptive strategies, and the double exponential transformation method.
Numerical results obtained for a set of 
2-loop self-energy, vertex and box diagrams are discussed in Section~\ref{2-loop-integral}; 3-loop massless and massive self-energy
diagrams are covered in Section~\ref{3-loop-self}, and 4-loop massless self-energy diagrams in Section ~\ref{4-loop-self}. 

Results from parallel distributed computations were obtained on the \emph{thor} cluster of the 
Center for High Performance Computing and Big Data at WMU, where we used
16-core cluster nodes with Intel(R) Xeon(R) E5-2670, 2.6\,GHz dual processors and 128\,GB of memory,
and the cluster's Infiniband interconnect for message passing via MPI.
Some sample sequential and parallel results were collected from runs on
Intel(R) Xeon(R) CPU E5-1660 3.30GHz, E5-2687W v3 3.10\,GHz, 
and on a 2.6\,GHz Intel(R) Core i7 Mac-Pro with 4 cores and 16\,GB memory under OS X.
For the inclusion of OpenMP~\cite{openmp} multi-threading compiler directives in the iterated
integration code (based on the Fortran version of {\sc Quadpack}), we used the (GNU) \emph{gfortran} compiler
and the Intel Fortran compiler, with the flags \emph{-fopenmp} and \emph{-openmp}, respectively.
{\sc ParInt} and its integrand functions were compiled with \emph{gcc (mpicc)}.
Besides Intel processors, we used POWER7(R) 3.83\,GHz on the KEKSC system A of the Computing Research Center at KEK (SR16000 model M1), with 
the HITACHI Fortran90 compiler that enables automatic parallelization with the flag ~{\it -parallel}.

\section{Feynman loop integrals and extrapolation}
\label{Feynman}
\subsection{General form of Feynman loop integrals}
\label{Feynman-gen}
{\color{black}Higher-order corrections} 
are required for accurate theoretical predictions of the {\color{black}cross section} for particle interactions.
Loop diagrams are taken into account, leading to the evaluation of {\color{black}loop integrals}.
The derivation of a closed analytic form is generally hard or impossible for 
higher-order loop integrals with arbitrary internal masses and external momenta.
Thus we resort to numerical calculations.

A scalar $L$-loop integral with $N$ internal lines can be represented in Feynman parameter
space by
\begin{equation}
{\color{black}\mathcal I\hspace*{-0.4mm} 
= {\mathcal I\hspace*{-0.4mm}}_N = (-1)^N \frac{\Gamma\left(N-\frac{{\color{black}n}L}{2}\right)}{(4\pi)^{{\color{black}n}L/2}}
\int_{0}^{1}\prod_{r=1}^{N}dx_{r}\, \delta(1-\sum x_{r})\,
\frac{1}{U^{n/2}(V-i{\color{black}\varrho})^{N-{\color{black}n}\,L/2}}},
\label{Lloop}
\end{equation}
%
%
where
\begin{equation}
V=M^2-\frac{1}{U}W,\qquad  M^2=\sum m_r^2 x_{r} \nonumber
\label{LloopB}
\end{equation}
and $m_r$ is the mass for the propagator associated with $x_r.$
Here
$U$ and $W$ are polynomials determined by the topology of the corresponding
diagram and physical parameters ($U = 1$ for 1-loop ($L = 1$) integrals), and
$n$ is the space-time dimension.
We further denote
\begin{equation}
{\mathcal I\hspace*{-0.4mm}}_N ~=~ 
 \frac{1}{(4\pi)^{\,n\,L/2}} \,I 
~~=~~ (-1)^N \frac{\Gamma\left(N-\frac{n\,L}{2}\right)}{(4\pi)^{\,n\,L/2}} \,J,
\label{LloopIJ}
\end{equation}
defining $I$ and $J$ as integrals with a factor different from that of ${\mathcal I\hspace*{-0.4mm}}_N,$
in order to draw comparisons with results in the literature.
We sometimes also use the following notation for Feynman parameters,
\begin{equation}
x_{j\,k\cdots}=x_j+x_k+\cdots\,. \nonumber
\label{LloopC}
\end{equation}

The integration in Eq~\eqref{Lloop} is taken over the $N$-dimensional unit cube. 
However, as a result of the $\delta$-function one of the $x_r$ can be expressed in terms 
of the other ones in view of $\sum_{j=1}^N x_j = 1,$
which reduces the integral dimension to $N-1$ and the domain to 
the $d = (N-1)$-dimensional unit simplex
\begin{equation}
{\mathcal S}_d = \{\,(x_1,x_2,\ldots,x_d)~\in~{\mathbb R}^d ~~|~
                      \sum_{j=1}^d x_j \le 1 \mbox{ and } x_j \ge 0\,\}.
\label{simplex}
\end{equation}

When the behavior of a singularity of the integrand is moderate, we can carry out the 
integration within the unit simplex domain without variable transformation. 
For the numerical integration where a steeper singularity appears, 
the unit simplex domain of Eq~\eqref{Lloop} can be transformed
to the $(N-1)$-dimensional unit cube, using
\begin{small}
\begin{align}
& x_1 = \tilde{x}_1 \nonumber\\
& x_2 = (1-x_1)\,\tilde{x}_2 \nonumber\\
& x_3 = (1-x_1-x_2)\,\tilde{x}_3 \label{cubetrans} \\
& \ldots  \nonumber\\
& x_{N-1} = (1-x_1-x_2-\ldots-x_{N-2})\,\tilde{x}_{N-1} \nonumber
\end{align}
\end{small}
with Jacobian ~$(1-x_1)\,(1-x_1-x_2)\ldots(1-x_1-x_2-\ldots-x_{N-2}),$ ~i.e.,
\begin{footnotesize}
\begin{align}
& \int_0^1 dx_1 \int_0^{1-x_1} dx_2 \int_0^{1-x_1-x_2-\ldots-x_{N-2}} \,dx_{N-1} \,f\,(x_1,x_2,\ldots,x_{N-1}) \label{transint} \\
& = \int_0^1 d\tilde{x}_1 ~(1-x_1) \int_0^1 d\tilde{x}_2 ~(1-x_1-x_2) \ldots \int_0^1 d\tilde{x}_{N-1} ~(1-x_1-x_2-\ldots-x_{N-2}) ~~f\,(x_1,(1-x_1)\,\tilde{x}_2,\ldots,(1-x_1-x_2-\ldots-x_{N-2})\,\tilde{x}_{N-1}) \nonumber.
\end{align}
\end{footnotesize}
We find that the approximations thus obtained are often more accurate than those generated with the
multivariate simplex rules in {\sc ParInt} (see Section~\ref{parint}), without the transformation.
Further, for some integrands with severe boundary singularities, 
we use the \emph{double-exponential} transformation by Takahasi and 
Mori~\cite{takahasi74,davis84,sugihara97}, which is 
given in Section~\ref{DE}. We show examples of its application in 
Sections~\ref{3-loop-massive-UV} and~\ref{uv-4ls}.  
Furthermore, we introduce another type of variable transformations related to the topology of Feynman diagrams 
to increase the accuracy of the results for some integrals in Sections~\ref{2ls},~\ref{uv-vertex} and~\ref{3-loop-massive-UV}. The integration domain is mapped to the
unit cube. Unlike the first two transformations, we determine
the latter using a heuristic approach.  
 
Loop integrals are notorious for singularities due to vanishing 
denominators, which may lead to divergence (e.g., IR or UV divergence) of the integral.
In the absence of IR and UV singularities, we have ${\color{black}n = 4}.$ 
For {\color{black}dimensional regularization} in case of IR singularities we set
${\color{black}n = 4+2\varepsilon}$ ~(cf.,~\cite{acat11}), and for UV singularities,
${\color{black}n = 4-2\varepsilon}.$ We apply the regularization by a 
numerical extrapolation as ${\color{black}\varepsilon \rightarrow 0}$
(cf., Section~\ref{extrap-exp}).

The term \,$i\,{\color{black}\hspace*{-0.3mm}\varrho}$\, prevents 
the integrand denominator in Eq~\eqref{Lloop} from vanishing in the interior of the domain, and can be used for regularization.
A regularization to keep the integral from diverging was achieved by extrapolation as
$\varrho \rightarrow 0$ in~\cite{edcpp03,eddacat03,acat07,acat08,ddacat10,cpc11y,jocs11}. 
Results given in~\cite{cpp10} applied iterated
integration with {\sc Quadpack} programs and a double extrapolation with respect to
$\varrho$ and $\varepsilon$ to deal with interior as well as IR singularities.

However, even for finite integrals,
setting $\varepsilon = 0$ or $\varrho = 0$ in the integrand of Eq~\eqref{Lloop} may not yield the
desired accuracy and it may be advantageous to extrapolate as $\varepsilon \rightarrow 0$ 
or $\varrho \rightarrow 0$
(cf., Section~\ref{3ls-massless-finite-integrals}).

\subsection{Numerical extrapolation}
\label{extrap-exp}
For an extrapolation with respect to the dimensional regularization parameter $\varepsilon,$
the integral of Eq~\eqref{Lloop} is evaluated as a sequence of 
${\cal I}(\varepsilon)$ for decreasing $\varepsilon = \varepsilon_\ell,$ which assumes
an asymptotic expansion of the form of Eq~\eqref{asymp} for $\gamma = \varepsilon.$
For example, the $\varphi_k(\varepsilon)$ functions in Eq~\eqref{asymp} may be integer powers of $\varepsilon,$
$\varphi_k(\varepsilon) = \varepsilon^k.$
Then for finite integrals, $\kappa = 0$ in Eq~\eqref{asymp} and the integral is represented by $C_0.$
 
{\color{black}Linear extrapolation} can be applied when the $\varphi_k(\varepsilon)$ functions are known.
In that case, ${\cal I}(\varepsilon)$ is approximated for decreasing values of $\varepsilon = \varepsilon_\ell,$
and Eq~\eqref{asymp} is truncated after $2, 3, \ldots, \nu$ terms to form linear systems
of increasing size in the $C_k$ variables. This is a generalized form of Richardson 
extrapolation~\cite{brezinski80,sidi03}.
If the integral approximation becomes harder with smaller $\varepsilon,$ we can use
slowly decreasing sequences $\{\varepsilon_\ell\},$ such as a geometric sequence with base $1/1.15.$
Another sequence of interest is based on the Bulirsch sequence $\{b_\ell\}: 1,2,3,4,6,8,12,16,24,\ldots$ (see~\cite{bulirsch64}); we employ
$\{1/b_j\}_{j\ge j_0},$ from a starting index $j_0$ in the Bulirsch sequence.
The stability of linear extrapolation using geometric, harmonic and Bulirsch type sequences
was studied by Lyness~\cite{lyness76} with respect to the mesh ratio 
of composite rules.
The condition of the system was found best for geometric and worse for the harmonic sequences,
with the Bulirsch sequence behavior in between.

We resort to non-linear extrapolation when the structure of the asymptotic expansion is not
known. In previous work we have made ample use of the $\epsilon$-algorithm~\cite{shanks55,wynn56,sidi96,sidi03,sidi11},
which can be applied with geometric sequences of $\varepsilon.$
The extrapolation results given in this paper are
achieved with a version of the $\epsilon$-algorithm code from {\sc Quadpack}~\cite{pi83}.
In between calls, the implementation retains the last two lower diagonals of the triangular
extrapolation table. When a new element $\mathcal I(\varepsilon_\ell)$ of the input sequence is supplied,
the algorithm calculates a new lower diagonal, together with an estimate or measure of the \emph{distance}
of each newly computed element from preceding neighboring elements.
With the location of the ``\emph{new}" element in the table relative to $e_0, ~e_1,
~e_2, ~e_3,$ pictured as:~~
\begin{tabular}{ccc}
 & $e_0$ & \\
$e_3$ & $e_1$ & \mbox{\emph{new}}\\
 & $e_2$ & 
\end{tabular}
 ~~~we have that $\mbox{\emph{new}} = e_1+1/(1/(e_1-e_3)+1/(e_2-e_1)-1/(e_1-e_0)),$
and the distance measure for the \emph{new}
element is set to $|e_2-e_1| + |e_1-e_0| + |e_2-\mbox{\emph{new}}|.$
The new lower diagonal element with the smallest value of the distance measure is then returned
as the result for this call to the extrapolation code.
Note that the accuracy of the extrapolated result is generally limited by the accuracy of the
input sequence.

For an extrapolation as $\varrho \rightarrow 0$ in Eq~\eqref{Lloop},
the integral is approximated by a sequence of numerical results for
${\cal I}(\varrho_\ell)$ with decreasing $\varrho_\ell.$ 
An asymptotic expansion of the form of Eq~\eqref{asymp}, 
${\cal I}(\varrho) \sim \sum_{k\,\ge \,\kappa} C_k\,\varphi_k(\varrho)$ 
 for $\varrho = \varrho_\ell$
is assumed, where the $\varphi_k(\varrho)$ functions are generally unknown, and
we perform non-linear extrapolation with the $\epsilon$-algorithm.



\section{Numerical Integration Methods}\label{integration}
Though various integration methods may be applicable in our approach,
we currently use three types of integration methods as presented in subsequent sections.
These are: numerical iterated integration, parallel adaptive integration and double-exponential
transformation methods.
\subsection{Numerical iterated integration}\label{multiter}
For iterated integration over a $d$-dimensional
product region we express Eq~\eqref{blackbox} as
\begin{equation}
{\mathcal I\hspace*{-0.4mm}}f = \int_{\alpha_1}^{\beta_1} dx_1 \int_{\alpha_2}^{\beta_2} dx_2 \ldots
     \int_{\alpha_d}^{\beta_d} dx_d ~f(x_1,x_2,\ldots,x_d),
\label{iterint}
\end{equation}
where the limits of integration are given by functions $\alpha_j =
\alpha_j\,(x_1,x_2,\ldots,x_{j-1})$ and $\beta_j = \beta_j\,(x_1,x_2,\ldots,x_{j-1}).$
In particular, the boundaries of the $d$-dimensional unit simplex ${\mathcal S}_d$
given by Eq~\eqref{simplex}
are $\alpha_j = 0$ and $\beta_j = 1-\sum_{k=1}^{j-1} x_k.$

For the numerical integration over the interval $[\alpha_j,\beta_j], ~1 \le j \le d$ in Eq~\eqref{iterint} we can apply, e.g., the
1D adaptive integration code {\sc Dqage} from
the {\sc Quadpack} package~\cite{pi83} in each coordinate direction, 
and select the $(K = 15)$-point Gauss-Kronrod rule pair via an input parameter,
for the integral (and error) approximation on each subinterval.
If an interval $[a,b]$ arises in the partitioning of \,$[\alpha_j,\beta_j],$
then the local integral approximation over $[a,b]$ is of the form
\begin{equation}
\int_a^b dx_j ~{\mathcal F}(c_1,\ldots,c_{j-1},x_j) ~\approx~ \sum_{k=1}^K w_k \,{\mathcal F}(c_1,\ldots,c_{j-1},x^{(k)}),
\label{localab}
\end{equation}
where the $w_k$ and $x^{(k)}, 1 \le k \le K,$ are the weights and abscissae of the local rule
scaled to the interval $[a,b]$ and applied in the $x_j$-direction. For $j = 1$ this is the
outer integration direction.
The function evaluation
\begin{equation}
{\mathcal F}(c_1,\ldots,c_{j-1},x^{(k)}) = \int_{\alpha_{j+1}}^{\beta_{j+1}} dx_{j+1} \ldots
           \int_{\alpha_d}^{\beta_d} dx_d ~f\,(c_1,\ldots,c_{j-1},x^{(k)},x_{j+1},\ldots,x_d) , ~~~~~1 \le k \le K,
\label{innerintegral}
\end{equation}
is itself an integral in the $x_{j+1},\ldots,x_d$\,-\,directions for $1 \le j < d,$ and is computed by the
method(s) for the inner integrations.
For $j = d,$ Eq~\eqref{innerintegral} is the evaluation of the integrand function
\begin{equation}
{\mathcal F}(c_1,\ldots,c_{d-1},x^{(k)}) = f\,(c_1,\ldots,c_{d-1},x^{(k)}). \nonumber
\end{equation}

Note that successive coordinate directions may be combined into layers in the iterated
integration scheme.
Furthermore, the error incurred in any inner integration
will contribute to the integration error in all of its subsequent
outer integrations~\cite{fritsch81,kahaner88,iccs10}.

Since the ${\mathcal F}(\,)$ evaluations on the right of Eq~\eqref{localab} are independent of one another
they can be evaluated in parallel.
Important benefits of this approach include that:\\
\emph{(i)} the granularity of the parallel integration is large, especially when the inner
integrals ${\mathcal F(\,)}$ are of dimension $\ge 2;$\\
\emph{(ii)} the points where the function ${\mathcal F}$ is evaluated in parallel
are the same as those of the sequential evaluation; i.e.,
apart from the order of the summation in Eq~\eqref{localab},
the parallel calculation is essentially the same as the sequential one.
This important property facilitates the debugging of parallel code.
As another characteristic, the parallelization does not increase the total amount of computational work.

In addition, the memory required for the procedure is determined by (the sum of)
the amounts of memory needed for the data pertaining to the subintervals incurred in each
coordinate direction (corresponding to the length of the recursion stack for a recursive implementation).
Consequently the total memory increases linearly as a function of the dimension $d.$
Note that successive coordinate directions may be combined into layers in the iterated
integration scheme.

To achieve the multi-threading, OpenMP~\cite{openmp} compiler directives were inserted in the iterated
integration code. For the Fortran version of {\sc Quadpack} we used the (GNU) \emph{gfortran} compiler
and the Intel Fortran compiler, with the flags -\emph{fopenmp} and -\emph{openmp}, respectively.

\subsection{{\sc ParInt} package}\label{parint}
Written in C and layered over MPI~\cite{openmpi}, the {\sc ParInt} methods
(parallel adaptive, quasi-Monte Carlo and Monte Carlo)
are implemented as tools for \emph{automatic}
integration, where the user defines the integrand function and the domain, and specifies
a relative and absolute error tolerance for the computation ($t_r$ and $t_a,$ respectively).
For {\sc Parint} the integrand is generally defined as a vector function 
with $m$ components,
\begin{equation}
\vec{f}: {\mathcal D}\subset {\mathbb R}^d \rightarrow {\mathbb R}^m,
\label{function}
\end{equation}
over a (finite) $d$-dimensional (hyper-rectangular or simplex) domain $\mathcal D.$
Denoting the exact integral by 
\begin{equation}
{\mathcal I\hspace*{-0.4mm}}\vec{f} = \int_{\mathcal D} \vec{f}\,(\vec{x})~d\vec{x},
\label{general}
\end{equation}
then the objective of Eq~\eqref{accuracy} is generalized to
returning an approximation ${\mathcal Q}\vec{f}$ and absolute error estimate $E_a\vec{f}$ such that
\begin{equation}
||\, {\mathcal Q}\vec{f}-{\mathcal I\hspace*{-0.4mm}}\vec{f} \,|| \le ||\, E_a\vec{f} \,|| \le \max\{\, t_a, \,t_r\, ||\, {\mathcal I\hspace*{-0.4mm}}\vec{f} \,|| \,\}
\label{acc}
\end{equation}
(in infinity norm).
In order to satisfy the error criterion of Eq~\eqref{acc} the program tests throughout whether
$$||\, E_a\vec{f} \,|| \le \max\,\{\, t_a, \,t_r\, ||\, {\mathcal Q}\vec{f} \,|| \,\}$$
is achieved.
We used the vector function integration capability in~\cite{ccp14} for a simultaneous 
computation of the entire entry sequence for extrapolation, obtained as the $m$ components 
of the integral ${\mathcal I\hspace*{-0.4mm}}\vec{f}.$


The available
cubature rules in {\sc ParInt} (to compute the integral approximation over the domain or
its subregions) include a set of rules for the $d$-dimensional cube~\cite{ge80,ge83,gnzbe91a},
the 1D (Gauss-Kronrod) rules used in {\sc Quadpack} and a set of rules for the $d$-dimensional simplex~\cite{gnz90,gm78,ddmathcomp79}.
Some results in this paper are computed over the $d$-dimensional simplex using
iterated integration with Gauss-Kronrod rules. 
In other cases, multivariate rules of polynomial degree 7 or 9 are used
over the $d$-dimensional unit cube.
A formula is said to be of a particular polynomial degree $k$
if it renders the exact value of the integral for integrands
that are polynomials of degree $\le k,$ and there are polynomials of degree $k+1$ for which the
formula is not exact.
The number of function evaluations per (sub)region is constant, and the total number of
subregions generated, or the number of function evaluations in the course of the integration,
is considered a measure of the computational effort.

\subsubsection{{\sc ParInt} adaptive methods}
\label{sect:parint-adaptive}
In the adaptive approach, the integration domain is divided initially among the workers.  Each on its own part of the
domain, the workers engage in an adaptive partitioning strategy similar to that of {\sc Dqage}
from {\sc Quadpack}~\cite{pi83} and of {\sc Dcuhre}~\cite{gnzbe91bn} by successive bisections.
The workers then each generate a local priority queue as a task pool of subregions.
\begin{wrapfigure}[9]{r}{2.1in}
\begin{small}
\vspace*{-7mm}
\begin{tabbing}
88\=This\= is the adaptive meta-algorit\=hm. \kill
\> Evaluate initial region and update results \\
\> Initialize priority queue with initial region \\
\> {\bf while} (eval. limit not reached and estim. err. $>$ tolerance) \\
\> \>    Retrieve region from priority queue \\
\> \>    Split region \\
\> \>    Evaluate new subregions and update results~~~~ \\
\> \>    Insert new subregions into priority queue
\end{tabbing}
\end{small}
\vspace*{-4mm}
\caption{Adaptive integration meta-algorithm}
\label{meta-alg}
\end{wrapfigure}
\indent
The priority queue is implemented as a max-heap 
keyed with the estimated integration errors over the subregions,
so that the subregion with the largest estimated error is stored in the root of the heap.
If the user specifies a maximum size for the heap structure on the worker,
the task pool is stored as a \emph{deap} or \emph{double-ended heap}, which allows
deleting of the maximum as well as the minimum element efficiently, in order to maintain a constant size
of the data structure once it reaches its maximum.

A task consists of the selection of the associated subregion and its subdivision
(generating two children regions), integration over the children, deletion of the parent region
(root of the heap) and insertion of the children into the heap (see Figure~\ref{meta-alg}).
%
The bisection of a region is performed perpendicularly to the coordinate direction in which the
integrand is found to vary the most, according to $4^{th}$-order differences computed in each
direction~\cite{gnzbe91bn}.
The subdivision procedure continues until 
the global error estimate falls below the tolerated error, or
the total number of function evaluations exceeds the user-specified maximum.

\subsubsection{Load balancing}
\label{sect:loadbalancing}
For a regular integrand behavior and $p$ MPI processes distributed evenly over homogeneous
processors, the computational load would ideally decrease by a factor of about $p.$
Otherwise it may be possible to improve the parallel time (and space) usage by load balancing,
to attempt keeping the loads on the worker task pools balanced.

The receiver-initiated, scheduler based load balancing strategy in {\sc ParInt} is
an important mechanism of the distributed integration algorithm~\cite{DGE96,cpp01,akdpdcs03,adkvpdcs04}.
The message passing is performed in a non-blocking and asynchronous manner, and permits overlapping of
computation and communication,
which benefits {\sc ParInt}'s efficiency
on a hybrid platform (multi-core and distributed) where multiple processes are assigned to each node.
As a result of the asynchronous processing and message passing on MPI, {\sc ParInt}
executes on a hybrid platform by assigning multiple processes across the nodes.
The user has the option of turning load balancing on or off,
as well as allowing or dis-allowing the controller to also act as a worker.

\subsubsection{Use of {\sc ParInt}}
\label{use}
{\sc ParInt} can be invoked from the command line, or by calling the {\verb pi_integrate() } function
in a program for computing an integral of the form of Eq~\eqref{general}. A user guide is provided in~\cite{parintweb}.
The call sequence passes a pointer to the integrand function, typed as a pointer to
a function that returns an integer, and where the parameters of the integrand function
correspond to the integral dimension, argument vector $\vec{x},$ number of component functions
{\verb nfuncs } (corresponding to $m$ in Eq~\eqref{function}) and the resulting 
component values of the function $\vec{f}\,(x).$
Apart from {\verb nfuncs }, further input parameters of {\verb pi_integrate() } are: an integer
identifying the cubature/quadrature rule to use, the
maximum number of function evaluations allowed, the region type (hyper-rectangle
or simplex) and specification. The output parameters are: the integral and error component approximations
{\verb result[] } and {\verb error[] }, and a user-declared pointer to a status structure.
The execution time is returned as part of the output
printed by the {\sc ParInt} \emph{pi\_print\_results()} function.

When {\sc ParInt} is used as a stand-alone executable, it uses the {\sc ParInt} Plug-in Library
(PPL) mechanism to specify integrand functions. The functions are written by the user, added to the
library (along with related attributes), and then compiled using a {\sc ParInt}-supplied compiler into \emph{plug-in
modules} (.ppl files). A single PPL file is loaded at runtime by the {\sc ParInt} executable.
Using a function library enables quick access to a predefined set of functions and lets {\sc ParInt}
users add and remove integrand functions dynamically without re-compiling the {\sc ParInt} binary.
Once these functions are stored in the library, they can be selected by name for integration.

For an execution on MPI, the MPI host file ({\verb myhostfile }) contains lines of the
form: {\verb"node_name slots=ppn"} where {\verb"ppn"} is the number of processes to be used on each
participating node.
A typical MPI run from the command line may be of the form\\
\begin{small}
{\verb"mpirun -np 64 --hostfile myhostfile ./parint -f fcn -lf 10000000 "}
{\verb"-ea 0.0 -er 5.0e-10"}\\
\end{small}
For example, with four  nodes listed in ~{\verb myhostfile }~ and ~{\verb"ppn = 16"},~
a total of 64 processes is requested on the specified nodes.
The integrand function of this run is named {\verb fcn } in the user's library; the maximum number of function
evaluations is 10000000, and the absolute and relative error tolerances are 0 and 5.0e-10, respectively.

Optionally the {\sc ParInt} installation can be configured to use long doubles instead of doubles.

\subsection{Double-exponential transformation}\label{DE}
The \emph{Double Exponential (tanh-sinh) formula}, referred to here as \emph{DE formula} in short, was proposed by Takahasi and Mori in 1974~\cite{takahasi74,davis84,sugihara97}.
It is an efficient method for the numerical approximation of an integral whose integrand is a holomorphic function
with end-point singularities. This formula transforms the integration variable in $\int_{0}^{1} f(x)\,dx$ to~
$x=\phi\,(t)=\frac{1}{2}(\rm{tanh\,(\frac{\pi}{2}\rm{sinh\,(t)})+1}).$
Then ~${\mathcal I\hspace*{-0.4mm}}=\int_{0}^{1} f(x) \,dx =\int_{-\infty}^{\infty} f\,(\phi(t))\,\phi'(t)\,dt$ ~with~
$\phi'(t)=\frac{\pi\,\rm{cosh\,(t)}}{4\,\rm{cosh^2(\frac{\pi}{2}\rm{sinh\,(t))}}}.$
After the transformation, the trapezoidal rule is applied leading to
\begin{equation}
I_{h}^{N_{eval}}=\sum_{k=-N_{-}}^{k=N_{+}}f\,(\phi\,(kh))\,\phi'(kh),
\label{DEsum}
\end{equation}
with mesh size $h$ and $N_{eval}=N_{-}+N_{+}+1$ function evaluations.
A major issue in numerical integration with the DE formula is the treatment of overflows 
at large $|t|$ and for
large $|N_{eval}|,$ and it is sometimes necessary to evaluate the integrand using multi-precision arithmetic even though it takes more CPU time than double precision.
This helps alleviating the loss of trailing digits in the evaluation of the integrand near the end-points.

For multi-dimensional loop integrals, we use the DE formula in a repeated integration scheme.
Apart from our sequential implementation we also developed code for multi-core systems using a parallel
library such as OpenMP or a compiler with auto-parallelization capabilities.
For an execution in a multi-precision environment, a dedicated accelerator system
consisting of multiple FPGA (Field Programmable Gate Array) boards was 
developed and its performance results were presented by Daisaka et al.~\cite{daisaka15}.

\section{2-loop integrals with massive internal lines}
\label{2-loop-integral}

In this section we calculate the integral $I$ of Eq~\eqref{LloopIJ} for $L=2$ and $n=4-2\varepsilon$
according to
\begin{eqnarray}
\label{UV}
I &=& (-1)^N {\Gamma\left(N-4+2\varepsilon \right)}
\int_{0}^{1}\prod_{r=1}^{N}dx_{r}\, \delta(1-\sum x_{r})\frac{1}{U^{2-\varepsilon}(V-i\varrho)^{N-4+2\varepsilon}}\\ 
  &=& (-1)^N {\Gamma\left(N-4+2\varepsilon \right)}
\int_{{\mathcal S}_{N-1}} \frac{1}{U^{2-\varepsilon}(V-i\varrho)^{N-4+2\varepsilon}}.\nonumber
\end{eqnarray}
IR divergence occurs through a singularity arising when $V$ 
vanishes at the boundaries of the domain. This problem can be addressed by dimensional regularization
with ${\color{black}n = 4+2\varepsilon},$ which we implemented numerically in~\cite{cpp10,acat11,jocs11}   
using an extrapolation as $\varepsilon\rightarrow 0$ ($\varepsilon > 0$). 
It is assumed that the denominator does not vanish in the interior of the integration domain,
so we can set $\varrho = 0.$

In this paper, we concentrate on UV divergence, which occurs when $U$ vanishes at the boundaries. The $\Gamma$-function in Eq~\eqref{UV} contributes to UV divergence when $N\le 4$.
We treat UV divergence by
a dimensional regularization with ${\color{black}n = 4-2\varepsilon},$ implemented by a numerical
extrapolation as $\varepsilon\rightarrow 0$ after either iterated integration 
with {\sc Dqags} from {\sc Quadpack}, 
multivariate adaptive integration with {\sc ParInt} or the DE formula.

\subsection{2-loop self-energy integrals}\label{2ls}
\begin{figure}
\begin{center}
\begin{subfigure}[ ]
\centering
\includegraphics[width=0.22\linewidth]{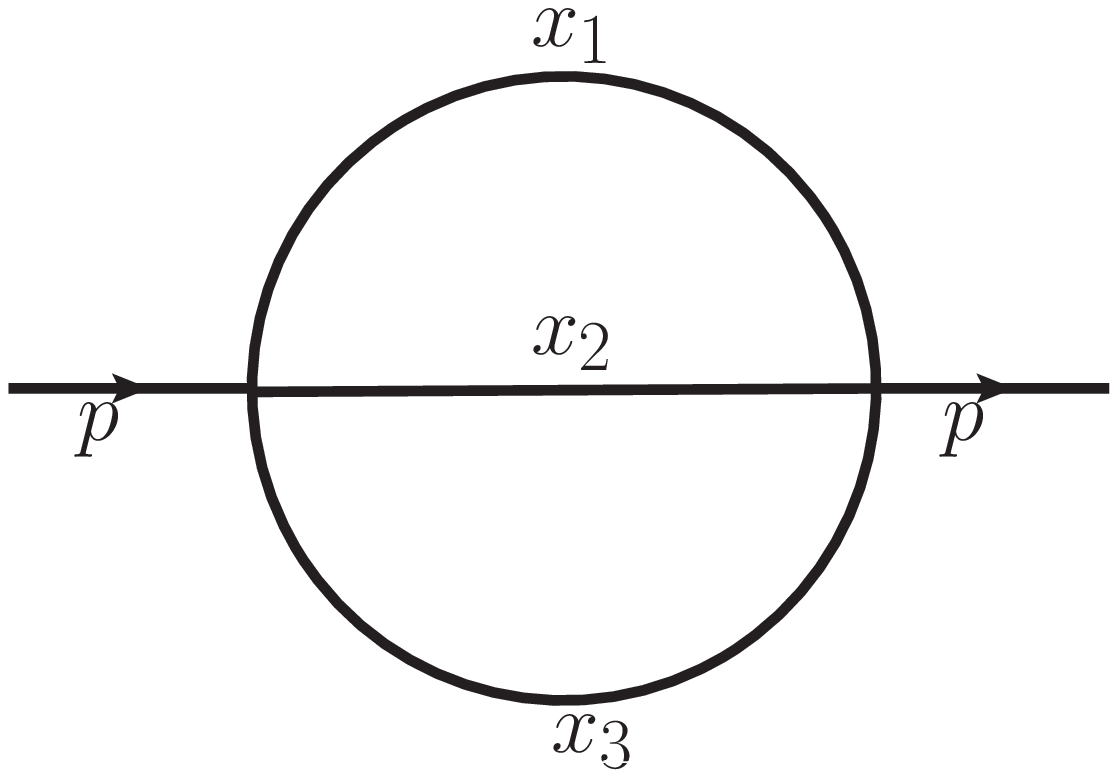}
\end{subfigure}
\begin{subfigure}[ ]
\centering
\includegraphics[width=0.22\linewidth]{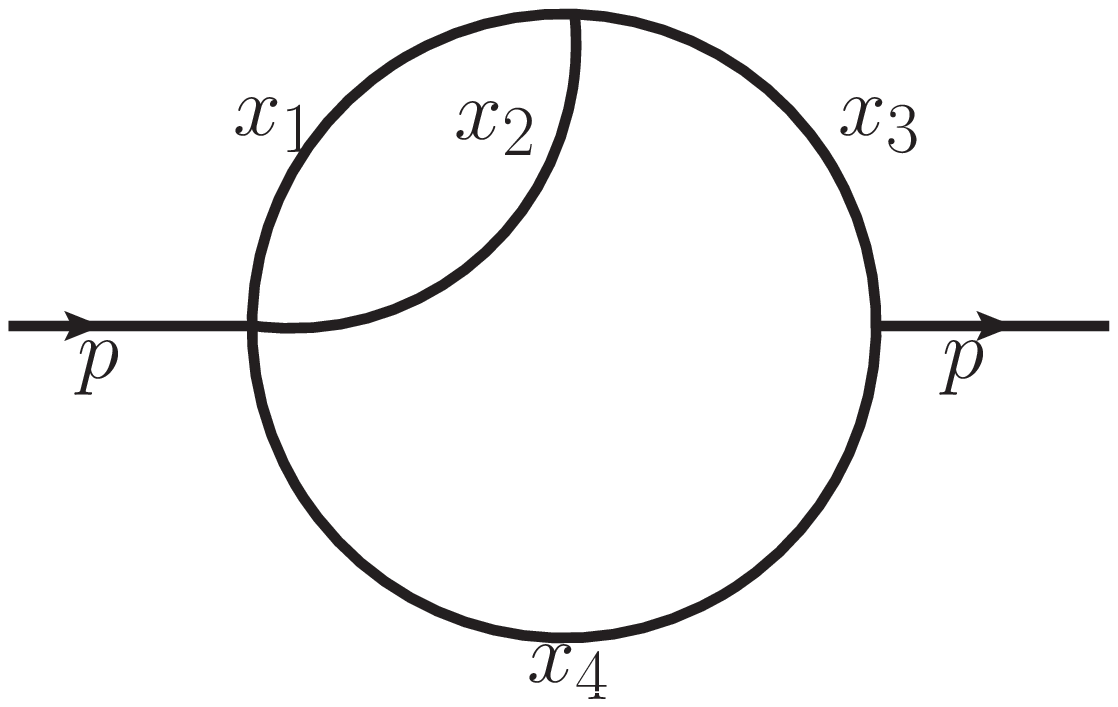}
\end{subfigure}
\begin{subfigure}[ ]
\centering
\includegraphics[width=0.22\linewidth]{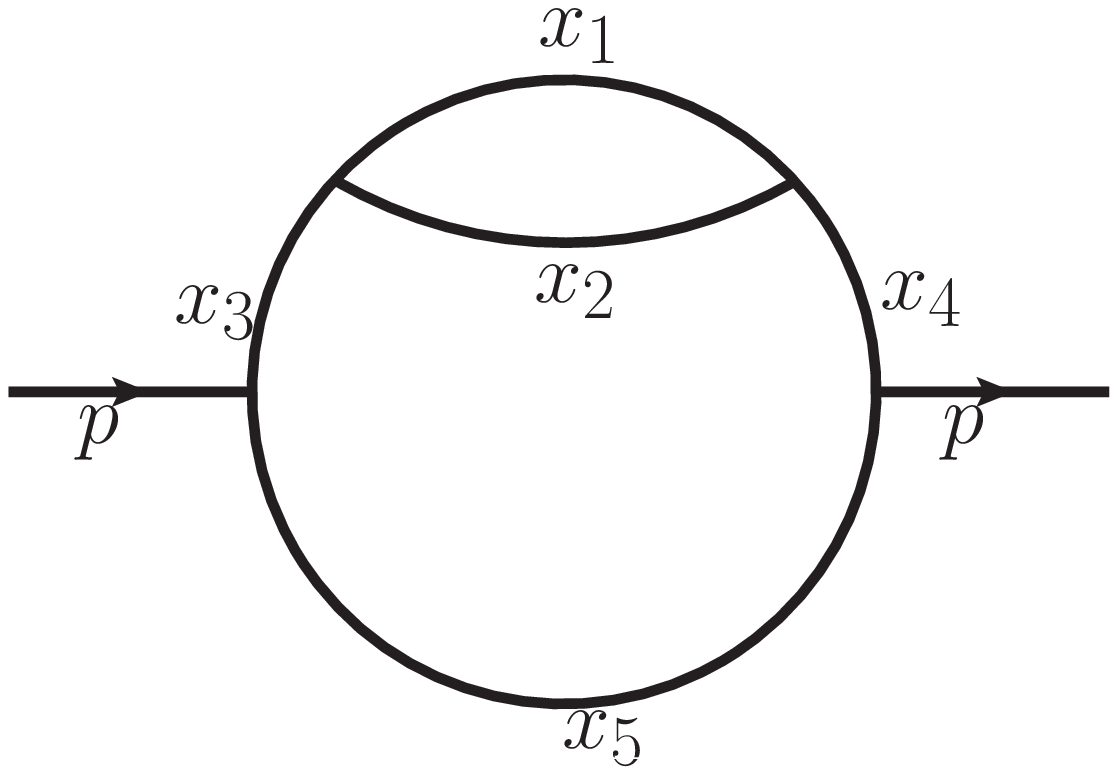}
\end{subfigure}
\vspace*{0.2in}
\begin{subfigure}[ ]
\centering
\includegraphics[width=0.22\linewidth]{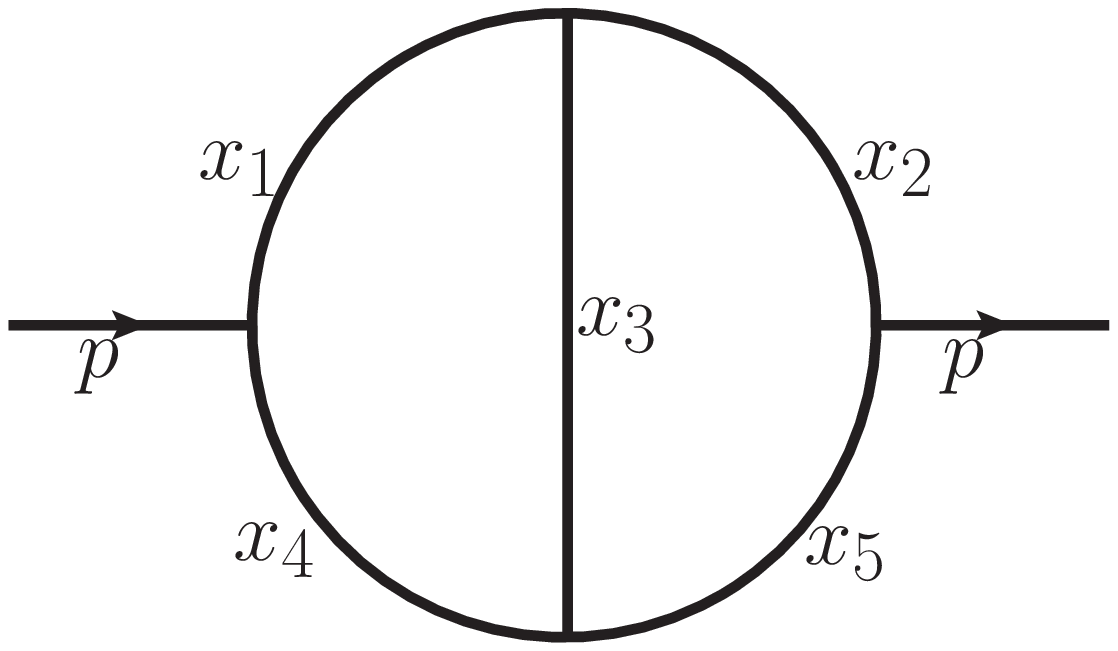}
\end{subfigure}
\caption{2-loop self-energy diagrams with massive internal lines:
 (a)~2-loop \emph{sunrise-sunset} $N = 3$ (Laporta~\cite{laporta01}\,, Fig 2(b)), 
 (b)~2-loop \emph{lemon} $N = 4$ (Laporta~\cite{laporta01}\,, Fig 2(c)), 
 (c)~2-loop \emph{half-boiled egg} $N = 5$, 
 (d)~2-loop \emph{Magdeburg} $N = 5$ ((Laporta~\cite{laporta01}\,, Fig 2(d))}.
\label{2-loop-UV-diagrams}
\end{center}
\end{figure}

Fig~\ref{2-loop-UV-diagrams} depicts 2-loop self-energy diagrams with $N = 3, \,4$ and 5
internal lines. We refer to the 2-loop self-energy diagrams (a-d) as the 
\emph{sunrise-sunset}, \emph{lemon}, \emph{half-boiled egg} and \emph{Magdeburg} diagrams, and to
the corresponding integrals as $I_a^{S2}$, $I_b^{S2}$, $I_c^{S2}$ and $I_d^{S2}$, respectively.
As show in in Fig~\ref{2-loop-UV-diagrams}, the entering momentum is $p,$ and we
denote $s=p^2$.

Analytic results for the 
integrals have been derived by many authors. We use the following formulas
for the functions $U$ and $V$ in Eq~\eqref{UV}:\\
$-$ for the \emph{\color{black}sunrise-sunset diagram}~(Fig~\ref{2-loop-UV-diagrams}(a)), 
$U = x_1 x_2 + x_2 x_3 + x_3 x_1, ~W/s = x_1 x_2 x_3;$ \\
$-$ for the \emph{\color{black}lemon diagram}~(Fig~\ref{2-loop-UV-diagrams}(b)), we have 
$U = x_{12} x_{34} + x_1 x_2, ~W/s = x_4\,(x_1 x_2 + x_2 x_3 + x_3 x_1);$ \\
$-$ for the \emph{\color{black}half-boiled egg diagram}~(Fig~\ref{2-loop-UV-diagrams}(c)), 
$U = x_{12} x_{345} + x_1 x_2, ~W/s = x_5\,(x_{12} x_{34} + x_1 x_2);$\\
$-$ for the \emph{Magdeburg diagram}~(Fig~\ref{2-loop-UV-diagrams}(d)), 
$U = x_{14} x_{25} + x_3 x_{1245}, ~W/s = x_{1}x_{4}x_{235} + x_{2}x_{5}x_{134} + x_3\,(x_1x_5 + x_2x_4).$\\

In the numerical evaluation, 
we take particular values for energy and masses, i.e., $s = p^2 = 1,$  and all masses $m_r=1$
in order to make comparisons with results in the references. 

The integrals are expanded with respect to the dimensional regularization parameter $\varepsilon.$
The integrals $I_a^{S2}, I_b^{S2}$ are divergent as $\displaystyle{1/\varepsilon^2},$ which is
the product of $\displaystyle{1/\varepsilon}$ from the $\Gamma$-function part and
 $\displaystyle{1/\varepsilon}$ from the integral part.
The integral $I_c^{S2}$ is divergent as $\displaystyle{1/\varepsilon},$ which 
comes from the integral part. The integral $I_d^{S2}$ is finite.
The expansions are of the form of Eq~\eqref{asymp}, 
\begin{equation}
{\color{black}S(\varepsilon) \sim \sum_{k\ge \kappa} C_k\,\varepsilon^k~~~~~~\mbox{as~~} \varepsilon\rightarrow 0},
\label{expand}
\end{equation}
and we use linear extrapolation to approximate the coefficients of the leading terms.
We multiply the integrals $I_{a}^{S2},$ $I_{b}^{S2}$ and $I_{d}^{S2}$ of Eq~\eqref{UV}
with the factor $ \Gamma(1+\varepsilon)^{-2}$ for comparison with the results
in Laporta~\cite{laporta01}, since the latter are computed with this factor. 
The \emph{\color{black}half-boiled egg diagram} is not covered in~\cite{laporta01}. 
We give the analytic formula for $J_c^{S2}$ in Appendix A of this paper.
\begin{eqnarray}
I_{a}^{S2}({\varepsilon}) ~\Gamma(1+{\varepsilon})^{-2} = \sum_{k\ge -2} C_k \,{\varepsilon^k} &=& -1.5\,{\varepsilon^{-2}}-4.25\,{\varepsilon^{-1}}-7.375-17.22197253479\,{\varepsilon}\ldots \label{Isexp}\\
\nonumber
I_{b}^{S2}({\varepsilon}) ~\Gamma(1+{\varepsilon})^{-2} = \sum_{k\ge -2} C_k \,{\varepsilon^k} &=& 0.5\,{\varepsilon^{-2}}+0.6862006357658\,{\varepsilon^{-1}}-0.6868398873414\\ 
&+&1.486398391913\,{\varepsilon}\ldots \label{Ilexp}\\
\nonumber
\\
\nonumber
J_{c}^{S2}({\varepsilon}) = \sum_{k\ge -1} C_k \,{\varepsilon^k} &=& 0.6045997880781\,{\varepsilon^{-1}}-0.1756970002260\,-0.2977242542666\,{\varepsilon}\\ 
&+&0.4140155361099\,{\varepsilon^2}\ldots \label{Ihexp}\\
\nonumber \\
\nonumber
I_{d}^{S2}({\varepsilon}) ~\Gamma(1+{\varepsilon})^{-2} = \sum_{k\ge 0} C_k \,{\varepsilon^k} &=& 0.9236318265199-1.284921671848\,{\varepsilon}+2.689507626490\,{\varepsilon^2}\\
&-&5.338399227511\,{\varepsilon^3}\ldots \label{Imexp}
\end{eqnarray}
Note that the value of $\kappa$ in Eq~\eqref{expand} corresponds with the index of the first coefficient 
$C_\kappa$ in the expansion. In that case we find that, if $\kappa$ is replaced by 
$\kappa-1$ for the extrapolation, then the first coefficient converges to $C_{\kappa-1} = 0.$

\begin{table}
\caption{\footnotesize{ Results UV \emph{sunrise-sunset} integral, $I_{a}^{S2} \,\Gamma(1+\varepsilon)^{-2}$ \,(on Mac Pro),~
rel. err. tol. $t_r = 10^{-9}$ (outer), $10^{-9}$ (inner), $T[s]$ = Time (elapsed user time in $s$),~ $\varepsilon = 1.2^{-10-\ell}$ (starting at $1.2^{-10}$),~~$E_r = $ outer integration estim. rel. error}}
\begin{scriptsize}
\begin{center}
\begin{tabular}{ccccccc}\hline
& \multicolumn{6}{l}{{\sc Integral ~$I_{a}^{S2} \,\Gamma(1+\varepsilon)^{-2}$ ~Extrapolation}} \\
 \hspace*{-1mm}$\ell$\hspace*{-1mm} & \hspace{-1mm}{ $E_r$ }\hspace{-1mm} & \hspace{-1mm}{\sc T[s]}
\hspace{-1mm} & {\sc Res.} ~$C_{-2}$ & \hspace{-1mm}{\sc Res.} ~$C_{-1}$\hspace{-1mm} & \hspace*{-0mm}{\sc Res.} ~$C_0$ & \hspace{-1mm}{\sc Res.} ~$C_1$ \\
\hline
 \hspace*{-1mm}$0$\hspace*{-1mm}& \hspace{-1mm}3.2e-10\hspace{-1mm} & \hspace{-1mm}0.015\hspace{-1mm} & \hspace{-1mm}\hspace{-1mm} & \hspace{-1mm}\hspace{-1mm} & \hspace*{-1.5mm}\hspace*{-1mm} & \hspace{-1mm}\hspace{-1mm} \\
 \hspace*{-1mm}$1$\hspace*{-1mm}& \hspace{-1mm}4.7e-10\hspace{-1mm} & \hspace{-1mm}0.013\hspace{-1mm} & \hspace{-1mm}-1.156740414\hspace{-1mm} & \hspace{-1mm}-8.2070492\hspace{-1mm} & \hspace*{-1.5mm}\hspace*{-1mm} & \hspace{-1mm}\hspace{-1mm} \\
 \hspace*{-1mm}$2$\hspace*{-1mm}& \hspace{-1mm}6.6e-10\hspace{-1mm} & \hspace{-1mm}0.013\hspace{-1mm} & \hspace{-1mm} -1.603088981\hspace{-1mm} & \hspace{-1mm}-2.1269693\hspace{-1mm} & \hspace*{-1.5mm}-20.5343\hspace*{-1mm} & \hspace{-1mm}\hspace{-1mm}  \\
 \hspace*{-1mm}$3$\hspace*{-1mm}& \hspace{-1mm}9.2e-10\hspace{-1mm} & \hspace{-1mm}0.013\hspace{-1mm} & \hspace{-1mm}-1.476861131\hspace{-1mm} & \hspace{-1mm}-4.9718825\hspace{-1mm} & \hspace*{-1.5mm}0.60362\hspace*{-1mm} & \hspace{-3mm}-51.7769\hspace{-1mm}  \\
 \hspace*{-1mm}$4$\hspace*{-1mm}& \hspace{-1mm}4.0e-11\hspace{-1mm} & \hspace{-1mm}0.028\hspace{-1mm} & \hspace{-1mm}-1.504342324\hspace{-1mm} & \hspace{-1mm}-4.0584835\hspace{-1mm} & \hspace*{-1.5mm}-10.6252\hspace*{-1mm} & \hspace{-3mm}\,8.73336\hspace{-1mm}  \\
 \hspace*{-1mm}$5$\hspace*{-1mm}& \hspace{-1mm}6.9e-11\hspace{-1mm} & \hspace{-1mm}0.029\hspace{-1mm} & \hspace{-1mm}-1.499360223\hspace{-1mm} & \hspace{-1mm}-4.2880409\hspace{-1mm} & \hspace*{-1.5mm}-6.46343\hspace*{-1mm} & \hspace{-3mm}-28.3731\hspace{-1mm}  \\
 \hspace*{-1mm}$6$\hspace*{-1mm}& \hspace{-1mm}1.3e-10\hspace{-1mm} & \hspace{-1mm}0.029\hspace{-1mm} & \hspace{-1mm}-1.500078550\hspace{-1mm} & \hspace{-1mm}-4.2438757\hspace{-1mm} & \hspace*{-1.5mm}-7.57341\hspace*{-1mm} & \hspace{-3mm}-13.7781\hspace{-1mm}  \\
 \hspace*{-1mm}$7$\hspace*{-1mm}& \hspace{-1mm}2.2e-10\hspace{-1mm} & \hspace{-1mm}0.028\hspace{-1mm} & \hspace{-1mm}-1.499992106\hspace{-1mm} & \hspace{-1mm}-4.2507887\hspace{-1mm} & \hspace*{-1.5mm}-7.34157\hspace*{-1mm} & \hspace{-3mm}-18.0041\hspace{-1mm}  \\
 \hspace*{-1mm}$8$\hspace*{-1mm}& \hspace{-1mm}3.6e-10\hspace{-1mm} & \hspace{-1mm}0.028\hspace{-1mm} & \hspace{-1mm}-1.500000763\hspace{-1mm} & \hspace{-1mm}-4.2499043\hspace{-1mm} & \hspace*{-1.5mm}-7.38015\hspace*{-1mm} & \hspace{-3mm}-17.0657\hspace{-1mm}  \\
 \hspace*{-1mm}$9$\hspace*{-1mm}& \hspace{-1mm}5.4e-10\hspace{-1mm} & \hspace{-1mm}0.029\hspace{-1mm} & \hspace{-1mm} -1.499999886\hspace{-1mm} & \hspace{-1mm}-4.2500174\hspace{-1mm} & \hspace*{-1.5mm}-7.37374\hspace*{-1mm} & \hspace{-3mm}-17.2650\hspace{-1mm}  \\
 \hspace*{-1mm}$10$\hspace*{-1mm}& \hspace{-1mm}7.8e-10\hspace{-1mm} & \hspace{-1mm}0.029\hspace{-1mm} & \hspace{-1mm}-1.500000026\hspace{-1mm} & \hspace{-1mm}-4.2499948\hspace{-1mm} & \hspace*{-1.5mm}-7.37544\hspace*{-1mm} & \hspace{-3mm}-17.2010\hspace{-1mm}  \\
\hline
\multicolumn{3}{r}{Eq~\eqref{Isexp}:} & -1.5 & -4.25 &\hspace*{-1.0mm} -7.375 & \hspace{-3mm}~-17.2220\\ \hline
\end{tabular}
\end{center}
\end{scriptsize}
\label{table1}
\end{table}

\begin{table}
\caption{\footnotesize{ {\color{black}Results UV \emph{lemon} integral}, $I_{b}^{S2} \,\Gamma(1+\varepsilon)^{-2}$ \,(on Mac Pro),~
rel. err. tol. $t_r = 10^{-10}$ (outer), $5\times 10^{-11}$ (inner two), $T[s]$ = Time (elapsed user time in $s$),~ $\varepsilon = 1/b_\ell$ (starting at 1/4),~~$E_r = $ outer integration estim. rel. error}}
\begin{scriptsize}
\begin{center}
\begin{tabular}{ccccccc}\hline
& \multicolumn{6}{l}{{\sc Integral ~$I_{b}^{S2} \,\Gamma(1+\varepsilon)^{-2}$~ Extrapolation}} \\
 \hspace*{-1mm}$b_\ell$\hspace*{-1mm} & \hspace{-1mm}{ $E_r$ }\hspace{-1mm} & \hspace{-1mm}{\sc T[s]}
\hspace{-1mm} & {\sc Res.} ~$C_{-2}$ & \hspace{-1mm}{\sc Res.} ~$C_{-1}$\hspace{-1mm} & \hspace*{-0mm}{\sc Res.} ~$C_0$ & \hspace{-1mm}{\sc Res.} ~$C_1$\hspace{-1mm} \\
\hline
 \hspace*{-1mm}$4$\hspace*{-1mm}& \hspace{-1mm}3.5e-11\hspace{-1mm} & \hspace{-1mm}0.36\hspace{-1mm} & \hspace{-1mm}\hspace{-1mm} & \hspace{-1mm}\hspace{-1mm} & \hspace*{-1.5mm}\hspace*{-1mm} & \hspace{-1mm}\hspace{-1mm} \\
 \hspace*{-1mm}$6$\hspace*{-1mm}& \hspace{-1mm}8.8e-11\hspace{-1mm} & \hspace{-1mm}0.34\hspace{-1mm} & \hspace{-1mm}0.5130221162587\hspace{-1mm} & \hspace{-1mm}0.52467607220\hspace{-1mm} & \hspace*{-1.5mm}\hspace*{-1mm} & \hspace{-1mm}\hspace{-1mm}  \\
 \hspace*{-1mm}$8$\hspace*{-1mm}& \hspace{-1mm}2.9e-12\hspace{-1mm} & \hspace{-1mm}0.40\hspace{-1mm} & \hspace{-1mm}0.5031467341833\hspace{-1mm} & \hspace{-1mm}0.62342989295\hspace{-1mm} & \hspace*{-1.5mm}-0.237009170\hspace*{-1mm} & \hspace{-1mm}\hspace{-1mm}  \\
 \hspace*{-1mm}$12$\hspace*{-1mm}& \hspace{-1mm}3.4e-12\hspace{-1mm} & \hspace{-1mm}0.41\hspace{-1mm} & \hspace{-1mm}0.5004379328119\hspace{-1mm} & \hspace{-1mm}0.67218831764\hspace{-1mm} & \hspace*{-1.5mm}-0.518724512\hspace*{-1mm} & \hspace{-1mm}0.52008986\hspace{-1mm}  \\
 \hspace*{-1mm}$16$\hspace*{-1mm}& \hspace{-1mm}1.5e-11\hspace{-1mm} & \hspace{-1mm}0.39\hspace{-1mm} & \hspace{-1mm}0.5000485801347\hspace{-1mm} & \hspace{-1mm}0.68386889795\hspace{-1mm} & \hspace*{-1.5mm}-0.643317369\hspace*{-1mm} & \hspace{-1mm}1.08075772\hspace{-1mm}  \\
 \hspace*{-1mm}$24$\hspace*{-1mm}& \hspace{-1mm}4.7e-11\hspace{-1mm} & \hspace{-1mm}0.38\hspace{-1mm} & \hspace{-1mm}0.5000037328535\hspace{-1mm} & \hspace{-1mm}0.68593187289\hspace{-1mm} & \hspace*{-1.5mm}-0.679195194\hspace*{-1mm} & \hspace{-1mm}1.37495588\hspace{-1mm}  \\
 \hspace*{-1mm}$32$\hspace*{-1mm}& \hspace{-1mm}4.1e-11\hspace{-1mm} & \hspace{-1mm}0.43\hspace{-1mm} & \hspace{-1mm}0.5000002195177\hspace{-1mm} & \hspace{-1mm}0.68617780639\hspace{-1mm} & \hspace*{-1.5mm}-0.685884585\hspace*{-1mm} & \hspace{-1mm}1.46545941\hspace{-1mm}  \\
 \hspace*{-1mm}$48$\hspace*{-1mm}& \hspace{-1mm}1.4e-11\hspace{-1mm} & \hspace{-1mm}0.44\hspace{-1mm} & \hspace{-1mm}0.5000000087538\hspace{-1mm} & \hspace{-1mm}0.68619930431\hspace{-1mm} & \hspace*{-1.5mm}-0.686757991\hspace*{-1mm} & \hspace{-1mm}1.48373011\hspace{-1mm}  \\
 \hspace*{-1mm}$64$\hspace*{-1mm}& \hspace{-1mm}1.3e-11\hspace{-1mm} & \hspace{-1mm}0.31\hspace{-1mm} & \hspace{-1mm}0.5000000002937\hspace{-1mm} & \hspace{-1mm}0.68620057333\hspace{-1mm} & \hspace*{-1.5mm}-0.686834471\hspace*{-1mm} & \hspace{-1mm}1.48614633\hspace{-1mm}  \\
 \hspace*{-1mm}$96$\hspace*{-1mm}& \hspace{-1mm}3.2e-11\hspace{-1mm} & \hspace{-1mm}0.31\hspace{-1mm} & \hspace{-1mm}0.5000000000039\hspace{-1mm} & \hspace{-1mm}0.68620063534\hspace{-1mm} & \hspace*{-1.5mm}-0.686839872\hspace*{-1mm} & \hspace{-1mm}1.48639673\hspace{-1mm}  \\
\hline
\multicolumn{3}{r} {{Eq}~\eqref{Ilexp}:} & 0.5 & 0.68620063577 &\hspace*{-1.0mm} -0.686839887 & 1.48639839\\ \hline
\end{tabular}
\end{center}
\end{scriptsize}
\label{table2}
\end{table}

%
\begin{table}
\caption{\footnotesize
Results UV \emph{half-boiled egg} integral, $J_{c}^{S2}$ \,(on Mac Pro),~
rel. err. tol. $t_r = 10^{-12}$ (outer), $5\times 10^{-13}$ (inner three), $T[s]$ = Time (elapsed user time in $s$),~
$\varepsilon = 1/b_\ell$ (starting at 1.0),~~$E_r = $ outer integration estim. rel. error}
\begin{scriptsize}
\begin{center}
\begin{tabular}{ccccccc}\hline
& \multicolumn{6}{l}{{\sc Integral ~$J_{c}^{S2}$~ Extrapolation}} \\
 \hspace*{-1mm}$b_\ell$\hspace*{-1mm} & \hspace{-1mm}{ $E_r$ }\hspace{-1mm} & \hspace{-1mm}{\sc T[s]}
\hspace{-1mm} & {\sc Res.} ~${C}_{-1}$ & \hspace{-1mm}{\sc Res.} ~${C}_{0}$\hspace{-1mm} & \hspace*{-0mm}{\sc Res.} ~${C}_1$ & \hspace{-1mm}{\sc Res.} ~${C}_2$\hspace{-1mm} \\
\hline
 \hspace*{-1mm}$1$\hspace*{-1mm}& \hspace{-1mm}4.2e-13\hspace{-1mm} & \hspace{-1mm}7.3\hspace{-1mm} & \hspace{-1mm}\hspace{-1mm} & \hspace{-1mm}\hspace{-1mm} & \hspace*{-1.5mm}\hspace*{-1mm} & \hspace{-1mm}\hspace{-1mm} \\
 \hspace*{-1mm}$2$\hspace*{-1mm}& \hspace{-1mm}3.9e-14\hspace{-1mm} & \hspace{-1mm}11.4\hspace{-1mm} & \hspace{-1mm}0.6121795323700\hspace{-1mm} & \hspace{-1mm}-0.26893337928\hspace{-1mm} & \hspace*{-1.5mm}\hspace*{-1mm} & \hspace{-1mm}\hspace{-1mm}  \\
 \hspace*{-1mm}$3$\hspace*{-1mm}& \hspace{-1mm}2.6e-13\hspace{-1mm} & \hspace{-1mm}10.8\hspace{-1mm} & \hspace{-1mm}0.6219220162954\hspace{-1mm} & \hspace{-1mm}-0.29816083105\hspace{-1mm} & \hspace*{-1.5mm}-0.019484967\hspace*{-1mm} & \hspace{-1mm}\hspace{-1mm}  \\
 \hspace*{-1mm}$4$\hspace*{-1mm}& \hspace{-1mm}5.2e-13\hspace{-1mm} & \hspace{-1mm}10.3\hspace{-1mm} & \hspace{-1mm}0.6084345649676\hspace{-1mm} & \hspace{-1mm}-0.21723612309\hspace{-1mm} & \hspace*{-1.5mm}-0.128876997\hspace*{-1mm} & \hspace{-1mm}0.08092471\hspace{-1mm}  \\
 \hspace*{-1mm}$6$\hspace*{-1mm}& \hspace{-1mm}1.1e-13\hspace{-1mm} & \hspace{-1mm}14.9\hspace{-1mm} & \hspace{-1mm}0.6050661595977\hspace{-1mm} & \hspace{-1mm}-0.18355206939\hspace{-1mm} & \hspace*{-1.5mm}-0.246771185\hspace*{-1mm} & \hspace{-1mm}0.24934498\hspace{-1mm}  \\
 \hspace*{-1mm}$8$\hspace*{-1mm}& \hspace{-1mm}7.9e-13\hspace{-1mm} & \hspace{-1mm}9.2\hspace{-1mm} & \hspace{-1mm}0.6046439346384\hspace{-1mm} & \hspace{-1mm}-0.17679647004\hspace{-1mm} & \hspace*{-1.5mm}-0.286882556\hspace*{-1mm} & \hspace{-1mm}0.35912347\hspace{-1mm}  \\
 \hspace*{-1mm}$12$\hspace*{-1mm}& \hspace{-1mm}9.6e-13\hspace{-1mm} & \hspace{-1mm}12.5\hspace{-1mm} & \hspace{-1mm}0.6046028724387\hspace{-1mm} & \hspace{-1mm}-0.17581097725\hspace{-1mm} & \hspace*{-1.5mm}-0.296039426\hspace*{-1mm} & \hspace{-1mm}0.40100691\hspace{-1mm}  \\
 \hspace*{-1mm}$16$\hspace*{-1mm}& \hspace{-1mm}2.3e-13\hspace{-1mm} & \hspace{-1mm}19.4\hspace{-1mm} & \hspace{-1mm}0.6045999612478\hspace{-1mm} & \hspace{-1mm}-0.17570617438\hspace{-1mm} & \hspace*{-1.5mm}-0.297527045\hspace*{-1mm} & \hspace{-1mm}0.41176667\hspace{-1mm}  \\
 \hspace*{-1mm}$24$\hspace*{-1mm}& \hspace{-1mm}8.7e-13\hspace{-1mm} & \hspace{-1mm}19.1\hspace{-1mm} & \hspace{-1mm}0.6045997948488\hspace{-1mm} & \hspace{-1mm}-0.17569752162\hspace{-1mm} & \hspace*{-1.5mm}-0.297707921\hspace*{-1mm} & \hspace{-1mm}0.41374216\hspace{-1mm}  \\
 \hspace*{-1mm}$32$\hspace*{-1mm}& \hspace{-1mm}7.7e-13\hspace{-1mm} & \hspace{-1mm}24.9\hspace{-1mm} & \hspace{-1mm}0.6045997882885\hspace{-1mm} & \hspace{-1mm}-0.17569702304\hspace{-1mm} & \hspace*{-1.5mm}-0.297723238\hspace*{-1mm} & \hspace{-1mm}0.41399119\hspace{-1mm}  \\
 \hspace*{-1mm}$48$\hspace*{-1mm}& \hspace{-1mm}4.1e-13\hspace{-1mm} & \hspace{-1mm}33.2\hspace{-1mm} & \hspace{-1mm}0.6045997880782\hspace{-1mm} & \hspace{-1mm}-0.17569700033\hspace{-1mm} & \hspace*{-1.5mm}-0.297724241\hspace*{-1mm} & \hspace{-1mm}0.41401488\hspace{-1mm}  \\
\hline
\multicolumn{3}{r} {{Eq}~\eqref{Ihexp}:} & 0.6045997880781 & -0.17569700023  &-0.297724254  &0.41401554 \\ \hline
\end{tabular}
\end{center}
\end{scriptsize}
\label{table3}
\end{table}
\begin{table}
\caption{\footnotesize 
Results UV \emph{Magdeburg} integral, $I_{d}^{S2} \,\Gamma(1+\varepsilon)^{-2}$ \,(by {\sc ParInt} on \emph{thor}
in \emph{long double} precision),~
rel. err. tol. $t_r = 10^{-13},$ max. \# evals = 1B, $T[s]$ = Time (elapsed user time in $s$),~
$\varepsilon = 2^{-\ell}$ (starting at 1.0),~~$E_r = $ estim. rel. error}
\begin{scriptsize}
\begin{center}
\begin{tabular}{ccccccc}\hline
& \multicolumn{6}{l}{{\sc Integral ~$I_{d}^{S2} \,\Gamma(1+\varepsilon)^{-2}$~ Extrapolation}} \\
 \hspace*{-1mm}$\ell$\hspace*{-1mm} & \hspace{-1mm}{ $E_r$ }\hspace{-1mm} & \hspace{-1mm}{\sc T[s]}
\hspace{-1mm} & {\sc Res.} ~$C_0$ & \hspace{-3mm}{\sc Res.} ~$C_1$\hspace{-1mm} & \hspace*{-0mm}{\sc Res.} ~$C_2$ & \hspace{-1mm}{\sc Res.} ~$C_3$\hspace{-1mm} \\
\hline
 \hspace*{-1mm}$0$\hspace*{-1mm}& \hspace{-1mm}8.5e-14\hspace{-1mm} & \hspace{-3mm}0.8\hspace{-4mm} & \hspace{-1mm}\hspace{-1mm} & \hspace{-1mm}\hspace{-1mm} & \hspace*{-1.5mm}\hspace*{-1mm} & \hspace{-1mm}\hspace{-1mm} \\
 \hspace*{-1mm}$1$\hspace*{-1mm}& \hspace{-1mm}1.0e-13\hspace{-1mm} & \hspace{-3mm}19.7\hspace{-4mm} & \hspace{-1mm}0.69130084611470\hspace{-1mm} & \hspace{-2mm}-0.142989490499\hspace{-1mm} & \hspace*{-1mm}\hspace*{-1mm} & \hspace{-1mm}\hspace{-1mm}  \\
 \hspace*{-1mm}$2$\hspace*{-1mm}& \hspace{-1mm}1.6e-13\hspace{-1mm} & \hspace{-3mm}7.5\hspace{-4mm} & \hspace{-1mm}0.84949643770104\hspace{-1mm} & \hspace{-2mm}-0.617576265258\hspace{-1mm} & \hspace*{-1mm}0.3163911832\hspace*{-1mm} & \hspace{-1mm}\hspace{-1mm}  \\
 \hspace*{-1mm}$3$\hspace*{-1mm}& \hspace{-1mm}4.8e-13\hspace{-1mm} & \hspace{-3mm}6.9\hspace{-4mm} & \hspace{-1mm}0.90878784906010\hspace{-1mm} & \hspace{-2mm}-1.032616144771\hspace{-1mm} & \hspace*{-1mm}1.1464709422\hspace*{-1mm} & \hspace{-2mm}-0.47433129\hspace{-1mm}  \\
 \hspace*{-1mm}$4$\hspace*{-1mm}& \hspace{-1mm}8.5e-13\hspace{-1mm} & \hspace{-3mm}6.6\hspace{-4mm} & \hspace{-1mm}0.92198476262012\hspace{-1mm} & \hspace{-2mm}-1.230569848171\hspace{-1mm} & \hspace*{-1mm}2.0702548914\hspace*{-1mm} & \hspace{-2mm}-2.05796092\hspace{-1mm}  \\
 \hspace*{-1mm}$5$\hspace*{-1mm}& \hspace{-1mm}1.2e-12\hspace{-1mm} & \hspace{-3mm}6.5\hspace{-4mm} & \hspace{-1mm}0.92353497740505\hspace{-1mm} & \hspace{-2mm}-1.278626506507\hspace{-1mm} & \hspace*{-1mm}2.5508214747\hspace*{-1mm} & \hspace{-2mm}-3.98022725\hspace{-1mm}  \\
 \hspace*{-1mm}$6$\hspace*{-1mm}& \hspace{-1mm}3.0e-12\hspace{-1mm} & \hspace{-3mm}6.6\hspace{-4mm} & \hspace{-1mm}0.92362889210499\hspace{-1mm} & \hspace{-2mm}-1.284543132600\hspace{-1mm} & \hspace*{-1mm}2.6730984140\hspace*{-1mm} & \hspace{-2mm}-5.02831530\hspace{-1mm}  \\
 \hspace*{-1mm}$7$\hspace*{-1mm}& \hspace{-1mm}2.4e-12\hspace{-1mm} & \hspace{-3mm}6.6\hspace{-4mm} & \hspace{-1mm}0.92363178137723\hspace{-1mm} & \hspace{-2mm}-1.284910070175\hspace{-1mm} & \hspace*{-1mm}2.6885097922\hspace*{-1mm} & \hspace{-2mm}-5.30131686\hspace{-1mm}  \\
 \hspace*{-1mm}$8$\hspace*{-1mm}& \hspace{-1mm}2.8e-12\hspace{-1mm} & \hspace{-3mm}6.6\hspace{-4mm} & \hspace{-1mm}0.92363182617006\hspace{-1mm} & \hspace{-2mm}-1.284921492347\hspace{-1mm} & \hspace*{-1mm}2.6894768694\hspace*{-1mm} & \hspace{-2mm}-5.33613164\hspace{-1mm}  \\
 \hspace*{-1mm}$9$\hspace*{-1mm}& \hspace{-1mm}2.8e-12\hspace{-1mm} & \hspace{-3mm}6.6\hspace{-4mm} & \hspace{-1mm}0.92363182651847\hspace{-1mm} & \hspace{-2mm}-1.284921670382\hspace{-1mm} & \hspace*{-1mm}2.6895071354\hspace*{-1mm} & \hspace{-2mm}-5.33832809\hspace{-1mm}  \\
 \hspace*{-1mm}$10$\hspace*{-1mm}& \hspace{-1mm}2.9e-12\hspace{-1mm} & \hspace{-3mm}6.6\hspace{-4mm} & \hspace{-1mm}0.92363182651995\hspace{-1mm} & \hspace{-2mm}-1.284921671903\hspace{-1mm} & \hspace*{-1mm}2.6895076534\hspace*{-1mm} & \hspace{-2mm}-5.33840357\hspace{-1mm}  \\
 \hspace*{-1mm}$11$\hspace*{-1mm}& \hspace{-1mm}2.9e-12\hspace{-1mm} & \hspace{-3mm}6.6\hspace{-4mm} & \hspace{-1mm}0.92363182651990\hspace{-1mm} & \hspace{-2mm}-1.284921671790\hspace{-1mm} & \hspace*{-1mm}2.6895075765\hspace*{-1mm} & \hspace{-2mm}-5.33838110\hspace{-1mm}  \\
 \hspace*{-1mm}$12$\hspace*{-1mm}& \hspace{-1mm}2.9e-12\hspace{-1mm} & \hspace{-3mm}6.6\hspace{-4mm} & \hspace{-1mm}0.92363182651992\hspace{-1mm} & \hspace{-2mm}-1.284921671898\hspace{-1mm} & \hspace*{-1mm}2.6895077252\hspace*{-1mm} & \hspace{-2mm}-5.33846839\hspace{-1mm}  \\
 \hspace*{-1mm}$13$\hspace*{-1mm}& \hspace{-1mm}3.7e-12\hspace{-1mm} & \hspace{-3mm}6.6\hspace{-4mm} & \hspace{-1mm}0.92363182651991\hspace{-1mm} & \hspace{-2mm}-1.284921671798\hspace{-1mm} & \hspace*{-1mm}2.6895075774\hspace*{-1mm} & \hspace{-2mm}-5.33835823\hspace{-1mm}  \\
 \hspace*{-1mm}$14$\hspace*{-1mm}& \hspace{-1mm}2.9e-12\hspace{-1mm} & \hspace{-3mm}6.6\hspace{-4mm} & \hspace{-1mm}0.92363182651991\hspace{-1mm} & \hspace{-2mm}-1.284921671840\hspace{-1mm} & \hspace*{-1mm}2.6895076182\hspace*{-1mm} & \hspace{-2mm}-5.33839951\hspace{-1mm}  \\
\hline
\multicolumn{3}{r} {{Eq}~\eqref{Imexp}:} & 0.9236318265199~ & \hspace*{-1mm}-1.284921671848 & 2.6895076265 & \hspace{-1mm}-5.33839923\\ \hline
\end{tabular}
\end{center}
\end{scriptsize}
\label{table-Magdeburg}
\end{table}
Tables~\ref{table1}, ~\ref{table2}, ~\ref{table3} and ~\ref{table-Magdeburg} show the convergence of the extrapolation method
for the integrals of Eqs~\eqref{Isexp}-\eqref{Imexp}.
While the integral $I_{d}^{S2}$ has no UV-divergent terms and starts from a finite term ($\kappa=0$), 
the coefficients ${C}_0$, ${C}_1$, ${C}_2$ and ${C}_3$ can be obtained using extrapolation.
To evaluate $I_{d}^{S2}$, we transform the variables as:
\begin{eqnarray}
\nonumber
x_1&=&y_{1m} y_{3m} y_4, ~~~~~~x_2~=~y_{1m} y_{3m} y_{4m},\\
x_3&=&y_1 y_{2m}, ~~~~~~x_4~=~y_{1m} y_3,\\
\nonumber
x_5&=&y_1 y_2
\end{eqnarray}
with $y_{im}=1-y_{i}$ and Jacobian $y_1 y_{1m}^2 y_{3m}$.
The accuracy and time of the calculation 
of the integral sequence for the extrapolation in Table~\ref{table-Magdeburg}
are improved considerably by the transformation.

For the first three integrals,
an iterated integration is applied with {\sc Dqags} from {\sc Quadpack},
on a Mac Pro, 2.6\,GHz Intel Core i7, with 16\,GB memory, under OS X.
The value of $E_r$ is (the absolute value of)
the estimated relative error returned by the outer integration
(not accounting for the inner integration error). It is listed for each integration,
as well as the elapsed user time $T[s]$ (in seconds). The time for
the extrapolation is negligible compared to that of the integration. We use a standard linear
system solver to solve very small systems (of sizes $2\times 2$ up to around $15\times 15$
for the cases in this paper).

Table~\ref{table-Magdeburg} illustrates an application of {\sc ParInt} for the Magdeburg integral,
on the \emph{thor} system of the High Performance Computing and Big Data Center at WMU,
with dual Intel Xeon E5-2670, 2.6\,GHz processors, 16 cores and 128\,GB of memory per node.
For the distributed computation with {\sc ParInt}, using 16 processes per node and 64 processes total,
the MPI host file has four lines of the form \emph{nx slots=16}
where \emph{nx} represents a selected node name.
The running time is reported (in seconds) from {\sc ParInt}, and comprises all computation
not inclusive of the process spawning and {\sc ParInt} initialization at the start of the program.
The cubature rule of degree 9 is used for integration over the subregions (see Section~\ref{parint}),
to an allowed maximum number of one billion (1B) integrand evaluations over all processes,
and a requested accuracy of $t_r = 10^{-13}$ in \emph{long double} precision.
The total estimated relative error is denoted by $E_r$ in Table~\ref{table-Magdeburg}.

The extrapolation parameter for Tables~\ref{table2} and~\ref{table3} adheres to
$\{1/b_\ell\}$ where $\{b_\ell\}$ is the Bulirsch sequence~\cite{bulirsch64} started
at an early index.
Tables~\ref{table1} and~\ref{table-Magdeburg} give results for a geometric sequence
of the extrapolation parameter $\varepsilon = \varepsilon_\ell.$
The convergence results in Tables~\ref{table1}-\ref{table2} and~\ref{table-Magdeburg} are compared with
with the expansion coefficients available from~\cite{laporta01} (see Eqs~\eqref{Isexp}-\eqref{Ilexp},~\eqref{Imexp}.
Table~\ref{table3} shows excellent approximations to the analytic result of Eq~\eqref{Ihexp}
derived in the Appendix.

Throughout the extrapolation we keep track of the difference with the previous result as
a measure of convergence. Increases of the distance between successive extrapolation results
are an indicator that the convergence is no longer improving and the procedure can be terminated.

\subsection{2-loop vertex integrals}\label{uv-vertex}

\begin{figure}
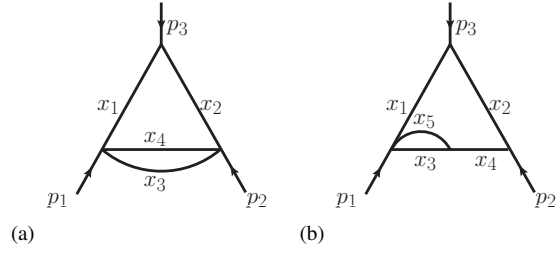

\begin{center}
\begin{subfigure}[ ]
\centering
\includegraphics[width=0.22\linewidth]{./figures/2lv-laporta3b-20161029.epsi}
\end{subfigure}
\begin{subfigure}[ ]
\centering
\includegraphics[width=0.22\linewidth]{./figures/2lv-laporta3c-20161029.epsi}
\end{subfigure}
\caption{\footnotesize 2-loop vertex (UV-divergent) diagram with massive internal lines:
 (a) $N = 4$ (Laporta~\cite{laporta01}\,, Fig 3(b) ), (b) $N = 5$ (Laporta~\cite{laporta01}\,, Fig 3(c) )}
\label{laporta-3ab}
\end{center}
\end{figure}

Fig~\ref{laporta-3ab}(a) and (b) depict 2-loop vertex diagrams with $N = 4$ and $5$ internal lines,
and the integrals of  Eq~\eqref{UV} are denoted by $I_a^{V2}$ and $I_b^{V2}$, respectively. 
The former is divergent as $\displaystyle{1/\varepsilon^2},$ which is
the product of $\displaystyle{1/\varepsilon}$ arising from the $\Gamma$-function factor and
 $\displaystyle{1/\varepsilon}$ from the integral.
The latter is divergent as $\displaystyle{1/\varepsilon}$ arising from the integral.\\
$-$ For  $I_a^{V2}$, we have
$U = x_{12}x_{34} + x_3 x_4$ and
$W = p_{1}^2 x_1 x_3 x_4 + p_{2}^2 x_2 x_3 x_4 + p_{3}^2 x_1 x_2 x_{34}.$ \\
$-$ For  $I_b^{V2}$, we have
$U=x_{124}x_3+x_{1234}x_5 $ and
$W=(p_1^2 x_1+p_2^2 x_2)\,(x_3x_{45}+x_4x_5)+p_3^2x_1x_2x_{35}$. \\

\begin{table}
\caption{\footnotesize
{\color{black}Results UV \emph{vertex} integral}, $I_{a}^{V2}$ \,(on Mac Pro),~
rel. err. tol. $t_r = 10^{-10}$ (outer), $5\times 10^{-11}$ (inner three), $T[s]$ = Time (elapsed user time in $s$),~
$\varepsilon = 1/b_\ell$ (starting at 1/3),~~$E_r = $ outer integration estim. rel. error}
\begin{scriptsize}
\begin{center}
\begin{tabular}{ccccccc}\hline
& \multicolumn{6}{l}{{\sc Integral ${I}_{a}^{V2} ~\Gamma(1+\varepsilon)^{-2}$~ Extrapolation}} \\
 \hspace*{-1mm}$b_\ell$\hspace*{-1mm} & \hspace{-1mm}{ $E_r$ }\hspace{-1mm} & \hspace{-1mm}{\sc T[s]}
\hspace{-1mm} & {\sc Res.} ~${C}_{-2}$ & \hspace{-1mm}{\sc Res.} ~${C}_{-1}$\hspace{-1mm} & \hspace*{-0mm}{\sc Res.} ~${C}_0$ & \hspace{-1mm}{\sc Res.} ~${C}_1$\hspace{-1mm} \\
\hline
 \hspace*{-1mm}$3$\hspace*{-1mm}&\hspace{-1mm} 8.7e-12     \hspace{-1mm} & \hspace{-1mm}$1.4$\hspace{-1mm} & \hspace{-1mm}\hspace{-1mm} & \hspace{-1mm}\hspace{-1mm} & \hspace*{-1.5mm}\hspace*{-1mm} & \hspace{-1mm}\hspace{-1mm} \\
 \hspace*{-1mm}$4$\hspace*{-1mm}&\hspace{-1mm} 2.3e-11     \hspace{-1mm} & \hspace{-1mm}$1.5$\hspace{-1mm} & \hspace{-1mm}0.51489021736\hspace{-1mm} & \hspace{-1mm}0.535680679\hspace{-1mm} & \hspace*{-1.5mm}\hspace*{-1mm} & \hspace{-1mm}\hspace{-1mm}  \\
 \hspace*{-1mm}$6$\hspace*{-1mm}&\hspace{-1mm} 3.1e-10     \hspace{-1mm} & \hspace{-1mm}$1.8$\hspace{-1mm} & \hspace{-1mm}0.50586162735\hspace{-1mm} & \hspace{-1mm}0.598880809\hspace{-1mm} & \hspace*{-1.5mm}-0.1083431\hspace*{-1mm} &\hspace{-1mm}\hspace{-1mm}  \\
 \hspace*{-1mm}$8$\hspace*{-1mm}&\hspace{-1mm} 8.2e-11     \hspace{-1mm} & \hspace{-1mm}$5.2$\hspace{-1mm} & \hspace{-1mm}0.50111160609\hspace{-1mm} & \hspace{-1mm}0.660631086\hspace{-1mm} & \hspace*{-1.5mm}-0.3648442\hspace*{-1mm} &\hspace{-2mm}0.342002\hspace{-1mm}  \\
 \hspace*{-1mm}$12$\hspace*{-1mm}&\hspace{-1mm} 1.2e-09     \hspace{-1mm} & \hspace{-1mm}$2.5$\hspace{-1mm} & \hspace{-1mm}0.50015901670\hspace{-1mm} & \hspace{-1mm}0.680635463\hspace{-1mm} & \hspace*{-1.5mm}-0.5153534\hspace*{-1mm} & \hspace{-2mm}0.822107\hspace{-1mm}  \\
 \hspace*{-1mm}$16$\hspace*{-1mm}&\hspace{-1mm} 1.7e-09     \hspace{-1mm} & \hspace{-1mm}$4.2$\hspace{-1mm} & \hspace{-1mm}0.50001764652\hspace{-1mm} & \hspace{-1mm}0.685300679\hspace{-1mm} & \hspace*{-1.5mm}-0.5733151\hspace*{-1mm} & \hspace{-2mm}1.161395\hspace{-1mm}  \\
 \hspace*{-1mm}$24$\hspace*{-1mm}&\hspace{-1mm} 4.4e-09     \hspace{-1mm} & \hspace{-1mm}$20.5$\hspace{-1mm} & \hspace{-1mm}0.50000135665\hspace{-1mm} & \hspace{-1mm}0.686098882\hspace{-1mm} & \hspace*{-1.5mm}-0.5885950\hspace*{-1mm} & \hspace{-2mm}1.307352\hspace{-1mm}  \\
 \hspace*{-1mm}$32$\hspace*{-1mm}&\hspace{-1mm} 7.7e-09     \hspace{-1mm} & \hspace{-1mm}$19.1$\hspace{-1mm} & \hspace{-1mm}0.50000007798\hspace{-1mm} & \hspace{-1mm}0.686192225\hspace{-1mm} & \hspace*{-1.5mm}-0.5912981\hspace*{-1mm} & \hspace{-2mm}1.347595\hspace{-1mm}  \\
 \hspace*{-1mm}$48$\hspace*{-1mm}&\hspace{-1mm} 3.8e-10     \hspace{-1mm} & \hspace{-1mm}$8.6$\hspace{-1mm} & \hspace{-1mm}0.50000000083\hspace{-1mm} & \hspace{-1mm}0.686200327\hspace{-1mm} & \hspace*{-1.5mm}-0.5916415\hspace*{-1mm} & \hspace{-2mm}1.355242\hspace{-1mm}  \\
\hline
\multicolumn{3}{r} {Eq}~\eqref{eq-3a}: & 0.5 & 0.686200636  & -0.5916667 & \hspace{-1mm}1.356197\\ \hline
\end{tabular}
\end{center}
\end{scriptsize}
\label{tab:vertex3a}
\end{table}
\begin{table}[h]
\caption{\footnotesize
Results UV \emph{vertex} integral, $I_{b}^{V2} \,\Gamma(1+\varepsilon)^{-2}$ \,(by {\sc ParInt} on \emph{thor} in \emph{long double} precision),~
rel. err. tol. $t_r = 10^{-13},$ max. \# evals = 10B, $T[s]$ = Time (elapsed user time in $s$),~
$\varepsilon = 1.2^{-\ell}$ (starting at $1.2^{-8}$),~~$E_r = $ estim. rel. error}
\begin{scriptsize}
\begin{center}
\begin{tabular}{ccccccc}\hline
& \multicolumn{6}{l}{{\sc Integral ~$I_{b}^{V2} \,\Gamma(1+\varepsilon)^{-2}$~ Extrapolation}} \\
 \hspace*{-1mm}$\ell$\hspace*{-1mm} & \hspace{-1mm}{ $E_r$ }\hspace{-1mm} & \hspace{-1mm}{\sc T[s]}
\hspace{-1mm} & {\sc Res.} ~$C_{-1}$ & \hspace{-3mm}{\sc Res.} ~$C_0$\hspace{-1mm} & \hspace*{-0mm}{\sc Res.} ~$C_1$ & \hspace{-1mm}{\sc Res.} ~$C_2$\hspace{-1mm} \\
\hline
 \hspace*{-1mm}$8$\hspace*{-1mm}& \hspace{-1mm}9.9e-14\hspace{-1mm} & \hspace{-3mm}0.7\hspace{-4mm} & \hspace{-1mm}\hspace{-1mm} & \hspace{-1mm}\hspace{-1mm} & \hspace*{-1.5mm}\hspace*{-1mm} & \hspace{-1mm}\hspace{-1mm} \\
 \hspace*{-1mm}$9$\hspace*{-1mm}& \hspace{-1mm}5.7e-14\hspace{-1mm} & \hspace{-3mm}0.9\hspace{-4mm} & \hspace{-1mm}0.653547537693\hspace{-1mm} & \hspace{-2mm}0.10873442398\hspace{-1mm} & \hspace*{-1mm}\hspace*{-1mm} & \hspace{-1mm}\hspace{-1mm}  \\
 \hspace*{-1mm}$10$\hspace*{-1mm}& \hspace{-1mm}8.7e-14\hspace{-1mm} & \hspace{-3mm}1.3\hspace{-4mm} & \hspace{-1mm}0.667737620294\hspace{-1mm} & \hspace{-2mm}-0.02549804326\hspace{-1mm} & \hspace*{-1mm}0.148227487\hspace*{-1mm} & \hspace{-1mm}\hspace{-1mm}  \\
 \hspace*{-1mm}$11$\hspace*{-1mm}& \hspace{-1mm}5.7e-14\hspace{-1mm} & \hspace{-3mm}1.7\hspace{-4mm} & \hspace{-1mm}0.670486635108\hspace{-1mm} & \hspace{-2mm}-0.06852379156\hspace{-1mm} & \hspace*{-1mm}0.536826159\hspace*{-1mm} & \hspace{-2mm}-0.37763368\hspace{-1mm}  \\
 \hspace*{-1mm}$12$\hspace*{-1mm}& \hspace{-1mm}9.8e-14\hspace{-1mm} & \hspace{-3mm}6.9\hspace{-4mm} & \hspace{-1mm}0.671112937626\hspace{-1mm} & \hspace{-2mm}-0.08297974141\hspace{-1mm} & \hspace*{-1mm}0.660237921\hspace*{-1mm} & \hspace{-2mm}-0.83947237\hspace{-1mm}  \\
 \hspace*{-1mm}$13$\hspace*{-1mm}& \hspace{-1mm}9.8e-14\hspace{-1mm} & \hspace{-3mm}8.4\hspace{-4mm} & \hspace{-1mm}0.671231024200\hspace{-1mm} & \hspace{-2mm}-0.08675821868\hspace{-1mm} & \hspace*{-1mm}0.707808436\hspace*{-1mm} & \hspace{-2mm}-1.13401646\hspace{-1mm}  \\
 \hspace*{-1mm}$14$\hspace*{-1mm}& \hspace{-1mm}9.6e-14\hspace{-1mm} & \hspace{-3mm}9.3\hspace{-4mm} & \hspace{-1mm}0.671250129509\hspace{-1mm} & \hspace{-2mm}-0.08757395495\hspace{-1mm} & \hspace*{-1mm}0.722045637\hspace*{-1mm} & \hspace{-2mm}-1.26401791\hspace{-1mm}  \\
 \hspace*{-1mm}$15$\hspace*{-1mm}& \hspace{-1mm}9.4e-14\hspace{-1mm} & \hspace{-3mm}10.3\hspace{-4mm} & \hspace{-1mm}0.671252763777\hspace{-1mm} & \hspace{-2mm}-0.08772025175\hspace{-1mm} & \hspace*{-1mm}0.725452770\hspace*{-1mm} & \hspace{-2mm}-1.30714663\hspace{-1mm}  \\
 \hspace*{-1mm}$16$\hspace*{-1mm}& \hspace{-1mm}1.0e-13\hspace{-1mm} & \hspace{-3mm}65.7\hspace{-4mm} & \hspace{-1mm}0.671253072371\hspace{-1mm} & \hspace{-2mm}-0.08774214433\hspace{-1mm} & \hspace*{-1mm}0.726115949\hspace*{-1mm} & \hspace{-2mm}-1.31834841\hspace{-1mm}  \\
 \hspace*{-1mm}$17$\hspace*{-1mm}& \hspace{-1mm}9.4e-14\hspace{-1mm} & \hspace{-3mm}13.8\hspace{-4mm} & \hspace{-1mm}0.671253102986\hspace{-1mm} & \hspace{-2mm}-0.08774488225\hspace{-1mm} & \hspace*{-1mm}0.726221896\hspace*{-1mm} & \hspace{-2mm}-1.32067610\hspace{-1mm}  \\
 \hspace*{-1mm}$18$\hspace*{-1mm}& \hspace{-1mm}9.9e-14\hspace{-1mm} & \hspace{-3mm}18.1\hspace{-4mm} & \hspace{-1mm}0.671253105554\hspace{-1mm} & \hspace{-2mm}-0.08774516889\hspace{-1mm} & \hspace*{-1mm}0.726235878\hspace*{-1mm} & \hspace{-2mm}-1.32106853\hspace{-1mm}  \\
 \hspace*{-1mm}$19$\hspace*{-1mm}& \hspace{-1mm}9.3e-14\hspace{-1mm} & \hspace{-3mm}39.5\hspace{-4mm} & \hspace{-1mm}0.671253105741\hspace{-1mm} & \hspace{-2mm}-0.08774519476\hspace{-1mm} & \hspace*{-1mm}0.726237453\hspace*{-1mm} & \hspace{-2mm}-1.32112425\hspace{-1mm}  \\
 \hspace*{-1mm}$20$\hspace*{-1mm}& \hspace{-1mm}1.0e-13\hspace{-1mm} & \hspace{-3mm}76.0\hspace{-4mm} & \hspace{-1mm}0.671253105743\hspace{-1mm} & \hspace{-2mm}-0.08774519512\hspace{-1mm} & \hspace*{-1mm}0.726237480\hspace*{-1mm} & \hspace{-2mm}-1.32112544\hspace{-1mm}  \\
 \hspace*{-1mm}$21$\hspace*{-1mm}& \hspace{-1mm}1.0e-13\hspace{-1mm} & \hspace{-3mm}77.4\hspace{-4mm} & \hspace{-1mm}0.671253105751\hspace{-1mm} & \hspace{-2mm}-0.08774519670\hspace{-1mm} & \hspace*{-1mm}0.726237628\hspace*{-1mm} & \hspace{-2mm}-1.32113350\hspace{-1mm}  \\
 \hspace*{-1mm}$22$\hspace*{-1mm}& \hspace{-1mm}1.1e-13\hspace{-1mm} & \hspace{-3mm}77.8\hspace{-4mm} & \hspace{-1mm}0.671253105748\hspace{-1mm} & \hspace{-2mm}-0.08774519610\hspace{-1mm} & \hspace*{-1mm}0.726237559\hspace*{-1mm} & \hspace{-2mm}-1.32112888\hspace{-1mm}  \\
\hline
\multicolumn{3}{r} {{Eq}~\eqref{eq-3b}:} & 0.671253105748& \hspace*{-1mm}-0.08774519609 & 0.726237563 & \hspace{-1mm}-1.32112949\\ \hline
\end{tabular}
\end{center}
\end{scriptsize}
\label{tab:vertex3b1.2}
\end{table}

\begin{table}
\caption{\footnotesize
Results UV \emph{vertex} integral, $I_{b}^{V2} \,\Gamma(1+\varepsilon)^{-2}$ \,(by {\sc ParInt} on \emph{thor} in \emph{long double} precision),~
rel. err. tol. $t_r = 10^{-13},$ max. \# evals = 10B, $T[s]$ = Time (elapsed user time in $s$),~
$\varepsilon = 1.5^{-\ell}$ (starting at 1.0),~~$E_r = $ estim. rel. error}
\begin{scriptsize}
\begin{center}
\begin{tabular}{ccccccc}\hline
& \multicolumn{6}{l}{{\sc Integral ~$I_{b}^{V2} \,\Gamma(1+\varepsilon)^{-2}$~ Extrapolation}} \\
 \hspace*{-1mm}$\ell$\hspace*{-1mm} & \hspace{-1mm}{ $E_r$ }\hspace{-1mm} & \hspace{-1mm}{\sc T[s]}
\hspace{-1mm} & {\sc Res.} ~$C_{-1}$ & \hspace{-3mm}{\sc Res.} ~$C_0$\hspace{-1mm} & \hspace*{-0mm}{\sc Res.} ~$C_1$ & \hspace{-1mm}{\sc Res.} ~$C_2$\hspace{-1mm} \\
\hline
 \hspace*{-1mm}$0$\hspace*{-1mm}& \hspace{-1mm}1.0e-13\hspace{-1mm} & \hspace{-3mm}0.31\hspace{-4mm} & \hspace{-1mm}\hspace{-1mm} & \hspace{-1mm}\hspace{-1mm} & \hspace*{-1.5mm}\hspace*{-1mm} & \hspace{-1mm}\hspace{-1mm} \\
 \hspace*{-1mm}$1$\hspace*{-1mm}& \hspace{-1mm}1.0e-13\hspace{-1mm} & \hspace{-3mm}0.77\hspace{-4mm} & \hspace{-1mm}0.552646148416\hspace{-1mm} & \hspace{-2mm}-0.03261093175\hspace{-1mm} & \hspace*{-1mm}\hspace*{-1mm} & \hspace{-1mm}\hspace{-1mm}  \\
 \hspace*{-1mm}$2$\hspace*{-1mm}& \hspace{-1mm}2.7e-14\hspace{-1mm} & \hspace{-3mm}0.26\hspace{-4mm} & \hspace{-1mm}0.645884613605\hspace{-1mm} & \hspace{-2mm}-0.09301315450\hspace{-1mm} & \hspace*{-1mm}0.139857698\hspace*{-1mm} & \hspace{-1mm}\hspace{-1mm}  \\
 \hspace*{-1mm}$3$\hspace*{-1mm}& \hspace{-1mm}3.9e-14\hspace{-1mm} & \hspace{-3mm}0.44\hspace{-4mm} & \hspace{-1mm}0.659977890791\hspace{-1mm} & \hspace{-2mm}-0.02607008786\hspace{-1mm} & \hspace*{-1mm}0.240272298\hspace*{-1mm} & \hspace{-2mm}-0.04756481\hspace{-1mm}  \\
 \hspace*{-1mm}$4$\hspace*{-1mm}& \hspace{-1mm}4.9e-14\hspace{-1mm} & \hspace{-3mm}0.92\hspace{-4mm} & \hspace{-1mm}0.668110702931\hspace{-1mm} & \hspace{-2mm}-0.04000901078\hspace{-1mm} & \hspace*{-1mm}0.428597729\hspace*{-1mm} & \hspace{-2mm}-0.02705818\hspace{-1mm}  \\
 \hspace*{-1mm}$5$\hspace*{-1mm}& \hspace{-1mm}1.2e-14\hspace{-1mm} & \hspace{-3mm}1.32\hspace{-4mm} & \hspace{-1mm}0.670596883473\hspace{-1mm} & \hspace{-2mm}-0.07279551667\hspace{-1mm} & \hspace*{-1mm}0.588431945\hspace*{-1mm} & \hspace{-2mm}-0.63020875\hspace{-1mm}  \\
 \hspace*{-1mm}$6$\hspace*{-1mm}& \hspace{-1mm}3.8e-14\hspace{-1mm} & \hspace{-3mm}1.96\hspace{-4mm} & \hspace{-1mm}0.671155085836\hspace{-1mm} & \hspace{-2mm}-0.08439565952\hspace{-1mm} & \hspace*{-1mm}0.680218075\hspace*{-1mm} & \hspace{-2mm}-0.98346458\hspace{-1mm}  \\
 \hspace*{-1mm}$7$\hspace*{-1mm}& \hspace{-1mm}9.5e-14\hspace{-1mm} & \hspace{-3mm}13.9\hspace{-4mm} & \hspace{-1mm}0.671242833695\hspace{-1mm} & \hspace{-2mm}-0.08721867268\hspace{-1mm} & \hspace*{-1mm}0.715417521\hspace*{-1mm} & \hspace{-2mm}-1.20334533\hspace{-1mm}  \\
 \hspace*{-1mm}$8$\hspace*{-1mm}& \hspace{-1mm}1.0e-13\hspace{-1mm} & \hspace{-3mm}76.5\hspace{-4mm} & \hspace{-1mm}0.671252362159\hspace{-1mm} & \hspace{-2mm}-0.08768802398\hspace{-1mm} & \hspace*{-1mm}0.724477468\hspace*{-1mm} & \hspace{-2mm}-1.29252918\hspace{-1mm}  \\
 \hspace*{-1mm}$9$\hspace*{-1mm}& \hspace{-1mm}1.0e-13\hspace{-1mm} & \hspace{-3mm}77.7\hspace{-4mm} & \hspace{-1mm}0.671253068976\hspace{-1mm} & \hspace{-2mm}-0.08774095520\hspace{-1mm} & \hspace*{-1mm}0.726041833\hspace*{-1mm} & \hspace{-2mm}-1.31636903\hspace{-1mm}  \\
 \hspace*{-1mm}$10$\hspace*{-1mm}& \hspace{-1mm}9.3e-15\hspace{-1mm} & \hspace{-3mm}26.8\hspace{-4mm} & \hspace{-1mm}0.671253104515\hspace{-1mm} & \hspace{-2mm}-0.08774498285\hspace{-1mm} & \hspace*{-1mm}0.726222803\hspace*{-1mm} & \hspace{-2mm}-1.32059155\hspace{-1mm}  \\
 \hspace*{-1mm}$11$\hspace*{-1mm}& \hspace{-1mm}2.1e-13\hspace{-1mm} & \hspace{-3mm}78.3\hspace{-4mm} & \hspace{-1mm}0.671253105720\hspace{-1mm} & \hspace{-2mm}-0.08774518885\hspace{-1mm} & \hspace*{-1mm}0.726236811\hspace*{-1mm} & \hspace{-2mm}-1.32108845\hspace{-1mm}  \\
 \hspace*{-1mm}$12$\hspace*{-1mm}& \hspace{-1mm}4.9e-13\hspace{-1mm} & \hspace{-3mm}79.5\hspace{-4mm} & \hspace{-1mm}0.671253105748\hspace{-1mm} & \hspace{-2mm}-0.08774519595\hspace{-1mm} & \hspace*{-1mm}0.726237540\hspace*{-1mm} & \hspace{-2mm}-1.32112754\hspace{-1mm}  \\
\hline
\multicolumn{3}{r} {{Eq}~\eqref{eq-3b}:} & 0.671253105748& \hspace*{-1mm}-0.08774519609 & 0.726237563 & \hspace{-1mm}-1.32112949\\ \hline
\end{tabular}
\end{center}
\end{scriptsize}
\label{tab:vertex3b}
\end{table}
In order to compare with Laporta's results~\cite{laporta01}, we put $m_r=1$ and $p_1^2=p_2^2=p_3^2=1,$
and multiply the integrals with a factor $\Gamma(1+{\varepsilon})^{-2}.$
The expansions from~\cite{laporta01} are:
\begin{eqnarray}
\nonumber
 I_{a}^{V2}({\varepsilon}) ~\Gamma(1+{\varepsilon})^{-2} = \sum_{k\ge -2} C_k \,{\varepsilon^k} &=& 0.5\,{\varepsilon^{-2}}+0.6862006357658\,{\varepsilon^{-1}}-0.5916667014024\\
&+&1.356196533114\,{\varepsilon}\ldots \label{eq-3a}\\
\nonumber \\
\nonumber
 I_{b}^{V2}({\varepsilon}) ~\Gamma(1+{\varepsilon})^{-2} = \sum_{k\ge -1} C_k \,{\varepsilon^k} &=& 0.671253105748\,{\varepsilon^{-1}}-0.08774519609257+0.7262375626947\,{\varepsilon}\\
&-&1.32112948587\,{\varepsilon^2}\ldots \label{eq-3b}
\end{eqnarray}

The numerical results for $I_a^{V2}$ obtained with iterated integration by {\sc Dqags},
and extrapolation using a Bulirsch sequence and
linear system solver are shown in Table~\ref{tab:vertex3a}.
Tables~\ref{tab:vertex3b1.2}-\ref{tab:vertex3b} show results for $I_{b}^{V2}$ with geometric sequences of
base 1/1.2 and 1/1.5, respectively,
achieved by {\sc ParInt} using 64 processes on four 16-core nodes of the \emph{thor} cluster.
Both deliver very accurate results, with
the final results in Table~\ref{tab:vertex3b1.2} slightly closer to the analytic values.
The extrapolation in Table~\ref{tab:vertex3b} converges somewhat faster.
For the computation of $I_{b}^{V2}$, we transform the variables as for $I_d^{S2}$
in Section~\ref{2ls}.
This transformation maps the integration domain to the 4-dimensional unit cube
and also guards against the loss of significant digits 
near the boundaries.

\subsection{2-loop box integrals}\label{2lb}

\begin{figure}
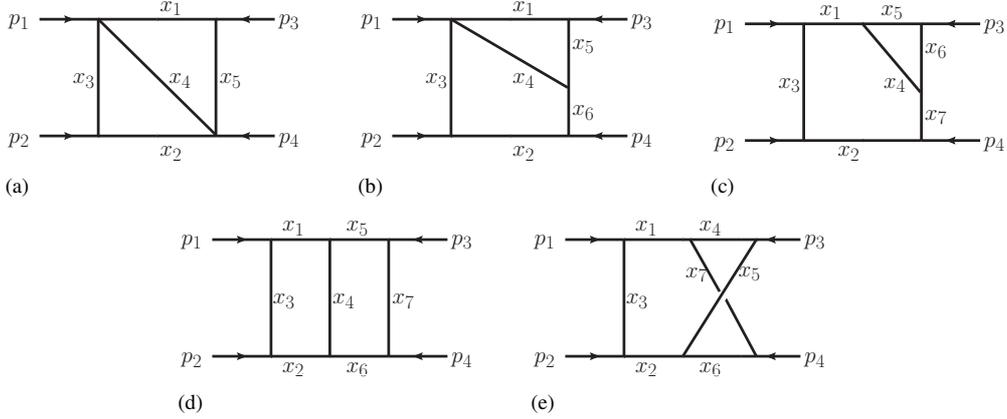

\begin{center}
\begin{subfigure}[ ]
\centering
\includegraphics[width=0.25\linewidth]{./figures/2lb-laporta4c-20161029.epsi} \quad
\end{subfigure}
\begin{subfigure}[ ]
\centering
\includegraphics[width=0.25\linewidth]{./figures/2lb-laporta4d-20161029.epsi} \quad
\end{subfigure}
\begin{subfigure}[ ]
\centering
\includegraphics[width=0.25\linewidth]{./figures/2lb-laporta4g-20161029.epsi} \\
\end{subfigure}
\begin{subfigure}[ ]
\centering
\includegraphics[width=0.25\linewidth]{./figures/2lb-laporta4h-20161029.epsi} \quad
\end{subfigure}
\begin{subfigure}[ ]
\centering
\includegraphics[width=0.25\linewidth]{./figures/2lb-laporta4i-20161029.epsi}
\end{subfigure}
\caption{\footnotesize 2-loop box diagrams with massive internal lines (finite diagrams)
 (a) {\it double-triangle} $N = 5$ (Laporta~\cite{laporta01},\, Fig 4(c)), 
 (b) {\it tetragon-triangle} $N = 6$ (Laporta~\cite{laporta01},\, Fig 4(d)), 
 (c) {\it pentagon-triangle} $N = 7$ (Laporta~\cite{laporta01},\, Fig 4(g)), 
 (d) {\it ladder} $N = 7$ (Laporta~\cite{laporta01},\, Fig 4(h)), 
 (e) {\it crossed ladder} $N = 7$ (Laporta~\cite{laporta01},\, Fig 4(i))}
\label{laporta-diagrams}
\end{center}
\end{figure}

The 2-loop box integrals according to Eq~\eqref{UV} for the diagrams in Fig~\ref{laporta-diagrams}
are all finite integrals and can be evaluated with $\varepsilon=0$.
Integral approximations obtained with {\sc ParInt} for the {\it double-triangle} ($N = 5$),
{\it tetragon-triangle} ($N = 6$), {\it pentagon-triangle} ($N = 7$), {\it ladder} and {\it crossed 
ladder} ($N = 7$) diagrams were presented in~\cite{ddacat13}. 
The $U$ and $W$ functions can also be found in the reference.
In the numerical evaluation, we set $s=t=1,\ p_j^2=1,\ m_r=1$ for simplicity
and for comparisons with other results in the literature.

This subsection provides timing results obtained with {\sc ParInt} on the \emph{thor} cluster,
corresponding to the five diagrams in Fig~\ref{laporta-diagrams}.
%
Table~\ref{parint-test-specs} gives a brief overview of pertinent test specifications, $T_1, ~T_p$
and the speedup $S_p = T_1/T_p$ for $p = 64.$
The times $T_p$ are expressed in seconds, as a function of the number of MPI processes $p, 1\le p\le 64.$
When referring to numbers of integrand evaluations, \emph{million} and
\emph{billion} are abbreviated by ``M" and ``B", respectively.
For instance, the times of the $N = 5$ double triangle diagram decrease
from 32.6 seconds at $p = 1,$ to \,0.74 seconds at $p = 64$ for $t_r = 10^{-10}$ (reaching a speedup of 44);
whereas the $N = 7$ crossed diagram times range between 27.6 seconds and 0.49 seconds for $t_r = 10^{-7}$
(with speedup exceeding 56).
The transformation of Eq~\eqref{cubetrans} was used.

For the two-loop crossed box problem as an example, we ran {\sc ParInt} in \emph{long double} precision. The results for
an increasing allowed maximum number of evaluations and increasingly strict (relative) error tolerance
$t_r$ (using 64 processes) are given in Table~\ref{parint-results-crossed}, as well as the corresponding double precision
results. Listed are: the integral approximation, relative error estimate $E_r,$ number of
function evaluations reached and time taken in \emph{long double} precision, followed by the
relative error estimate, number of function evaluations reached and running time in \emph{double} precision.
For a comparable number of function evaluations, the time using long doubles is slightly less than twice
the time taken using doubles. The \emph{iflag} parameter returns 0 when the requested accuracy is assumed
to be achieved, and 1 otherwise. Reaching the maximum number of evaluations results in abnormal termination
with \emph{iflag} = 1. The integral approximation for the \emph{double} computation is not listed
in Table~\ref{parint-results-crossed}; it
is consistent with the \emph{long double} result within the estimated error (which appears to be over-estimated).
Using doubles the program terminates abnormally for the requested relative accuracy of $t_r = 10^{-10}.$
Normal termination is achieved in this case around 239B evaluations with long doubles.

\begin{figure}[t]
\begin{center}
\begin{subfigure}[ ]
\centering
\includegraphics[width=0.40\linewidth]{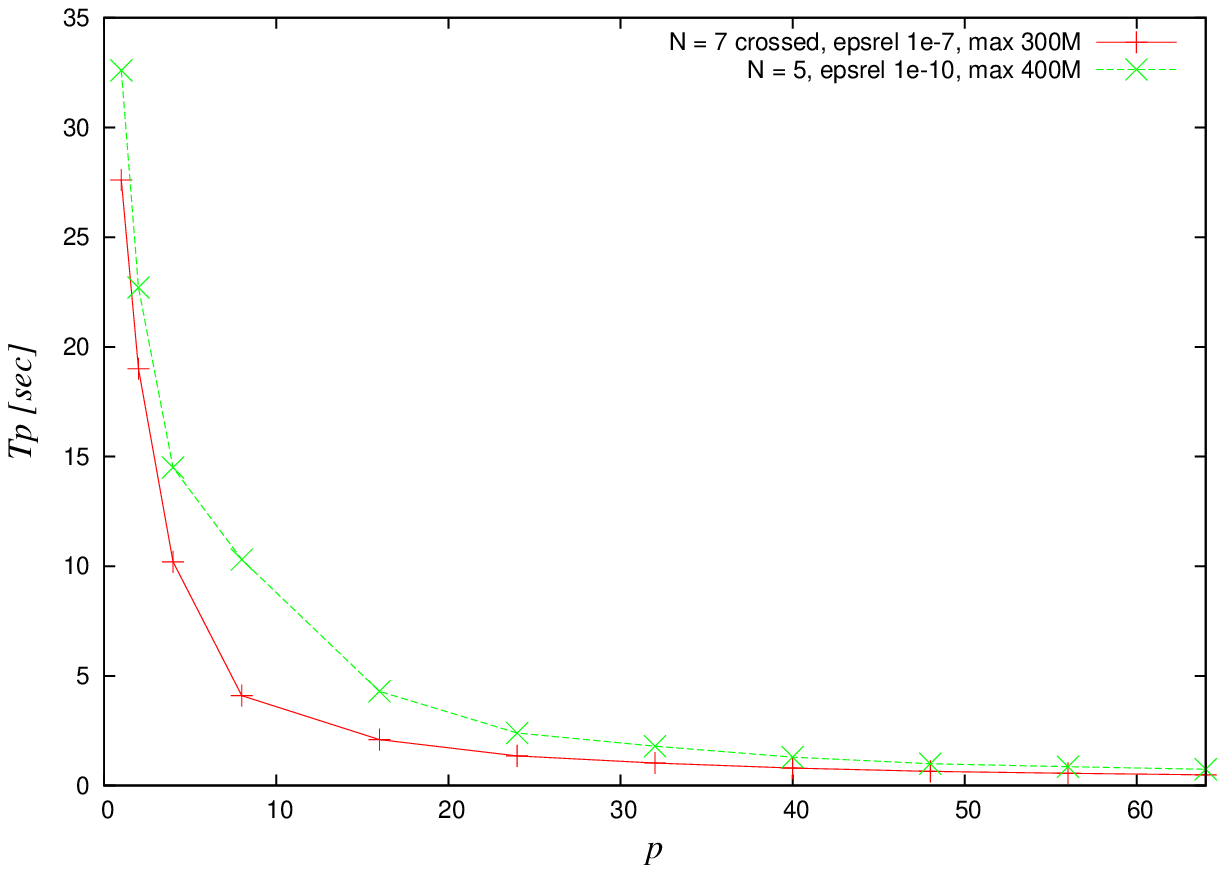} \quad
\end{subfigure}
\begin{subfigure}[ ]
\centering
\includegraphics[width=0.40\linewidth]{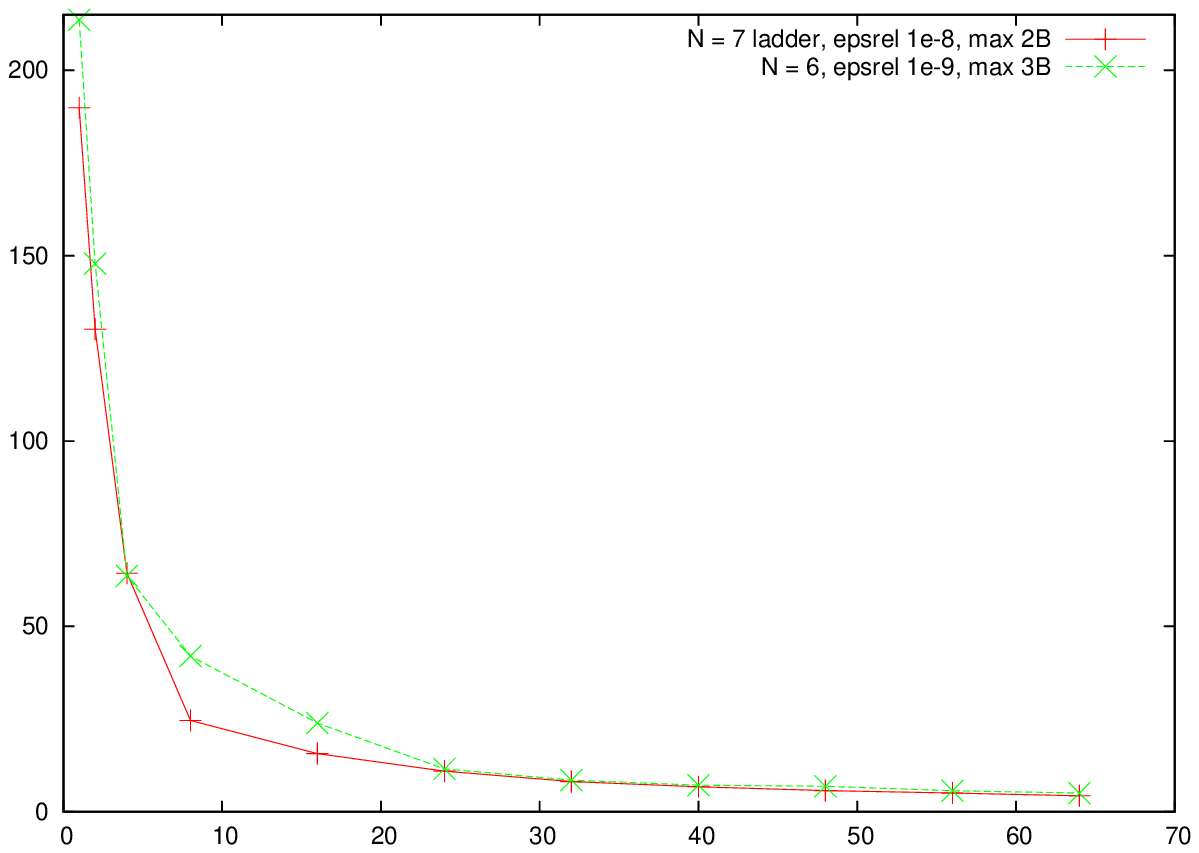}
\end{subfigure}
\begin{subfigure}[ ]
\centering
\includegraphics[width=0.41\linewidth]{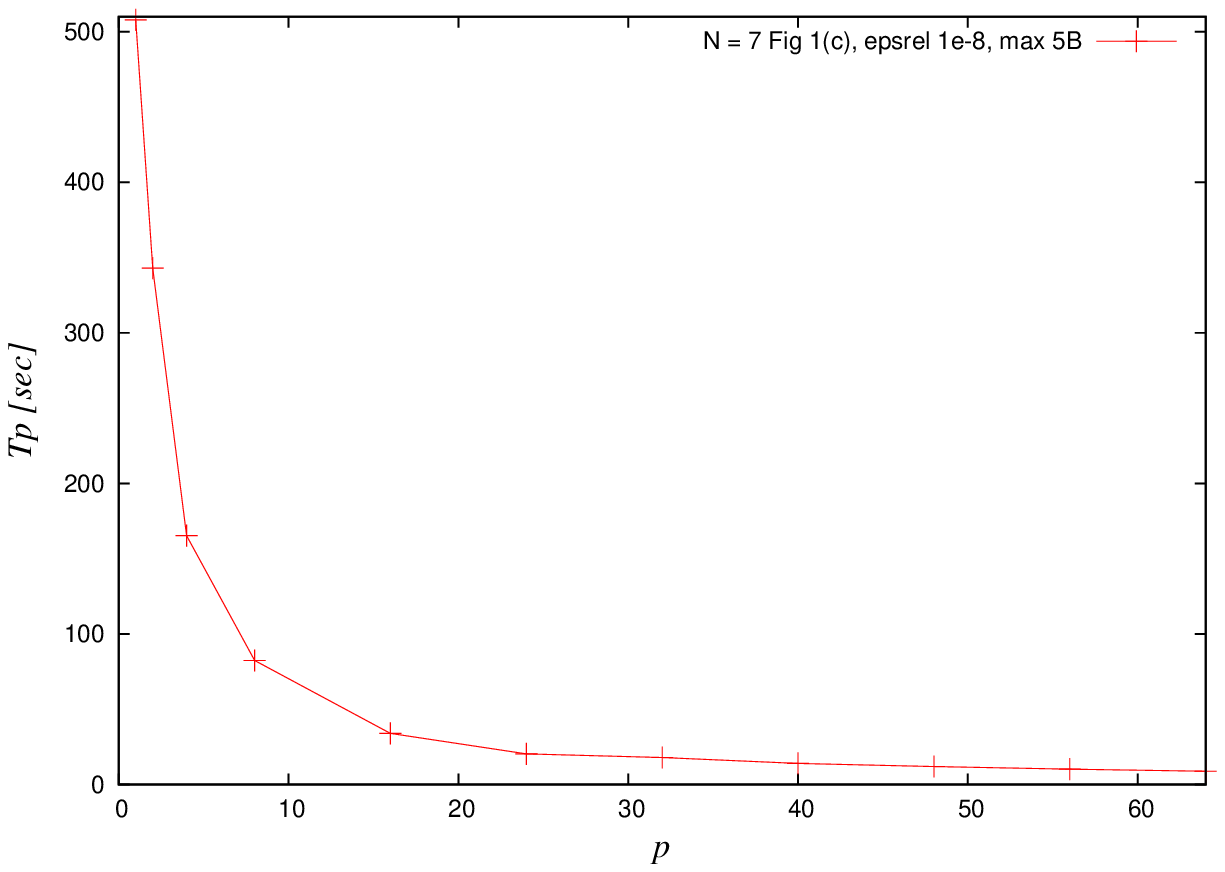} \quad
\end{subfigure}
\begin{subfigure}[ ]
\centering
\includegraphics[width=0.41\linewidth]{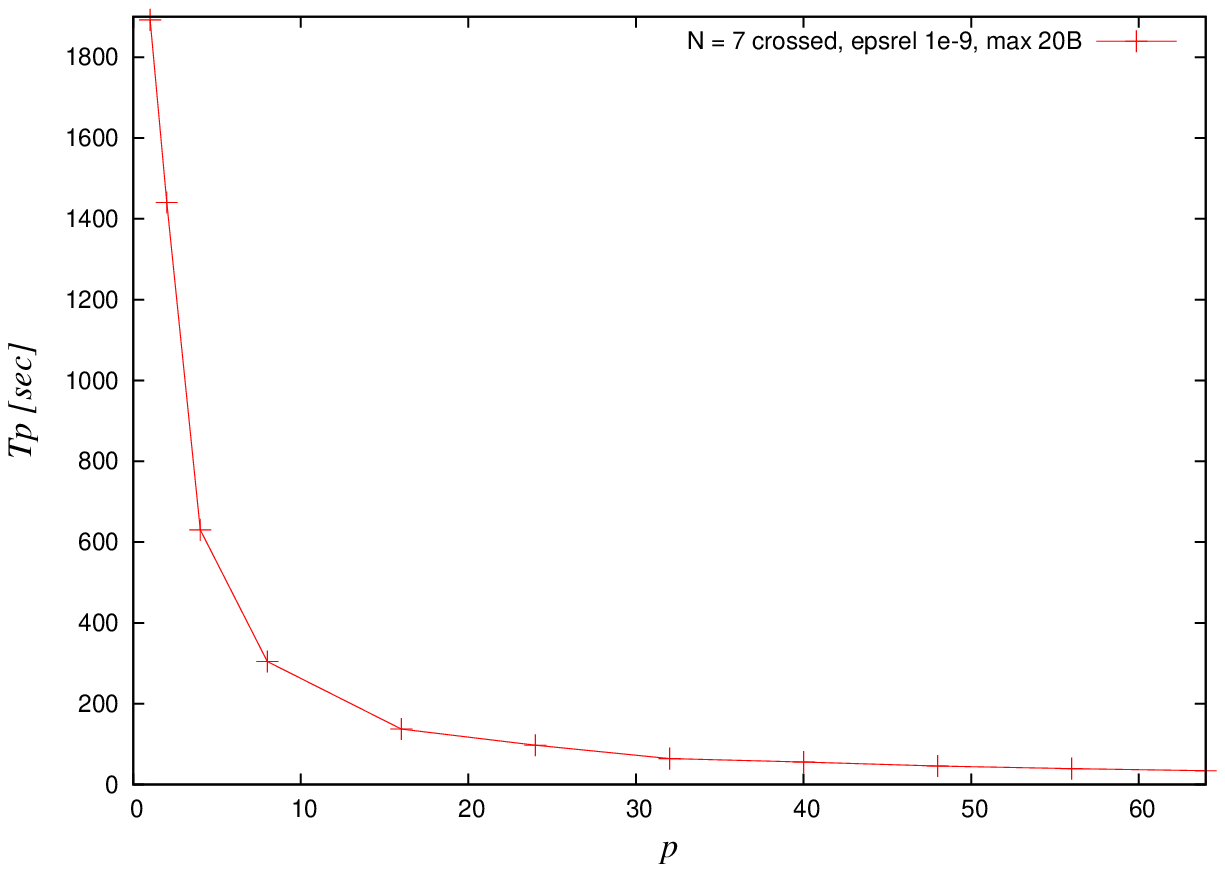}
\end{subfigure}
\caption{{\sc ParInt} timing plots for Fig~\ref{laporta-diagrams} diagrams
as a function of the number of procs. $p.$ (a) $N = 5$ [Fig~\ref{laporta-diagrams}(a)], $t_r = 10^{-10}$ and $N = 7$ [Fig~\ref{laporta-diagrams}(e)], $t_r = 10^{-7};$ (b) $N = 6$ [Fig~\ref{laporta-diagrams}(b)], $t_r = 10^{-8}$ and $N = 7$ [Fig~\ref{laporta-diagrams}(d)], $t_r = 10^{-8};$ (c) $N = 7$ [Fig~\ref{laporta-diagrams}(c)], $t_r = 10^{-8};$ (d) $N = 7$ [Fig~\ref{laporta-diagrams}(e)], $t_r = 10^{-9}$}
\label{parint-timings}
\end{center}
\end{figure}

Fig~\ref{parint-timings} shows {\sc ParInt} timing plots for the diagrams of 
Fig~\ref{laporta-diagrams}, depicting a considerable time decrease as a function of the number of processes $p.$ 
Plots with similar orders of the time are grouped in Fig~\ref{parint-timings}(a) and in 
Fig~\ref{parint-timings}(b).

Timing results obtained with parallel (multi-threaded) iterated integration were given in~\cite{ddacat13}.
\begin{table}
\centering
\caption{\footnotesize Test specifications and range of times in Fig~\ref{parint-timings}(a)-(d)}
\begin{footnotesize}
\begin{tabular}{lccccccc}
 & & & & & & & \\
\hline
 {\sc Diagram} & {\sc Figure/Timing Plot} & $N$ & {\sc Rel Tol} & {\sc Max evals} & $T_1[s]$ & $T_{64}[s]$  & {\sc Speedup}    \\
    &      &     &  $E_r$    &             &          &              &   $S_p$ for $p = 64$   \\
\hline
 double triangle & Fig~\ref{laporta-diagrams}(a) / Fig~\ref{parint-timings}(a) & 5 & $10^{-10}$ & 400M & 32.6 & 0.74 & 44.1 \\
 crossed ladder & Fig~\ref{laporta-diagrams}(e) / Fig~\ref{parint-timings}(a) & 7 & $10^{-7}$  & 300M & 27.6 & 0.49 & 56.3 \\
 tetragon triangle & Fig~\ref{laporta-diagrams}(b) / Fig~\ref{parint-timings}(b) & 6 & $10^{-9}$  & 3B & 213.6 & 5.06 & 42.2 \\
 ladder & Fig~\ref{laporta-diagrams}(d) / Fig~\ref{parint-timings}(b) & 7 & $10^{-8}$  & 2B & 189.9 & 4.33 & 43.9 \\
 pentagon triangle & Fig~\ref{laporta-diagrams}(c) / Fig~\ref{parint-timings}(c) & 7 & $10^{-8}$  & 5B & 507.9 & 8.83 & 57.5 \\
 crossed ladder & Fig~\ref{laporta-diagrams}(e) / Fig~\ref{parint-timings}(d) & 7 & $10^{-9}$  & 20B & 1892.5 & 34.6 & 54.7 \\ \hline
\end{tabular}
\end{footnotesize}
\label{parint-test-specs}
\end{table}
\begin{table}
\centering
\caption{\footnotesize {\sc ParInt} long double and double results for crossed diagram Fig~\ref{laporta-diagrams}(e)}
\begin{scriptsize}
\begin{tabular}{crcccrrccrr}
 & & & & & & & & & & \\
\hline
$t_r$ & Max & \multicolumn{5}{c}{\emph{long double} precision} & \multicolumn{4}{c}{\emph{double} precision} \\
 & Evals & {\sc Integral Approx} & $E_r$ & \emph{iflag} & \#{\sc Evals} & {\sc Time}[s] & $E_r$ &\emph{iflag} & \#{\sc Evals} & {\sc Time} \\
\hline
$10^{-08}$ & 600M & 0.085351397048123 & 3.5e-07 & 1 &   600000793 & 1.7
                                      & 3.4E-07 & 1 &   600001115 & 0.95 \\
           &   1B & 0.085351397753978 & 1.7e-07 & 1 &  1000000141 & 2.9
                                      & 1.6e-07 & 1 &  1000000141 & 1.6 \\
           &   2B & 0.085351398064779 & 2.9e-08 & 1 &  2000000443 & 6.0
                                      & 2.9e-08 & 1 &  2000000765 & 3.3 \\
           &   6B & 0.085351398130559 & 5.6e-09 & 0 &  4164032999 & 14.3
                                      & 8.0e-09 & 0 &  4424455329 & 8.6 \\
$10^{-09}$ &  10B & 0.085351398143465 & 5.5e-09 & 1 & 10000001571 & 29.7
                                      & 5.5e-09 & 1 & 10000000927 & 16.4 \\
           &  50B & 0.085351398152623 & 9.9e-10 & 0 & 35701321579 & 124.4
                                      & 4.5e-10 & 0 &  9799638359 & 16.0 \\
$10^{-10}$ &  80B & 0.085351398153315 & 5.5e-10 & 1 & 80000001137 & 240.2
                                      & 5.5e-10 & 1 & 80000000171 & 133.5 \\
           & 100B & 0.085351398153507 & 4.1e-10 & 1 & 100000000093 & 302.0
                                      & 4.1e-10 & 1 & 100000000093 & 168.3 \\
           & 300B & 0.085351398153798 & 9.1e-11 & 0 & 238854968513 & 642.3
                                      & 1.3e-10 & 1 & 300000000279 & 587.2  \\ \hline
\end{tabular}
\end{scriptsize}
\label{parint-results-crossed}
\end{table}
%

\section{3-loop self-energy integrals} \label{3-loop-self}

In this section we deal with the integral determined by Eq~\eqref{LloopIJ} for 
$L=3$ and $n=4-2\varepsilon,$
\begin{equation}
I = (-1)^N {\Gamma\left(N-6+3\varepsilon \right)}
\int_{0}^{1}\prod_{r=1}^{N}dx_{r}\, 
\delta(1-\sum x_{r})\frac{1}{U^{2-\varepsilon}(V-i\varrho)^{N-6+3\varepsilon}}.
\label{threeLOOP}
\end{equation}
\noindent
UV divergence occurs when $U$ vanishes at the boundaries. The $\Gamma$-function in Eq~\eqref{threeLOOP} contributes to UV divergence when $N\le 6$.
As shown in the figures, the entering momentum is $p,$ and we denote $s=p^2$.

We calculate integrals adhering to Eq~\eqref{threeLOOP},
denoted by $I_a^{S3},\, I_b^{S3},\,\cdots, I_j^{S3},$
for the diagrams of Fig~\ref{3ls-self}(a)-(j), respectively.
In the following four subsections, 
we consider massless/massive internal lines and
UV finite/divergent cases.

\subsection{3-loop finite integrals with massless internal lines}\label{3ls-massless-finite-integrals}

The integrals $I_a^{S3},\, I_b^{S3},\, I_c^{S3},\, I_d^{S3}$ are finite,
with the $U,\ W$ functions given in Eqs~\eqref{SEthrLa}-\eqref{SEthrLd} below.

\begin{equation}
\mathrm{(a)}\quad
\left\{
\begin{array}{l}
U=
  x_4 x_7 x_{12356} + x_{12}   x_{47}   x_{56} 
  +x_3\, ( x_{12}  x_4+ x_{56}  x_7+ x_{12}   x_{56}  )   
\\
W/s=
  x_4 x_7  x_{15}  x_{236} 
   +x_4 \,(x_1 x_2  x_{356} + x_{12}   x_{36}  x_5) \\
   \quad +x_7 \,(x_5 x_6  x_{123} +x_1  x_{23}   x_{56} )
   +x_3 \,(x_1 x_2  x_{56} + x_{12}  x_5 x_6)    
\end{array}
\right.
\label{SEthrLa}
\end{equation}

\begin{equation}
\mathrm{(b)}\quad
\left\{
\begin{array}{l}
U=
  x_7 \,( x_{12}  x_{3456} + x_{34}   x_{56} )
   + x_{13}   x_{24}   x_{56}   + x_1 x_2  x_{34}  +  x_{12}  x_3 x_4   
\\
W/s=
  x_7 \,( x_{15}   x_{26}   x_{34} +x_1 x_2  x_{56} + x_{12}  x_5 x_6)
   +(x_1 x_3+ x_{13}  x_5) \,(x_2 x_4+x_2 x_6+x_4 x_6)   
\end{array}
\right.
\label{SEthrLb}
\end{equation}

\begin{equation}
\mathrm{(c)}\quad
\left\{
\begin{array}{l}
U=
  x_5 x_8 x_{123467}
   +x_5  x_{12}  x_{3467}  + x_8 x_{1234}   x_{67} 
   + x_{12}   x_{34}   x_{67}    
\\
W/s=
  x_5 x_8  x_{136}   x_{247} 
  + x_5 \,(x_1 x_2 x_{3467} + x_{12}   x_{36}   x_{47} ) \\
 \quad + x_8 \,(x_6 x_7 x_{1234} + x_{13}   x_{24}   x_{67} )
  + x_1 x_2  x_{34}   x_{67} + x_{12} x_3 x_4   x_{67} + x_{12}   x_{34} x_6 x_7    
\end{array}
\right.
\label{SEthrLc}
\end{equation}

\begin{equation}
\mathrm{(d)}\quad
\left\{
\begin{array}{l}
U=
  x_5 x_8 x_{123467}
   +x_5  x_{124}   x_{367} 
   +x_8  x_{123}   x_{467} 
   + x_{12}  x_3  x_{467} + x_{123}  x_4  x_{67}    
\\
W/s=
  x_5 x_8  x_{136}   x_{247} 
   +x_5 \,( x_{17}   x_{24}   x_{36} +x_1 x_7 x_{2346} ) \\
 \quad  +x_8 \,( x_{26}   x_{13}   x_{47} +x_2 x_6 x_{1347} )
   + x_{34}  \,(x_1 x_2  x_{67} + x_{12}  x_6 x_7)
   +x_3 x_4  x_{17}   x_{26}    
\end{array}
\right.
\label{SEthrLd}
\end{equation}

In this subsection we take $m_r=0$ for the internal lines.
In the absence of divergences we set $\varrho=\varepsilon=0$ in
Eq~\eqref{threeLOOP} for  $I_a^{S3},\, I_b^{S3},\, I_c^{S3},$
and show the results in Table~\ref{table-rst-eps0}.
However, the integral $I_d^{S3}$ is problematic with $\varrho=\varepsilon=0$
in view of integrand singularities, and we use the extrapolation method 
with either $\varepsilon=0$ and in the limit as $\rho \rightarrow 0,$ 
or with $\varrho=0$ and in the limit as $\varepsilon \rightarrow 0.$
The analytic result is the same for the four diagrams (see~\cite{baikov10}),
\begin{equation}\label{BC-3loop}
I_a^{S3}=I_b^{S3}=I_c^{S3}=I_d^{S3}=
20 \,\zeta_5 = 20.738555102867\ldots\,. \nonumber
\end{equation}

\begin{figure}[h]
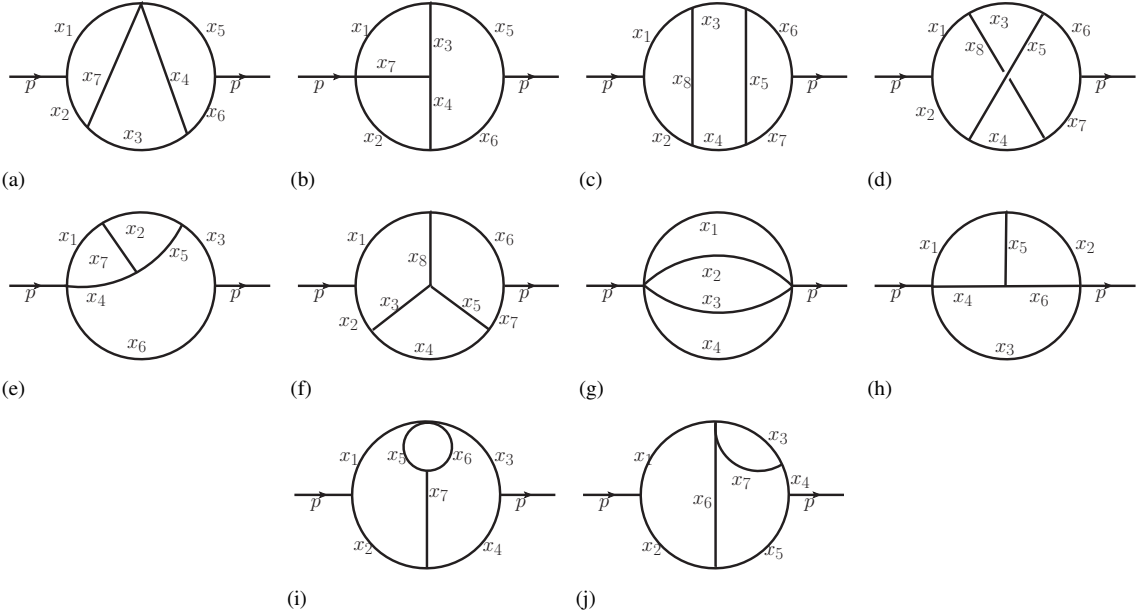

\begin{center}
\begin{subfigure}[ ]
\centering
\includegraphics[width=0.22\linewidth]{./figures/3ls-fig6a-20161029.epsi}
\end{subfigure}
\begin{subfigure}[ ]
\centering
\includegraphics[width=0.22\linewidth]{./figures/3ls-fig6b-20161029.epsi}
\end{subfigure}
\begin{subfigure}[ ]
\centering
\includegraphics[width=0.22\linewidth]{./figures/3ls-fig6c-20161029.epsi}
\end{subfigure}
\begin{subfigure}[ ]
\centering
\includegraphics[width=0.22\linewidth]{./figures/3ls-fig6d-20161029.epsi}
\end{subfigure}
\begin{subfigure}[ ]
\centering
\includegraphics[width=0.22\linewidth]{./figures/3ls-fig6e-20161029.epsi}
\end{subfigure}
\begin{subfigure}[ ]
\centering
\includegraphics[width=0.22\linewidth]{./figures/3ls-fig6f-20161029.epsi}
\end{subfigure}
\begin{subfigure}[ ]
\centering
\includegraphics[width=0.22\linewidth]{./figures/3ls-fig6g-20161029.epsi}
\end{subfigure}
\begin{subfigure}[ ]
\centering
\includegraphics[width=0.22\linewidth]{./figures/3ls-fig6h-20161029.epsi}
\end{subfigure}
\begin{subfigure}[ ]
\centering
\includegraphics[width=0.22\linewidth]{./figures/3ls-fig6i-20161029.epsi}
\end{subfigure}
\begin{subfigure}[ ]
\centering
\includegraphics[width=0.22\linewidth]{./figures/3ls-fig6j-20161029.epsi}
\end{subfigure}
\caption{3-loop self-energy diagrams with massive and massless internal lines 
(finite and UV-divergent diagrams), cf. Laporta~\cite{laporta01}, Baikov and Chetyrkin~\cite{baikov10}: 
(a) $N=7$, (b) $N=7$, (c) $N=8$, (d) $N=8$, (e) $N=7$, 
(f) $N=8$, (g) $N=4$, (h) $N=6$, (i) $N=7$, (j) $N=7$}
\label{3ls-self}
\end{center}
\end{figure}

\begin{figure}[h]
\begin{center}
\begin{subfigure}[ ]
\centering
\includegraphics[width=0.32\linewidth]{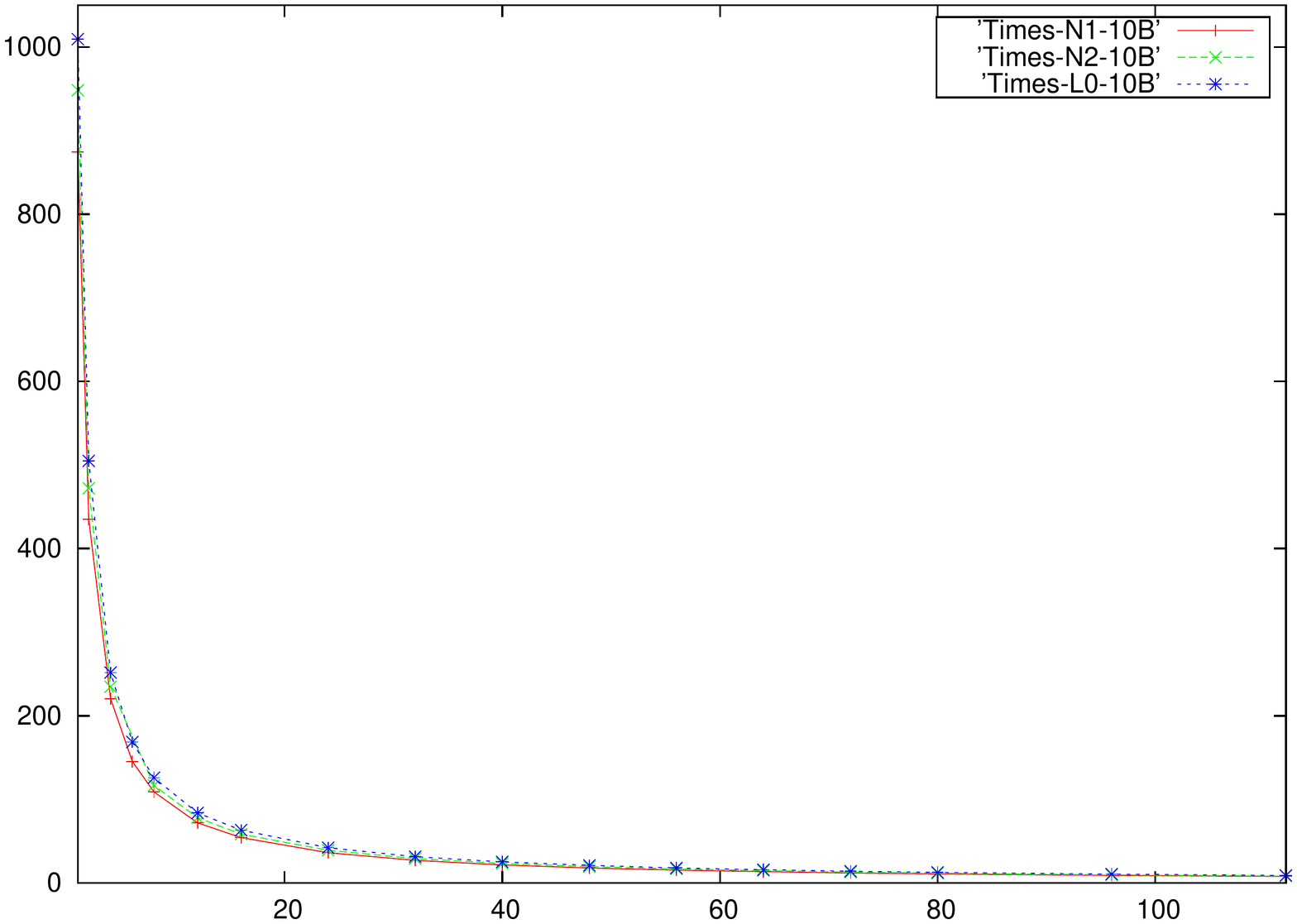}
\end{subfigure}
\begin{subfigure}[ ]
\centering
\includegraphics[width=0.32\linewidth]{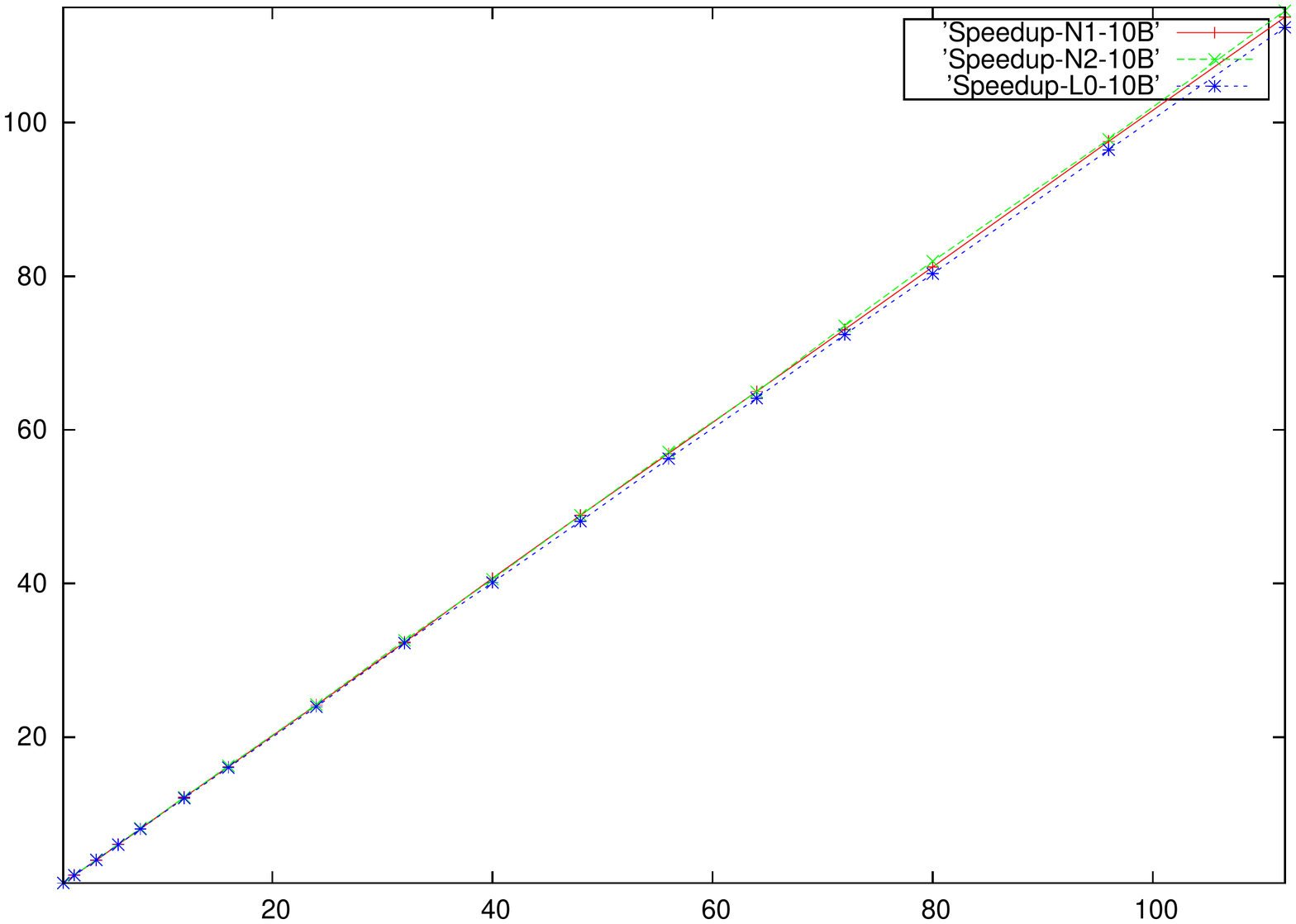} \\%
\end{subfigure}
\caption{{\sc ParInt} parallel performance for computation
of integrals (Fig~\ref{3ls-self}(a-c)) using 10B evaluations:
(a) Computation times (in seconds), (b) $S_{peedup}=T_1/T_p$}
\label{fig:performance}
\end{center}
\end{figure}

{\sc ParInt} numerical results and timings were given in~\cite{iccs15} for the 3-loop diagrams of Fig~\ref{3ls-self}(a)-(d),
from runs on 16-core nodes 
of the \emph{thor} cluster. 
For the integration over each subregion, the rule of degree 9 is used (see Section~\ref{parint}), 
which evaluates the 6D integrand at 453 points and the 7D integrand at 717 points.
These integrals are transformed to the unit cube according to Eq~\eqref{cubetrans}.
Table~\ref{table-rst-eps0} lists the results obtained directly with $\varrho = \varepsilon = 0$
for Fig~\ref{3ls-self}(a)-(c) by {\sc ParInt} using 48 processes: integral approximation,
relative error estimate $E_r,$ and time in seconds for various total numbers 
of function evaluations in double and long double precision.
Fig~\ref{fig:performance} shows times and speedups
as a function of the number of processes $p,$ for a computation
of the integrals of Fig~\ref{3ls-self}(a)-(c) using 10B integrand evaluations
in double precision.
Denoting the time in seconds for $p$ processes by $T_p[s],$ the corresponding speedup
given by $S_p = T_1/T_p$ is nearly optimal --
which would coincide with the diagonal in the graph, or slightly
superlinear ($ > p$) over the given range of $p.$

\begin{table}
\caption{\footnotesize {\sc ParInt} accuracy and times with 48 procs.~for loop integrals of Fig~\ref{3ls-self}(a)-(c) with massless internal lines,
using \,$\varrho = 0,$\, and various numbers of function evaluations; $E_r = $ 
integration estim. rel. error}
\begin{scriptsize}
\begin{center}
\begin{tabular}{cccccccc}\hline
 & & \multicolumn{3}{c}{\sc double precision}
 & \multicolumn{3}{c}{\sc long double precision} \\
Diagram & \# {\sc Fcn.} \hspace*{-1mm} & {\sc Integral} \hspace{-1mm} & \hspace{-1mm}{\sc Rel. err. }\hspace{-1mm} & \hspace{-2mm}{\sc Time}
  \hspace*{0mm} & {\sc Integral} \hspace{-1mm} & \hspace{-1mm}{\sc Rel. err. }\hspace{-1mm} & \hspace{-2mm}{\sc Time} \\
 & {\sc Evals.} \hspace*{-1mm} & {\sc Result} \hspace{-1mm} & \hspace{-1mm}{{\sc Est.} $E_r$ }\hspace{-1mm} & \hspace{-2mm}{\sc T[s]}\hspace{-1mm} & \hspace{-1mm}{\sc Result} \hspace{-1mm} & \hspace{-1mm}{{\sc Est.} $E_r$ }\hspace{-1mm} & \hspace{-1mm}{\sc T[s]} \\
 ~~\\
\hline
 ~~\\
\emph{Exact:}&  & 20.73855510 & & & 20.73855510 & & \\
\hline
 ~~\\
Fig~\ref{3ls-self} (a) & 5B & 20.73871652 & 2.21e-05 & 9.0 & 20.73871522 & 2.5e-05 & 16.0 \\
    & 10B & 20.73856839 & 3.50e-06 & 17.9 & 20.73856878 & 3.42e-06 & 32.1 \\
    & 25B & 20.73855535 & 3.79e-07 & 44.9 & 20.73855539 & 3.71e-07 & 80.5 \\
    & 50B & 20.73855508 & 9.07e-08 & 90.3 & 20.73855508 & 8.94e-08 & 161.1 \\
    & 75B & 20.73855507 & 4.26e-08 & 135.6 & 20.73855507 & 4.23e-08 & 242.1 \\
    & 100B & 20.73855508 & 2.59e-08 & 180.8 & 20.73855508 & 2.56e-08 & 323.2 \\
\hline
~~\\
Fig~\ref{3ls-self} (b)& 5B & 20.73933800 & 3.69e-05 & 9.7 & 20.73933292 & 3.63e-05 & 17.5 \\
    & 10B & 20.73872210 & 6.64e-06 & 19.4 & 20.73872078 & 6.61e-06 & 35.1 \\
    & 25B & 20.73857098 & 8.32e-07 & 48.6 & 20.73857018 & 8.12e-07 & 87.9 \\
    & 50B & 20.73855716 & 1.96e-07 & 98.2 & 20.73855718 & 1.95e-07 & 175.9 \\
    & 75B & 20.73855576 & 9.28e-08 & 146.4 & 20.73855575 & 9.12e-08 & 264.4 \\
    & 100B & 20.73855540 & 5.68e-08 & 196.7 & 20.73855540 & 5.43e-08 & 352.6 \\
\hline
 ~~\\
Fig~\ref{3ls-self} (c)& 5B & 20.74194961 & 1.19e-03 & 10.3 & 20.74196270 & 1.19e-03 & 19.6 \\
    & 10B & 20.73886434 & 3.64e-04 & 20.7 & 20.73880437 & 3.51e-04 & 39.2 \\
    & 25B & 20.73827908 & 5.04e-05 & 51.7 & 20.73827607 & 5.04e-05 & 98.2 \\
    & 50B & 20.73841802 & 1.09e-05 & 103.4 & 20.73842495 & 9.79e-06 & 196.7 \\
    & 75B & 20.73848662 & 3.66e-06 & 146.4 & 20.73848624 & 3.70e-06 & 295.0 \\
    & 100B & 20.73851402 & 1.96e-06 & 207.5 & 20.73851338 & 1.98e-06 & 393.9 \\
\hline
\end{tabular}
\end{center}
\end{scriptsize}
\label{table-rst-eps0}
\end{table}

\begin{table}
\begin{center}
\caption{{\footnotesize Integration with {\sc ParInt} using 64 procs., max. \# evals = 150B, $\varrho=\varrho_\ell = 2^{-\ell}, \ell=20,21,\ldots$ and extrapolation with $\epsilon$-algorithm for Fig~\ref{3ls-self}(d) integral with massless internal lines}}
\label{table:extrap-v}
\begin{scriptsize}
 ~~\\
\begin{tabular}{cccccc}\hline
 & \multicolumn{3}{c}{{\sc Integral Fig}~\ref{3ls-self}(d)}
 & \multicolumn{2}{c}{\sc Extrapolation} \\
 {\hspace*{-5mm}$\ell$}\hspace*{-5mm} & {\sc Integral} \hspace{-1mm} & \hspace{-1mm}{ $E_r$ }\hspace{-1mm} & \hspace{-1mm}{\sc T[s]}
\hspace{-1mm} & \hspace{-1mm}{\sc Last} & Selected \\
\hline
 & & & & & \\
 \hspace*{-1mm} 20\hspace*{-1mm}& \hspace{-1mm} 19.69036128576084 \hspace{-1mm} & \hspace{-1mm}1.44e-07\hspace{-1mm} & \hspace{-1mm}474.5 & &  \\
 \hspace*{-1mm} 21\hspace*{-1mm}& \hspace{-1mm} 19.91633676759658 \hspace{-1mm} & \hspace{-1mm}1.64e-07\hspace{-1mm} & \hspace{-1mm}474.6 & &  \\
 \hspace*{-1mm} 22\hspace*{-1mm}& \hspace{-1mm} 20.09256513888053 \hspace{-1mm} & \hspace{-1mm}1.84e-07\hspace{-1mm} & \hspace{-1mm}474.7 & 20.71685142 & 20.71685142 \\
 \hspace*{-1mm} 23\hspace*{-1mm}& \hspace{-1mm} 20.23033834092921 \hspace{-1mm} & \hspace{-1mm}1.94e-07\hspace{-1mm} & \hspace{-1mm}474.7 & 20.72393791 & 20.72393792 \\
 \hspace*{-1mm} 24\hspace*{-1mm}& \hspace{-1mm} 20.33827222472266 \hspace{-1mm} & \hspace{-1mm}2.16e-07\hspace{-1mm} & \hspace{-1mm}474.7 & 20.73801873 & 20.73801873 \\
 \hspace*{-1mm} 25\hspace*{-1mm}& \hspace{-1mm} 20.42297783943613 \hspace{-1mm} & \hspace{-1mm}2.36e-07\hspace{-1mm} & \hspace{-1mm}474.7 & 20.73801873 & 20.73801873 \\
 \hspace*{-1mm} 26\hspace*{-1mm}& \hspace{-1mm} 20.48955227010659 \hspace{-1mm} & \hspace{-1mm}2.53e-07\hspace{-1mm} & \hspace{-1mm}474.7 & 20.73815511 & 20.73815511 \\
 \hspace*{-1mm} 27\hspace*{-1mm}& \hspace{-1mm} 20.54194208640818 \hspace{-1mm} & \hspace{-1mm}2.74e-07\hspace{-1mm} & \hspace{-1mm}474.7 & 20.73825979 & 20.73825979 \\
 \hspace*{-1mm} 28\hspace*{-1mm}& \hspace{-1mm} 20.58321345028519 \hspace{-1mm} & \hspace{-1mm}2.95e-07\hspace{-1mm} & \hspace{-1mm}474.7 & 20.73811441 & 20.73811441 \\
 \hspace*{-1mm} 29\hspace*{-1mm}& \hspace{-1mm} 20.61575568045697 \hspace{-1mm} & \hspace{-1mm}3.14e-07\hspace{-1mm} & \hspace{-1mm}474.8 & 20.74104946 & 20.73840576 \\
 \hspace*{-1mm} 30\hspace*{-1mm}& \hspace{-1mm} 20.64143527978811 \hspace{-1mm} & \hspace{-1mm}3.34e-07\hspace{-1mm} & \hspace{-1mm}474.6 & 20.73854289 & 20.73833347 \\
 \hspace*{-1mm} 31\hspace*{-1mm}& \hspace{-1mm} 20.66171281606668 \hspace{-1mm} & \hspace{-1mm}3.51e-07\hspace{-1mm} & \hspace{-1mm}474.7 & 20.73861767 & 20.73855952 \\
 \hspace*{-1mm} 32\hspace*{-1mm}& \hspace{-1mm} 20.67774379215522 \hspace{-1mm} & \hspace{-1mm}4.16e-07\hspace{-1mm} & \hspace{-1mm}474.6 & 20.73855567 & 20.73847582 \\
\hline
 & & & {Eq}~\eqref{BC-3loop}: & 20.73855510 & 20.73855510 \\
\hline
\end{tabular}
\end{scriptsize}
\end{center}
\end{table}


With $\varrho = 0$ the integrand has boundary singularities.
For example, the integrals of Fig~\ref{3ls-self}(a)-(d) have a zero denominator
with $U = 0$ \,at\, $x_2 = 1$ and the other variables 0
(on the boundary of the unit simplex).
Thus the integrand program codes test for zero denominators.
However some of the computations overflow by integrand evaluations in the vicinity of the singularities, which is found to occur for Fig~\ref{3ls-self} (d) in double precision at 1B\,=\,$10^9$ evaluations or higher.

\begin{table}
\caption{{\footnotesize Integration with {\sc ParInt} using 64 procs., max. \# evals = 100B, $\varepsilon=\varepsilon_\ell = 2^{-\ell}, \ell=8,9,\ldots$ and extrapolation with $\epsilon$-algorithm for Fig~\ref{3ls-self}(d) integral with massless internal lines}}
\label{table:dimreg-r}
\begin{scriptsize}
\begin{center}
\begin{tabular}{cccccc}\hline
 & \multicolumn{3}{c}{{\sc Integral Fig}~\ref{3ls-self}(d)}
 & \multicolumn{2}{c}{\sc Extrapolation} \\
 {\hspace*{-5mm}$\ell$}\hspace*{-5mm} & {\sc Integral} \hspace{-1mm} & \hspace{-1mm}{ $E_r$ }\hspace{-1mm} & \hspace{-1mm}{\sc T[s]}
\hspace{-1mm} & \hspace{-1mm}{\sc Last} & Selected \\
\hline
 & & & & & \\
 \hspace*{-1mm} 8\hspace*{-1mm}& \hspace{-1mm} 21.21987706233486 \hspace{-1mm} & \hspace{-1mm}8.84e-08 \hspace{-1mm} & \hspace{-1mm}648.5 & &  \\
 \hspace*{-1mm} 9\hspace*{-1mm}& \hspace{-1mm} 20.97727482739239 \hspace{-1mm} & \hspace{-1mm}8.69e-08 \hspace{-1mm} & \hspace{-1mm}648.0 & &  \\
 \hspace*{-1mm} 10\hspace*{-1mm}& \hspace{-1mm} 20.85743468356065 \hspace{-1mm} & \hspace{-1mm}8.61e-08\hspace{-1mm} & \hspace{-1mm}649.0 & 20.74044694 & 20.74044693 \\
 \hspace*{-1mm} 11\hspace*{-1mm}& \hspace{-1mm} 20.79787566374119 \hspace{-1mm} & \hspace{-1mm}8.56e-08\hspace{-1mm} & \hspace{-1mm}647.7 & 20.73903010 & 20.73903010 \\
 \hspace*{-1mm} 12\hspace*{-1mm}& \hspace{-1mm} 20.76818590083646 \hspace{-1mm} & \hspace{-1mm}8.53e-08\hspace{-1mm} & \hspace{-1mm}648.1 & 20.73855734 & 20.73855734 \\
 \hspace*{-1mm} 13\hspace*{-1mm}& \hspace{-1mm} 20.75336343266232 \hspace{-1mm} & \hspace{-1mm}8.39e-08\hspace{-1mm} & \hspace{-1mm}648.3 & 20.73855626 & 20.73855626 \\
 \hspace*{-1mm} 14\hspace*{-1mm}& \hspace{-1mm} 20.74595780282081 \hspace{-1mm} & \hspace{-1mm}8.29e-08\hspace{-1mm} & \hspace{-1mm}647.6 & 20.73855580 & 20.73855580 \\
 \hspace*{-1mm} 15\hspace*{-1mm}& \hspace{-1mm} 20.74225639032920 \hspace{-1mm} & \hspace{-1mm}8.21e-08\hspace{-1mm} & \hspace{-1mm}647.6 & 20.73855592 & 20.73855592 \\
\hline
 & & & {Eq}~\eqref{BC-3loop}: & 20.73855510 & 20.73855510 \\
\hline
\end{tabular}
\end{center}
\end{scriptsize}
\end{table}

For the integral $I_d^{S3},$ we first take $\varepsilon=0$ and use the following form with non-zero
$\varrho:$
\begin{equation}
\Re{e}\,I_d^{S3} =  {\Gamma\left(2 \right)}
\int_{0}^{1}\prod_{r=1}^{8}dx_{r}\, \delta(1-\sum x_{r})
\frac{V^2-\varrho^2}{U^2(V^2+\varrho^2)^2}.
\label{ReN=8integral}
\end{equation}
Table~\ref{table:extrap-v}
shows an extrapolation as $\varrho\rightarrow 0$ using the $\epsilon$-algorithm of
Wynn~\cite{shanks55,wynn56,sidi96,sidi03,sidi11} (see Section~\ref{extrap-exp}).
The $\varrho_\ell$ geometric sequence is computed with base 2, \,$\varrho_\ell = 2^{-\ell}$
and the integration is performed in long double precision using 150B evaluation points.
The \emph{Selected} column lists the element along the new lower diagonal that is presumed
the best, based on its distance from the neighboring elements as computed by the $\epsilon$-algorithm
function from {\sc Quadpack}.
The \emph{Last} column lists the final (utmost right) element computed in the lower
diagonal. Overall the $\epsilon$-algorithm function from {\sc Quadpack} appears successful at
selecting a competitive element as its result for the iteration.

For an extrapolation as $\varepsilon \rightarrow 0$ we set $\varrho=0,$ so that
\begin{equation}
I_d^{S3} = {\Gamma\left(2+3\varepsilon \right)}
\int_{0}^{1}\prod_{r=1}^{8}dx_{r}\, 
\delta(1-\sum x_{r})
\frac{1}{U^{2-\varepsilon}V^{2+3\varepsilon}}.
\label{IDS3}
\end{equation}
Baikov and Chetyrkin~\cite{baikov10} derive asymptotic expansions in integer powers of 
\,$\varepsilon$\, for 3- and 4-loop integrals arising from diagrams with massless propagators.
Table~\ref{table:dimreg-r} gives an extrapolation in $\varepsilon$
for the integral of the Fig~\ref{3ls-self}(d) diagram,
using 100B evaluations for the integrations in long double precision.
The results show good agreement with the literature~\cite{baikov10,smirnov10}.
The integrand with $\varrho = 0$ in Eq~\eqref{IDS3} has a singular behavior 
with $UV = 0$ at the boundaries of the domain. 
The extrapolation converges faster than that with respect to $\varrho$ in
Table~\ref{table:extrap-v}.
The times are larger compared to those of
Table~\ref{table:extrap-v},
likely by calling the \emph{pow} function (in the C programming language) for each integrand evaluation, whereas the integrand
of Eq~\eqref{ReN=8integral}
for the $\varrho$ extrapolation can be calculated using only multiplications, divisions,
additions and subtractions.

\subsection{3-loop finite integrals with massive internal lines}
\begin{table}
\caption{\footnotesize Parallel performance of {\sc ParInt} for 3-loop diagrams of Fig~\ref{3ls-self} (a)-(f) with massive internal lines, abs. tolerance
$t_a = 5\times 10^{-10},$ and max. number of evaluations = 10B}
\centering
\begin{scriptsize}
\begin{tabular}{cccllccc}
 & & & & & & & \\
\hline
3-loop & $N$ & Result & Result & Result & $T_1[s]$ & $T_{64}[s]$  & $S_{64}$   \\
 diag. &     & Laporta~\cite{laporta01} & $p = 1$ & $p = 64$ &          &              &            \\
\hline
 & & & & & & & \\
 {Fig~\ref{3ls-self}~(a)} & 7 & 2.00250004111 & {\color{black} 2.0025000411}3 & {\color{black} 2.002500041}2 & 879.3 & 13.4 & 65.6 \\
 {Fig~\ref{3ls-self}~(b)} & 7 & 1.34139924145 & {\color{black} 1.3413992414}7 & {\color{black} 1.341399241}6 & 1026.2 & 14.4 & 71.3 \\ 
 {Fig~\ref{3ls-self}~(c)} & 8 & 0.27960892328 & {\color{black} 0.27960892}27 & {\color{black} 0.27960892}0 & 1019.7 & 15.9 & 64.1 \\
 {Fig~\ref{3ls-self}~(d)}  & 8 & 0.14801330396 & {\color{black} 0.148013303}6 & {\color{black} 0.14801330}26 & 976.6 & 16.4 & 59.5 \\
 {Fig~\ref{3ls-self}~(e)} & 7 & 1.32644820827 & {\color{black} 1.32644820}6 & {\color{black} 1.326448}19 & 902.7 & 15.8 & 57.1 \\
 {Fig~\ref{3ls-self}~(f)} & 8 & 0.18262723754 & {\color{black} 0.182627237}2 & {\color{black} 0.18262723}68 & 1018.3 & 15.8 & 64.4 \\ \hline
\end{tabular}
\end{scriptsize}
\label{parint-test3l-specs}
\end{table}
For a set of 3-loop self-energy diagrams with massive internal lines given in Fig~\ref{3ls-self} (a)-(f),
corresponding numerical results and {\sc ParInt} performance results are shown in Table~\ref{parint-test3l-specs}.
The $U,\ W$ functions for (a)-(d) are given in the previous subsection and
those for (e) and (f) are listed in Eqs~\eqref{SEthrLe}-\eqref{SEthrLf} below.

\begin{equation}
\mathrm{(e)}\quad
\left\{
\begin{array}{l}
U=
  x_6 x_7 x_{1245} + x_7  x_{12}   x_{345} 
  +(x_1 x_2+x_3 x_7)  \,x_{45} + x_{14}   x_{25}   x_{36} 
  + x_{12}  x_4 x_5   
\\
W/s=
  x_6 \,(x_7  x_{12}   x_{345} +(x_1 x_2+x_3 x_7)  \,x_{45} 
   + x_{12}  x_4 x_5 +x_3  x_{14}   x_{25} )    
\end{array}
\right.
\label{SEthrLe}
\end{equation}

\begin{equation}
\mathrm{(f)}\quad
\left\{
\begin{array}{l}
U=
  (x_3 x_5+x_5 x_8+x_3 x_8) \,x_{12467}
  +x_5  x_{12}   x_{467}   +x_3  x_{124}   x_{67} 
  +x_4  x_{12}   x_{67}  + x_4 x_8 x_{1267}    
\\
W/s=
  (x_3 x_5+x_5 x_8+x_3 x_8)  \,x_{16}   x_{247} 
  +x_5 \,(x_1 x_2  x_{467} + x_{12}   x_{47}  x_6) \\
 \quad +x_3 \,(x_1 x_6  x_{247} + x_{16}   x_{24}  x_7)
  +x_4 \,(x_8  x_{16}   x_{27}  + x_1 x_2  x_{67} + x_{12}x_6 x_7  )   
\end{array}
\right.
\label{SEthrLf}
\end{equation}

In order to compare our integral approximations with Laporta's~\cite{laporta01},
we set all masses $m_r = 1$ and $s = 1,$ and furthermore divide the integral by $~\Gamma(1+\varepsilon)^3.$
The integrals are transformed from the (unit) simplex to the (unit) cube
according to the transformation of Eqs~\eqref{cubetrans} and~\eqref{transint}
and the integration is taken over the cube, using a basic integration rule of polynomial
degree 9 (see Section~\ref{parint}) and a maximum total of 10B evaluations.
The function evaluations are distributed over all the processes.
The absolute tolerance is $t_a = 5\times10^{-10}$ and the maximum number of 
integrand evaluations is $10B = 10^{10}$ (which is reached in producing 
the results of Table~\ref{parint-test3l-specs}). 

The results in Table~\ref{parint-test3l-specs} are given for $p = 1$ and for $p = 64$ {\sc MPI} processes.
$T_1$ is the time with one process and $T_{64}$ is the parallel time on the \emph{thor} cluster with $p = 64$ processes, distributed over four 16-core, 2.6\,GHz compute nodes
and using the Infiniband interconnect for message passing via MPI. 
The speedup $S_{64} = T_1/T_{64}$ 
indicates good scalability of the parallel implementation (see also~\cite{acat14,ccp14}).
%
Note that superlinear speedups ($S_{64}$) are obtained in some cases, where the 
speedup exceeds the number of processes. This is partially due to the fact that the 
timing is done within {\sc ParInt} after the processes are started. 
It may also be noted that the adaptive partitioning reaches somewhat more accuracy sequentially.
Each process has its own priority queue, keyed with the absolute error estimates over their
region. This may lead to unnecessary work by the processes locally, 
which increases with the number of processes. 

Tables~\ref{table1-3ls} and~\ref{table4-3ls} are computed with consecutive calls to {\verb pi_integrate() } in
a loop and linear extrapolation, for the functions depicted in Fig~\ref{3ls-self}(b) and (f), respectively.
The values of $C_0, C_1$ and $C_2$ are listed ($\kappa = 0$ in Eq~\eqref{asymp}).
Results from extrapolation with the $\epsilon$-algorithm are shown in Table~\ref{table2-3ls}
for the diagram in Fig~\ref{3ls-self}(b). More extrapolations
are needed with the $\epsilon$-algorithm than with linear extrapolation. In this case
the latter is more accurate and efficient but utilizes knowledge of the structure of the 
asymptotic expansion, i.e., that $\varphi_k(\varepsilon) = \varepsilon^k;$ 
this is not assumed for the non-linear extrapolation with the $\epsilon$-algorithm.

Tables~\ref{table3-3ls} and~\ref{table5-3ls} illustrate the vector function integration capability
of {\sc ParInt} to calculate the entry sequence for the extrapolation (as a vector integral result - see~\eqref{general}) with one call to 
{\verb pi_integrate()}. This procedure delivers excellent accuracy and efficiency. 
Note that the integration of the vector function in Table~\ref{table3-3ls} took 2403.4 seconds, 
compared to the total time of 3766.6 seconds for the integrations listed in Table~\ref{table1-3ls}. 
With regard to Table~\ref{table5-3ls}, the time for integrating the vector function was 
2180.2 seconds, vs. the total (sum) of 3426.3 seconds for the iterations in Table~\ref{table4-3ls}. 

\begin{table}
\caption{\footnotesize Integral and leading expansion coefficients using {\sc ParInt} with 16 procs., and linear extrapolation for 3-loop integral of Fig~\ref{3ls-self}(b), ~err. tol. $t_a = 10^{-12}$; max. \# evals = 20B, $\varepsilon=\varepsilon_\ell=2^{-\ell}, \ell=3,4,\ldots$, $T[s]$ = elapsed time (s); $E_a = $ integration estim. abs. error}
\begin{scriptsize}
\begin{center}
\begin{tabular}{ccccccc}\hline
& \multicolumn{2}{c}{{\sc Integral Fig}~\ref{3ls-self}(b)}
& \multicolumn{4}{c}{{\sc Extrapolation}} \\
 \raisebox{1mm}{\hspace*{-5mm}$\ell$}\hspace*{-5mm} & {\sc Integral} \hspace{-1mm} & \hspace{-1mm}{ $E_a$ }\hspace{-1mm} & \hspace{-1mm}{\sc T[s]}
\hspace{-1mm} & \hspace{-1mm}{\sc Result} ~$C_0$\hspace{-1mm} & \hspace*{-0mm}{\sc Result} ~$C_1$ & \hspace{-1mm}{\sc Result} ~$C_2$\hspace{-1mm} \\
\hline
 \hspace*{-1mm}${3}$\hspace*{-1mm}& \hspace{-1mm}0.89462319318517\hspace{-1mm} & \hspace{-1mm}6.33e-10\hspace{-1mm} & \hspace{-1mm}356.9\hspace{-1mm} & \hspace{-1mm}\hspace{-1mm} & \hspace*{-1.5mm}\hspace*{-1mm} & \hspace{-1mm}\hspace{-1mm}  \\
 \hspace*{-1mm}${4}$\hspace*{-1mm}& \hspace{-1mm}1.07605987265074\hspace{-1mm} & \hspace{-1mm}8.56e-10\hspace{-1mm} & \hspace{-1mm}426.3\hspace{-1mm} & \hspace{-1mm}1.257496552116\hspace{-1mm} & \hspace*{-1.5mm} -2.9029868714\hspace*{-1mm} & \hspace{-1mm}\hspace{-1mm}  \\
 \hspace*{-1mm}${5}$\hspace*{-1mm}& \hspace{-1mm}1.19524813881849\hspace{-1mm} & \hspace{-1mm}1.15e-09\hspace{-1mm} & \hspace{-1mm}426.2\hspace{-1mm} & \hspace{-1mm}1.333416355943\hspace{-1mm} & \hspace*{-1.5mm}-4.7250621633\hspace*{-1mm} & \hspace{-1mm}9.7177349\hspace{-1mm}  \\
 \hspace*{-1mm}${6}$\hspace*{-1mm}& \hspace{-1mm}1.26445377191768\hspace{-1mm} & \hspace{-1mm}1.34e-09\hspace{-1mm} & \hspace{-1mm}426.2\hspace{-1mm} & \hspace{-1mm}1.341017173944\hspace{-1mm} & \hspace*{-1.5mm}-5.1507079713\hspace*{-1mm} & \hspace{-1mm}16.5280678\hspace{-1mm}  \\
 \hspace*{-1mm}${7}$\hspace*{-1mm}& \hspace{-1mm}1.30188593759114\hspace{-1mm} & \hspace{-1mm}1.46e-09\hspace{-1mm} & \hspace{-1mm}426.2\hspace{-1mm} & \hspace{-1mm}1.341390110905\hspace{-1mm} & \hspace*{-1.5mm}-5.1954604067\hspace*{-1mm} & \hspace{-1mm}18.1988254\hspace{-1mm}  \\
 \hspace*{-1mm}${8}$\hspace*{-1mm}& \hspace{-1mm}1.32137252564963\hspace{-1mm} & \hspace{-1mm}1.52e-09\hspace{-1mm} & \hspace{-1mm}426.2\hspace{-1mm} & \hspace{-1mm}1.341399132800\hspace{-1mm} & \hspace*{-1.5mm}-5.1976978366\hspace*{-1mm} & \hspace{-1mm}18.3778198\hspace{-1mm}  \\
 \hspace*{-1mm}${9}$\hspace*{-1mm}& \hspace{-1mm}1.33131707386056\hspace{-1mm} & \hspace{-1mm}1.55e-09\hspace{-1mm} & \hspace{-1mm}426.2\hspace{-1mm} & \hspace{-1mm}1.341399240859\hspace{-1mm} & \hspace*{-1.5mm}-5.1977522983\hspace*{-1mm} & \hspace{-1mm}18.3868241\hspace{-1mm}  \\
 \hspace*{-1mm}${10}$\hspace*{-1mm}& \hspace{-1mm}1.33634079003748\hspace{-1mm} & \hspace{-1mm}1.57e-09\hspace{-1mm} & \hspace{-1mm}426.2\hspace{-1mm} & \hspace{-1mm}1.341399241503\hspace{-1mm} & \hspace*{-1.5mm}-5.1977529523\hspace*{-1mm} & \hspace{-1mm}18.3870438\hspace{-1mm}  \\
\hspace*{-1mm}${11}$\hspace*{-1mm}& \hspace{-1mm}1.33886565298449\hspace{-1mm} & \hspace{-1mm}1.58e-09\hspace{-1mm} & \hspace{-1mm}426.2\hspace{-1mm} & \hspace{-1mm}1.341399241505\hspace{-1mm} & \hspace*{-1.5mm}-5.1977529584\hspace*{-1mm} & \hspace{-1mm}18.3870480\hspace{-1mm}  \\
\hline
\multicolumn{4}{r}{Laporta~\cite{laporta01}:} & 1.341399241447 &\hspace*{-1.0mm} -5.1977529559 & 18.3870466 \\ \hline
\end{tabular}
\end{center}
\end{scriptsize}
\label{table1-3ls}
\end{table}
\begin{table}
\caption{\footnotesize{Integral and leading expansion coefficients using {\sc ParInt} with 16 procs., and extrapolation with $\epsilon$-algorithm for 3-loop integral of Fig~\ref{3ls-self}(b), ~err. tol. $t_a = 10^{-12}$; max. \# evals = 20B, $\varepsilon=\varepsilon_\ell=2^{-\ell}, \ell=3,4, \ldots$, $T[s]$ = elapsed time (s); $E_a = $ integration estim. abs. error}}
\begin{scriptsize}
\begin{center}
\begin{tabular}{ccccccc}\hline
& \multicolumn{2}{c}{{\sc Integral Fig}~\ref{3ls-self}(b)}
& \multicolumn{4}{c}{{\sc Extrapolation}} \\
 \raisebox{1mm}{\hspace*{-5mm}$\ell$}\hspace*{-5mm} & {\sc Integral} \hspace{-1mm} & \hspace{-1mm}{ $E_a$ }\hspace{-1mm} & \hspace{-1mm}{\sc T[s]}
\hspace{-1mm} & \hspace{-1mm}{\sc Result} ~$C_0$\hspace{-1mm} & \hspace*{-0mm}{\sc Result} ~$C_1$ & \hspace{-1mm}{\sc Result} ~$C_2$\hspace{-1mm} \\
\hline
 \hspace*{-1mm}${3}$\hspace*{-1mm}& \hspace{-1mm}0.89462319318517\hspace{-1mm} & \hspace{-1mm}6.33e-10\hspace{-1mm} & \hspace{-1mm}356.9\hspace{-1mm} & \hspace{-1mm}\hspace{-1mm} & \hspace*{-1.5mm}\hspace*{-1mm} & \hspace{-1mm}\hspace{-1mm}  \\
 \hspace*{-1mm}${4}$\hspace*{-1mm}& \hspace{-1mm}1.07605987265074\hspace{-1mm} & \hspace{-1mm}8.56e-10\hspace{-1mm} & \hspace{-1mm}426.3\hspace{-1mm} & \hspace{-1mm}\hspace{-1mm} & \hspace*{-1.5mm} \hspace*{-1mm} & \hspace{-1mm}\hspace{-1mm}  \\
 \hspace*{-1mm}${5}$\hspace*{-1mm}& \hspace{-1mm}1.19524813881849\hspace{-1mm} & \hspace{-1mm}1.15e-09\hspace{-1mm} & \hspace{-1mm}426.2\hspace{-1mm} & \hspace{-1mm}1.423460265674\hspace{-1mm} & \hspace*{-1.5mm}-2.0676022044\hspace*{-1mm} & \hspace{-1mm}-7.9521322\hspace{-1mm}  \\
 \hspace*{-1mm}${6}$\hspace*{-1mm}& \hspace{-1mm}1.26445377191768\hspace{-1mm} & \hspace{-1mm}1.34e-09\hspace{-1mm} & \hspace{-1mm}426.2\hspace{-1mm} & \hspace{-1mm}1.360275447540\hspace{-1mm} & \hspace*{-1.5mm}-3.2163693548\hspace*{-1mm} & \hspace{-1mm}-23.710490\hspace{-1mm}  \\
 \hspace*{-1mm}${7}$\hspace*{-1mm}& \hspace{-1mm}1.30188593759114\hspace{-1mm} & \hspace{-1mm}1.46e-09\hspace{-1mm} & \hspace{-1mm}426.2\hspace{-1mm} & \hspace{-1mm}1.339480501116\hspace{-1mm} & \hspace*{-1.5mm}-6.2931480198\hspace*{-1mm} & \hspace{-1mm}5.5420653\hspace{-1mm}  \\
 \hspace*{-1mm}${8}$\hspace*{-1mm}& \hspace{-1mm}1.32137252564963\hspace{-1mm} & \hspace{-1mm}1.52e-09\hspace{-1mm} & \hspace{-1mm}426.2\hspace{-1mm} & \hspace{-1mm}1.341163816983\hspace{-1mm} & \hspace*{-1.5mm}-5.4086365404\hspace*{-1mm} & \hspace{-1mm}9.8856275\hspace{-1mm}  \\
 \hspace*{-1mm}${9}$\hspace*{-1mm}& \hspace{-1mm}1.33131707386056\hspace{-1mm} & \hspace{-1mm}1.55e-09\hspace{-1mm} & \hspace{-1mm}426.2\hspace{-1mm} & \hspace{-1mm}1.341410985041\hspace{-1mm} & \hspace*{-1.5mm}-5.1746466051\hspace*{-1mm} & \hspace{-1mm}25.0634683\hspace{-1mm}  \\
 \hspace*{-1mm}${10}$\hspace*{-1mm}& \hspace{-1mm}1.33634079003748\hspace{-1mm} & \hspace{-1mm}1.57e-09\hspace{-1mm} & \hspace{-1mm}426.2\hspace{-1mm} & \hspace{-1mm}1.341399965444\hspace{-1mm} & \hspace*{-1.5mm}-5.1950222482\hspace*{-1mm} & \hspace{-1mm}19.5001958\hspace{-1mm}  \\
\hspace*{-1mm}${11}$\hspace*{-1mm}& \hspace{-1mm}1.33886565298449\hspace{-1mm} & \hspace{-1mm}1.58e-09\hspace{-1mm} & \hspace{-1mm}426.2\hspace{-1mm} & \hspace{-1mm}1.341399223875\hspace{-1mm} & \hspace*{-1.5mm}-5.5197895049\hspace*{-1mm} & \hspace{-1mm}18.6327973\hspace{-1mm}  \\
\hspace*{-1mm}${12}$\hspace*{-1mm}& \hspace{-1mm}1.34013135392416\hspace{-1mm} & \hspace{-1mm}1.58e-09\hspace{-1mm} & \hspace{-1mm}426.2\hspace{-1mm} & \hspace{-1mm}1.341399240952\hspace{-1mm} & \hspace*{-1.5mm}-5.1977615908\hspace*{-1mm} & \hspace{-1mm}18.3726055\hspace{-1mm}  \\
\hspace*{-1mm}${13}$\hspace*{-1mm}& \hspace{-1mm}1.34076502405465\hspace{-1mm} & \hspace{-1mm}1.59e-09\hspace{-1mm} & \hspace{-1mm}426.2\hspace{-1mm} & \hspace{-1mm}1.341399241506\hspace{-1mm} & \hspace*{-1.5mm}-5.1977527356\hspace*{-1mm} & \hspace{-1mm}18.3878520\hspace{-1mm}  \\
\hline
\multicolumn{4}{r}{Laporta~\cite{laporta01}:} & 1.341399241447 &\hspace*{-1.0mm} -5.1977529559 & 18.3870466 \\ \hline
\end{tabular}
\end{center}
\end{scriptsize}
\label{table2-3ls}
\end{table}
\begin{table}
\caption{\footnotesize{Integral and leading expansion coefficients using one call to {\sc ParInt} with vector function, 16 procs. and linear extrapolation for
3-loop integral of Fig~\ref{3ls-self}(b),
 ~err. tol. $t_a = 10^{-12}$; max. \# evals = 20B, $\varepsilon=\varepsilon_\ell=2^{-\ell}, \ell=3,4, \ldots$;
$E_a = $ integration estim. abs. error}}
\begin{scriptsize}
\begin{center}
\begin{tabular}{cccccc}\hline
& \multicolumn{1}{c}{{\sc Integral Fig}~\ref{3ls-self}(b)}
& \multicolumn{4}{c}{{\sc Extrapolation}} \\
 \raisebox{1mm}{\hspace*{-5mm}$\ell$}\hspace*{-5mm} & {\sc Integral} \hspace{-1mm} & \hspace{-1mm}{ $E_a$ }\hspace{-1mm} & 
\hspace{-1mm}{\sc Result} ~$C_0$\hspace{-1mm} & \hspace*{-0mm}{\sc Result} ~$C_1$ & \hspace{-1mm}{\sc Result} ~$C_2$\hspace{-1mm} \\
\hline
 \hspace*{-1mm}${3}$\hspace*{-1mm}& \hspace{-1mm}0.89462319317861\hspace{-1mm} & \hspace{-1mm}6.69e-10\hspace{-1mm} & \hspace{-1mm}\hspace{-1mm} & \hspace*{-1.5mm}\hspace*{-1mm} & \hspace{-1mm}\hspace{-1mm}  \\
 \hspace*{-1mm}${4}$\hspace*{-1mm}& \hspace{-1mm}1.07605987265229\hspace{-1mm} & \hspace{-1mm}1.11e-09\hspace{-1mm}  & \hspace{-1mm}1.257496552126\hspace{-1mm} & \hspace*{-1.5mm}-2.9029868716\hspace*{-1mm} & \hspace{-1mm}\hspace{-1mm}  \\
 \hspace*{-1mm}${5}$\hspace*{-1mm}& \hspace{-1mm}1.19524813882721\hspace{-1mm} & \hspace{-1mm}1.49e-09\hspace{-1mm}  & \hspace{-1mm}1.333416355961\hspace{-1mm} & \hspace*{-1.5mm}-4.7250621636\hspace*{-1mm} & \hspace{-1mm}9.7177349\hspace{-1mm}  \\
 \hspace*{-1mm}${6}$\hspace*{-1mm}& \hspace{-1mm}1.26445377193259\hspace{-1mm} & \hspace{-1mm}1.74e-09\hspace{-1mm}  & \hspace{-1mm}1.341017173967\hspace{-1mm} & \hspace*{-1.5mm}-5.1507079720\hspace*{-1mm} & \hspace{-1mm}16.5280678\hspace{-1mm}  \\
 \hspace*{-1mm}${7}$\hspace*{-1mm}& \hspace{-1mm}1.30188593761059\hspace{-1mm} & \hspace{-1mm}1.88e-09\hspace{-1mm}  & \hspace{-1mm}1.341390110931\hspace{-1mm} & \hspace*{-1.5mm}-5.1954604075\hspace*{-1mm} & \hspace{-1mm}18.1988254\hspace{-1mm}  \\
 \hspace*{-1mm}${8}$\hspace*{-1mm}& \hspace{-1mm}1.32137252566889\hspace{-1mm} & \hspace{-1mm}1.96e-09\hspace{-1mm}  & \hspace{-1mm}1.341399132816\hspace{-1mm} & \hspace*{-1.5mm}-5.1976978351\hspace*{-1mm} & \hspace{-1mm}18.3778196\hspace{-1mm}  \\
 \hspace*{-1mm}${9}$\hspace*{-1mm}& \hspace{-1mm}1.33131707388060\hspace{-1mm} & \hspace{-1mm}2.01e-09\hspace{-1mm}   & \hspace{-1mm}1.341399240882\hspace{-1mm} & \hspace*{-1.5mm}-5.1977523004\hspace*{-1mm} & \hspace{-1mm}18.3868246\hspace{-1mm}  \\
 \hspace*{-1mm}${10}$\hspace*{-1mm}& \hspace{-1mm}1.33634079005812\hspace{-1mm} & \hspace{-1mm}2.03e-09\hspace{-1mm} & \hspace{-1mm}1.341399241524\hspace{-1mm} & \hspace*{-1.5mm}-5.1977529529\hspace*{-1mm} & \hspace{-1mm}18.3870438\hspace{-1mm}  \\
\hspace*{-1mm}${11}$\hspace*{-1mm}& \hspace{-1mm}1.33886565300527\hspace{-1mm} & \hspace{-1mm}2.04e-09\hspace{-1mm}  & \hspace{-1mm}1.341399241526\hspace{-1mm} & \hspace*{-1.5mm}-5.1977529577\hspace*{-1mm} & \hspace{-1mm}18.3870471\hspace{-1mm}  \\
\hline
\multicolumn{3}{r}{Laporta~\cite{laporta01}:} & 1.341399241447 &\hspace*{-1.0mm} -5.1977529559 & 18.3870466 \\ \hline
\end{tabular}
\end{center}
\end{scriptsize}
\label{table3-3ls}
\end{table}
\begin{table}
\caption{\footnotesize{Integral and leading expansion coefficients using {\sc ParInt} with 16 procs., and linear extrapolation for
3-loop integral of Fig~\ref{3ls-self}(f),
 ~err. tol. $t_a = 10^{-12}$; max. \# evals = 20B, $\varepsilon=\varepsilon_\ell=2^{-\ell}, \ell=3,4, \ldots$, $T[s]$ = elapsed time (s);
$E_a = $ integration estim. abs. error}}
\begin{scriptsize}
\begin{center}
\begin{tabular}{ccccccc}\hline
& \multicolumn{2}{c}{{\sc Integral Fig}~\ref{3ls-self}(f)}
& \multicolumn{4}{c}{{\sc Extrapolation}} \\
 \raisebox{1mm}{\hspace*{-5mm}$\ell$}\hspace*{-5mm} & {\sc Integral} \hspace{-1mm} & \hspace{-1mm}{ $E_a$ }\hspace{-1mm} & \hspace{-1mm}{\sc T[s]}
\hspace{-1mm} & \hspace{-1mm}{\sc Result} ~$C_0$\hspace{-1mm} & \hspace*{-0mm}{\sc Result} ~$C_1$ & \hspace{-1mm}{\sc Result} ~$C_2$\hspace{-1mm} \\
\hline
 \hspace*{-1mm}${3}$\hspace*{-1mm}& \hspace{-1mm}0.176698722960541\hspace{-1mm} & \hspace{-1mm}1.01e-09\hspace{-1mm} & \hspace{-1mm}363.8\hspace{-1mm} & \hspace{-1mm}\hspace{-1mm} & \hspace*{-1.5mm}\hspace*{-1mm} & \hspace{-1mm}\hspace{-1mm}  \\
 \hspace*{-1mm}${4}$\hspace*{-1mm}& \hspace{-1mm}0.179083545661235\hspace{-1mm} & \hspace{-1mm}9.17e-10\hspace{-1mm} & \hspace{-1mm}437.2\hspace{-1mm} & \hspace{-1mm}0.181468368362\hspace{-1mm} & \hspace*{-1.5mm}-0.0381571632\hspace*{-1mm} & \hspace{-1mm}\hspace{-1mm}  \\
 \hspace*{-1mm}${5}$\hspace*{-1mm}& \hspace{-1mm}0.180693790881676\hspace{-1mm} & \hspace{-1mm}9.24e-10\hspace{-1mm} & \hspace{-1mm}437.3\hspace{-1mm} & \hspace{-1mm}0.182582592016\hspace{-1mm} & \hspace*{-1.5mm}-0.0648985309\hspace*{-1mm} & \hspace{-1mm}0.14262063\hspace{-1mm}  \\
 \hspace*{-1mm}${6}$\hspace*{-1mm}& \hspace{-1mm}0.181617679292608\hspace{-1mm} & \hspace{-1mm}9.34E-10\hspace{-1mm} & \hspace{-1mm}437.2\hspace{-1mm} & \hspace{-1mm}0.182626195317\hspace{-1mm} & \hspace*{-1.5mm}-0.0673403158\hspace*{-1mm} & \hspace{-1mm}0.18168919\hspace{-1mm}  \\
 \hspace*{-1mm}${7}$\hspace*{-1mm}& \hspace{-1mm}0.182111420125171\hspace{-1mm} & \hspace{-1mm}9.79e-10\hspace{-1mm} & \hspace{-1mm}437.7\hspace{-1mm} & \hspace{-1mm}0.182627225039\hspace{-1mm} & \hspace*{-1.5mm}-0.0674638824\hspace*{-1mm} & \hspace{-1mm}0.18630234\hspace{-1mm}  \\
 \hspace*{-1mm}${8}$\hspace*{-1mm}& \hspace{-1mm}0.18236652627441\hspace{-1mm} & \hspace{-1mm}9.84e-10\hspace{-1mm} & \hspace{-1mm}437.6\hspace{-1mm} & \hspace{-1mm}0.182627237225\hspace{-1mm} & \hspace*{-1.5mm}-0.0674669046\hspace*{-1mm} & \hspace{-1mm}0.18654412\hspace{-1mm}  \\
 \hspace*{-1mm}${9}$\hspace*{-1mm}& \hspace{-1mm}0.182496175680214\hspace{-1mm} & \hspace{-1mm}9.86e-10\hspace{-1mm} & \hspace{-1mm}437.7\hspace{-1mm} & \hspace{-1mm}0.182627237219\hspace{-1mm} & \hspace*{-1.5mm}-0.0674669014\hspace*{-1mm} & \hspace{-1mm}0.18654358\hspace{-1mm}  \\
 \hspace*{-1mm}${10}$\hspace*{-1mm}& \hspace{-1mm}0.182561529243301\hspace{-1mm} & \hspace{-1mm}9.86e-10\hspace{-1mm} & \hspace{-1mm}437.8\hspace{-1mm} & \hspace{-1mm}0.182627237221\hspace{-1mm} & \hspace*{-1.5mm}-0.0674669038\hspace*{-1mm} & \hspace{-1mm}0.18654439\hspace{-1mm}  \\
\hline
\multicolumn{4}{r}{Laporta~\cite{laporta01}:} & 0.182627237539 &\hspace*{-1.0mm} -0.0674669097 & 0.18654624 \\ \hline
\end{tabular}
\end{center}
\end{scriptsize}
\label{table4-3ls}
\end{table}
\begin{table}
\caption{\footnotesize{Integral and leading expansion coefficients using one call to {\sc ParInt} with vector function, 16 procs. and linear extrapolation for
3-loop integral of Fig~\ref{3ls-self}(f),
 ~err. tol. $t_a = 10^{-12}$; max. \# evals = 20B, $\varepsilon=\varepsilon_\ell=2^{-\ell}, \ell=3,4, \ldots$;
$E_a = $ integration estim. abs. error}}
\begin{scriptsize}
\begin{center}
\begin{tabular}{cccccc}\hline
& \multicolumn{2}{c}{{\sc Integral Fig}~\ref{3ls-self}(f)}
& \multicolumn{3}{c}{{\sc Extrapolation}} \\
 \raisebox{1mm}{\hspace*{-5mm}$\ell$}\hspace*{-5mm} & {\sc Integral} \hspace{-1mm} & \hspace{-1mm}{ $E_a$ }\hspace{-1mm} & \hspace{-1mm}
{\sc Result} ~$C_0$\hspace{-1mm} & \hspace*{-0mm}{\sc Result} ~$C_1$ & \hspace{-1mm}{\sc Result} ~$C_2$\hspace{-1mm} \\
\hline
 \hspace*{-1mm}${3}$\hspace*{-1mm}& \hspace{-1mm}0.176698722966533\hspace{-1mm} & \hspace{-1mm}1.57e-09\hspace{-1mm} & \hspace{-1mm}\hspace{-1mm} & \hspace*{-1.5mm}\hspace*{-1mm} & \hspace{-1mm}\hspace{-1mm}  \\
 \hspace*{-1mm}${4}$\hspace*{-1mm}& \hspace{-1mm}0.179083545593469\hspace{-1mm} & \hspace{-1mm}1.63e-09\hspace{-1mm} & \hspace{-1mm}0.181468368220\hspace{-1mm} & \hspace*{-1.5mm}-0.0381571620\hspace*{-1mm} & \hspace{-1mm}\hspace{-1mm}  \\
 \hspace*{-1mm}${5}$\hspace*{-1mm}& \hspace{-1mm}0.1806937908054~~\hspace{-1mm} & \hspace{-1mm}1.71e-09\hspace{-1mm} & \hspace{-1mm}0.182582591950\hspace{-1mm} & \hspace*{-1.5mm}-0.0648985315\hspace*{-1mm} & \hspace{-1mm}0.1426206455\hspace{-1mm}  \\
 \hspace*{-1mm}${6}$\hspace*{-1mm}& \hspace{-1mm}0.181617679222644\hspace{-1mm} & \hspace{-1mm}1.76e-09\hspace{-1mm} & \hspace{-1mm}0.182626195261\hspace{-1mm} & \hspace*{-1.5mm}-0.0673403170\hspace*{-1mm} & \hspace{-1mm}0.1816892047\hspace{-1mm}  \\
 \hspace*{-1mm}${7}$\hspace*{-1mm}& \hspace{-1mm}0.182111420130547\hspace{-1mm} & \hspace{-1mm}1.80e-09\hspace{-1mm} & \hspace{-1mm}0.182627225208\hspace{-1mm} & \hspace*{-1.5mm}-0.0674639106\hspace*{-1mm} & \hspace{-1mm}0.1863033669\hspace{-1mm}  \\
 \hspace*{-1mm}${8}$\hspace*{-1mm}& \hspace{-1mm}0.182366526276538\hspace{-1mm} & \hspace{-1mm}1.81e-09\hspace{-1mm} & \hspace{-1mm}0.182627237153\hspace{-1mm} & \hspace*{-1.5mm}-0.0674668729\hspace*{-1mm} & \hspace{-1mm}0.1865403500\hspace{-1mm}  \\
 \hspace*{-1mm}${9}$\hspace*{-1mm}& \hspace{-1mm}0.182496175679009\hspace{-1mm} & \hspace{-1mm}1.82e-09\hspace{-1mm} & \hspace{-1mm}0.182627237223\hspace{-1mm} & \hspace*{-1.5mm}-0.0674669081\hspace*{-1mm} & \hspace{-1mm}0.1865461701\hspace{-1mm}  \\
 \hspace*{-1mm}${10}$\hspace*{-1mm}& \hspace{-1mm}0.182561529242407\hspace{-1mm} & \hspace{-1mm}1.83e-09\hspace{-1mm} & \hspace{-1mm}0.182627237223\hspace{-1mm} & \hspace*{-1.5mm}-0.0674669083\hspace*{-1mm} & \hspace{-1mm}0.1865462415\hspace{-1mm}  \\
\hline
\multicolumn{3}{r}{Laporta~\cite{laporta01}:} & 0.182627237539 &\hspace*{-1.0mm} -0.0674669097 & 0.1865462421 \\ \hline
\end{tabular}
\end{center}
\end{scriptsize}
\label{table5-3ls}
\end{table}


\subsection{3-loop UV-divergent integrals with massless internal lines}\label{3-loop-massless-UV}
This section handles the integral associated with the massless diagram of 
Fig~\ref{3ls-self}(g) (the \emph{3-loop sunrise-sunset} diagram named $P_{3}$ in~\cite{baikov10}), 
which has a $1/\varepsilon$ singularity in the dimensional regularization parameter,
arising from the $\Gamma$-function factor \,($\Gamma(N-6+3\varepsilon)$)\, 
in Eq~\eqref{threeLOOP}.

The polynomials $U$ and $W$ for \,$I_g^{S3}$\, are given by

\begin{equation}
\mathrm{(g)}\quad
\left\{
\begin{array}{l}
U= x_1 x_2 x_3 + x_1 x_2 x_4 + x_1 x_3 x_4 + x_2 x_3 x_4
\\
W/s=x_1 x_2 x_3 x_4
\end{array}
\right.
\label{SEthrLg}
\end{equation}
We take $\varrho = 0$ in Eq~\eqref{threeLOOP}, and the numerical evaluation is done with $s=1$.
In order to compare the result to that of Baikov and Chetyrkin~\cite{baikov10}, we
multiply with the factor $n(\varepsilon)^L$ where $n(\varepsilon)$ is defined
(in their footnote 11, p. 193) as
\begin{equation}
n(\varepsilon) = \frac{\Gamma(2-2\varepsilon)}{\Gamma(1+\varepsilon)~\Gamma(1-\varepsilon)^2}\,,
\label{factorn}
\end{equation}
leading to the expansion
\begin{align}
& 
 ~n(\varepsilon)^3 \,I_g^{S3} = \frac{1}{36}\,\frac{1}{\varepsilon}+\frac{35}{216}+\frac{991}{1296}\,\varepsilon
  +\,\ldots \nonumber \\
& = \,0.027777777777\,\frac{1}{\varepsilon}+0.162037037037+0.764660493827\,\varepsilon
  +\,\ldots \label{p3-act} 
\end{align}

\begin{table}
\caption{\footnotesize
{\color{black}Results UV \emph{3-loop} integral}, $n(\varepsilon)^3 \,I_g^{S3}$ (on 4 nodes/64 procs thor cluster),
abs. err. tol. $t_a = 10^{-12},$ $T[s]$ = Time (elapsed user time in s);
$\varepsilon = \varepsilon_\ell = 2^{-\ell},~ \ell = 8,9,\ldots,$ ~~$E_a = $ integration estim. abs. error}
\label{p3ext}
\begin{scriptsize}
\begin{center}
\begin{tabular}{cccccc}\hline
& \multicolumn{2}{c}{{\sc Integral Fig}~\ref{3ls-self}(g) }
& \multicolumn{3}{c}{{\sc Extrapolation}} \\
 \raisebox{1mm}{\hspace*{-5mm}$\ell$}\hspace*{-5mm} & \hspace{-1mm}{ $E_a$ }\hspace{-1mm} & \hspace{-1mm}{\sc T[s]}
\hspace{-1mm} & {\sc Result} ~${C}_{-1}$ & \hspace{-1mm}{\sc Result} ~${C}_{0}$\hspace{-1mm} & \hspace*{-0mm}{\sc Result} ~${C}_1$ \\
\hline
\hspace*{-1mm}$8$\hspace*{-1mm}& \hspace{-1mm}2.8e-14\hspace{-1mm} & \hspace{-1mm}0.37\hspace{-1mm} & \hspace{-1mm}\hspace{-1mm} & \hspace{-1mm}\hspace{-1mm} & \hspace*{-1.5mm}\hspace*{-1mm} \\
\hspace*{-1mm}$9$\hspace*{-1mm}& \hspace{-1mm}1.4e-13\hspace{-1mm} & \hspace{-1mm}0.65\hspace{-1mm} & \hspace{-1mm}0.0277718241800302\hspace{-1mm} & \hspace{-1mm}0.166588879051\hspace{-1mm} & \hspace*{-1.5mm}\hspace*{-1mm} \\
\hspace*{-1mm}$10$\hspace*{-1mm}& \hspace{-1mm}1.7e-13\hspace{-1mm} & \hspace{-1mm}0.96\hspace{-1mm} & \hspace{-1mm}0.0277777978854937\hspace{-1mm} & \hspace{-1mm}0.162001073255\hspace{-1mm} & \hspace*{-1.5mm}0.78298552 \hspace*{-1mm} \\
\hspace*{-1mm}$11$\hspace*{-1mm}& \hspace{-1mm}1.6e-13\hspace{-1mm} & \hspace{-1mm}0.56\hspace{-1mm} & \hspace{-1mm}0.0277777777439553\hspace{-1mm} & \hspace{-1mm}0.162037166892\hspace{-1mm} & \hspace*{-1.5mm}0.76450558 \hspace*{-1mm} \\
\hspace*{-1mm}$12$\hspace*{-1mm}& \hspace{-1mm}1.7e-13\hspace{-1mm} & \hspace{-1mm}1.01\hspace{-1mm} & \hspace{-1mm}0.0277777777777756\hspace{-1mm} & \hspace{-1mm}0.162037037022\hspace{-1mm} & \hspace*{-1.5mm}0.76466073 \hspace*{-1mm} \\
\hline
\multicolumn{3}{r} \emph{Eq}~\eqref{p3-act}: & 0.0277777777777777 & 0.162037037037 &\hspace*{-1.0mm} \hspace*{-1.5mm}0.76466049\hspace*{-1mm} \\ 
\hline
\end{tabular}
\end{center}
\end{scriptsize}
\end{table}

Based on integrations with {\sc ParInt}, a maximum of 10B function evaluations 
and an absolute error tolerance of $10^{-12}$ 
(on the computation of the integral
${\mathcal I}_g^{S3}/\Gamma(-2+3\varepsilon)$), 
the results in 
Table~\ref{p3ext} are produced using linear extrapolation. 
{\sc ParInt} returns a 0 error flag for the integrals in the input sequence to the
extrapolation, indicating that a successful termination
is assumed according to Eqs~\eqref{accuracy} or~\eqref{acc} for the requested accuracy.

\subsection{3-loop UV-divergent integrals with massive internal lines}
\label{3-loop-massive-UV}

In this subsection we calculate the integrals corresponding to massive diagrams of 
Fig~\ref{3ls-self}(h)-(j).
The integral $I_{h}^{S3}$ is divergent as $\displaystyle{1/\varepsilon},$ resulting from
the $\Gamma$-function factor.
On the other hand, the integrals $I_{i}^{S3}$ and $I_{j}^{S3}$ are divergent as $\displaystyle{1/\varepsilon}$ due to the integral part.
The $U, W$ functions for $I_h^{S3},\, I_i^{S3},\, and\, I_j^{S3}$ are listed in Eqs~\eqref{SEthrLh}-\eqref{SEthrLj} below. 
%
\begin{equation}
\mathrm{(h)}\quad
\left\{
\begin{array}{l}
U=
  x_5 \,( x_{12}   x_{346} +x_3  x_{46} )
  +x_3  x_{14}   x_{26} +x_1 x_2  x_{46} + x_{12}  x_4 x_6    
\\
W/s=
  x_3 \,(  x_5  x_{12}   x_{46} 
   +x_1 x_2  x_{46} + x_{12}  x_4 x_6 )  
\end{array}
\right.
\label{SEthrLh}
\end{equation}

\begin{equation}
\mathrm{(i)}\quad
\left\{
\begin{array}{l}
%
U=x_{56}\,(x_{12}x_{34} + x_{1234}x_7) + x_5x_6 x_{1234}
\\
%
W/s=x_{56}\,(x_1 x_2 x_3 + x_1 x_2 x_4 + x_1 x_3 x_4 + x_2 x_3 x_4 + x_{13} x_{24} x_7 )
+x_5x_6  x_{13} x_{24}
\end{array}
\right.
\label{SEthrLi}
\end{equation}

\begin{equation}
\mathrm{(j)}\quad
\left\{
\begin{array}{l}
U=x_{37}\,(x_{12}x_{45} + x_{1245}x_6) + x_3x_7 x_{126}
\\
W/s=x_{37}\,(x_1 x_2 x_4 + x_1 x_2 x_5 + x_1 x_4 x_5 + x_2 x_4 x_5 + x_{14} x_{25} x_6 )\\ 
 \quad +x_3x_7 \,( x_1 x_{25} +x_2x_5 + x_{25} x_6 )
\end{array}
\right.
\label{SEthrLj}
\end{equation}

\noindent
Expansions for these integrals from \cite{laporta01} are:
 
\begin{align}
 &I_{h}^{S3}({\varepsilon}) ~\Gamma(1+{\varepsilon})^{-3} = \sum_{k\ge -1} C_k {\varepsilon^k} = 2.404113806319\,{\varepsilon^{-1}}-9.7634244476+34.99888166\,{\varepsilon}-116.0420478\,\varepsilon^2\ldots \label{I6hexp}
\end{align}

\begin{align}
 &I_{i}^{S3}({\varepsilon}) ~\Gamma(1+{\varepsilon})^{-3} = \sum_{k\ge -1} C_k {\varepsilon^k} = 0.923631826520\,{\varepsilon^{-1}}-2.4234916344+8.3813497101\,{\varepsilon}-26.99362122\,\varepsilon^2\ldots \label{Ioexp}
\end{align}

\begin{align}
 &I_{j}^{S3}({\varepsilon}) ~\Gamma(1+{\varepsilon})^{-3} = \sum_{k\ge -1} C_k {\varepsilon^k} = 0.923631826520\,{\varepsilon^{-1}}-2.1161697185+6.9295446853\,{\varepsilon}-21.50327838\,\varepsilon^2\ldots \label{Ipexp}
\end{align}

The evaluation is performed with $s=1,\, m_r=1$.
Numerical results by {\sc ParInt} on the \emph{thor} cluster,
for the asymptotic expansion coefficients of $I_{h}^{S3}\,\Gamma(1+{\varepsilon})^{-3}$
in Eq~\eqref{I6hexp}, are listed in Table~\ref{table5}.
A geometric sequence in base $2^{-1}$ is used for $\varepsilon.$

For the computation of $I_{i}^{S3}$ and $I_{j}^{S3}$, the variables are transformed as:
\begin{eqnarray}\label{KK-trans}
\nonumber
x_1&=&y_{1m} y_2 y_4 y_5, ~~~~x_2~=~y_{1m} y_2 y_4 y_{5m},\\
\nonumber
x_3&=&y_{1m} y_2 y_{4m} y_6, ~~~x_4~=~y_{1m} y_2 y_{4m} y_{6m},\\
x_5&=&y_{1} y_3, ~~~~x_6~=~y_{1} y_{3m},\\
\nonumber
x_7&=&y_{1m} y_{2m}
\end{eqnarray}
with $y_{im}=1-y_{i}$ and Jacobian $y_1 y_{1m}^4 y_2^3 y_4 y_{4m}$ for the former, and 

\begin{eqnarray}\label{TI-trans}
\nonumber
x_1&=&y_1 y_2 y_4, ~~~~x_2=y_1 y_2 y_{4m},  \\
\nonumber
x_3&=&y_{1m} y_3 y_6, ~~~x_4=y_1 y_{2m} y_5,  \\
x_5&=&y_1 y_{2m} y_{5m}, ~~~x_6=y_{1m} y_{3m}, \\
\nonumber
x_7&=&y_{1m} y_3 y_{6m}
\end{eqnarray}
with $y_{im}=1-y_{i}$ and Jacobian $y_1^{3} y_{1m}^2 y_{2m} y_2 y_3$ for the latter.
These variable transformations are beneficial to smoothen the integrand 
boundary singularities.
The above two variable transformations can be adopted for both $I_{i}^{S3}$ and $I_{j}^{S3}$. 
However, the transformation \eqref{KK-trans} works better for $I_{i}^{S3}$ and \eqref{TI-trans} works better for $I_{j}^{S3}.$ 


\begin{table}
\caption{\footnotesize
{\color{black}Results UV \emph{3-loop} integral} of Fig~\ref{3ls-self}(h) (on 4 nodes with 16 procs per node of \emph{thor} cluster),
err. tol. $t_a = 10^{-12},$ $T[s]$ = Time (elapsed user time in s);
$\varepsilon = \varepsilon_\ell = 2^{-\ell},~ \ell = 2,3,\ldots,$ ~~$E_r = $ integration estim. abs. error}
\begin{scriptsize}
\begin{center}
\begin{tabular}{ccccccc}\hline
& \multicolumn{2}{c}{{\sc Integral Fig~\ref{3ls-self}}(h)}
& \multicolumn{4}{c}{{\sc Extrapolation}} \\
 \raisebox{1mm}{\hspace*{-5mm}$\ell$}\hspace*{-5mm} & \hspace{-1mm}{ $E_r$ }\hspace{-1mm} & \hspace{-1mm}{\sc T[s]}
\hspace{-1mm} & {\sc Result} ~${C}_{-1}$ & \hspace{-1mm}{\sc Result} ~${C}_{0}$\hspace{-1mm} & \hspace*{-0mm}{\sc Result} ~${C}_1$ & \hspace{-1mm}{\sc Result} ~${C}_2$\hspace{-1mm} \\
\hline
\hspace*{-1mm}$2$\hspace*{-1mm}& \hspace{-1mm}5.7e-10\hspace{-1mm} & \hspace{-1mm}17.6\hspace{-1mm} & \hspace{-1mm}\hspace{-1mm} & \hspace{-1mm}\hspace{-1mm} & \hspace*{-1.5mm}\hspace*{-1mm} & \hspace{-1mm}\hspace{-1mm} \\
\hspace*{-1mm}$3$\hspace*{-1mm}& \hspace{-1mm}1.8e-09\hspace{-1mm} & \hspace{-1mm}28.3\hspace{-1mm} & \hspace{-1mm}1.990363419875\hspace{-1mm} & \hspace{-1mm}-3.3722810037\hspace{-1mm} & \hspace*{-1.5mm}\hspace*{-1mm} & \hspace{-1mm}\hspace{-1mm}  \\
\hspace*{-1mm}$4$\hspace*{-1mm}& \hspace{-1mm}3.2e-09\hspace{-1mm} & \hspace{-1mm}28.3\hspace{-1mm} & \hspace{-1mm}2.330127359814\hspace{-1mm} & \hspace{-1mm}-7.4494482830\hspace{-1mm} & \hspace*{-1.5mm}10.87244608\hspace*{-1mm} & \hspace{-1mm}\hspace{-1mm}  \\
\hspace*{-1mm}$5$\hspace*{-1mm}& \hspace{-1mm}4.9e-09\hspace{-1mm} & \hspace{-1mm}28.3\hspace{-1mm} & \hspace{-1mm}2.397223358974\hspace{-1mm} & \hspace{-1mm}-9.3281362595\hspace{-1mm} & \hspace*{-1.5mm}25.90194989\hspace*{-1mm} & \hspace{-1mm}\hspace{-1mm} -34.353152 \\
\hspace*{-1mm}$6$\hspace*{-1mm}& \hspace{-1mm}6.3e-09\hspace{-1mm} & \hspace{-1mm}28.3\hspace{-1mm} & \hspace{-1mm}2.403788052525\hspace{-1mm} & \hspace{-1mm}-9.7220178725\hspace{-1mm} & \hspace*{-1.5mm}33.25440667\hspace*{-1mm} & \hspace{-1mm}\hspace{-1mm} -84.769999 \\
\hspace*{-1mm}$7$\hspace*{-1mm}& \hspace{-1mm}7.0e-09\hspace{-1mm} & \hspace{-1mm}28.3\hspace{-1mm} & \hspace{-1mm}2.404106076736\hspace{-1mm} & \hspace{-1mm}-9.7634244078\hspace{-1mm} & \hspace*{-1.5mm}34.83180676\hspace*{-1mm} & \hspace{-1mm}\hspace{-1mm} -110.00840 \\
\hspace*{-1mm}$8$\hspace*{-1mm}& \hspace{-1mm}7.4e-09\hspace{-1mm} & \hspace{-1mm}28.3\hspace{-1mm} & \hspace{-1mm}2.404113714590\hspace{-1mm} & \hspace{-1mm}-9.7633776138\hspace{-1mm} & \hspace*{-1.5mm}34.99091852\hspace*{-1mm} & \hspace{-1mm}\hspace{-1mm} -115.46366 \\
\hspace*{-1mm}$9$\hspace*{-1mm}& \hspace{-1mm}7.7e-09\hspace{-1mm} & \hspace{-1mm}28.3\hspace{-1mm} & \hspace{-1mm}2.404113805535\hspace{-1mm} & \hspace{-1mm}-9.7634238138\hspace{-1mm} & \hspace*{-1.5mm}34.99868012\hspace*{-1mm} & \hspace{-1mm}\hspace{-1mm} -116.01362 \\
\hspace*{-1mm}$10$\hspace*{-1mm}& \hspace{-1mm}8.5e-09\hspace{-1mm} & \hspace{-1mm}28.3\hspace{-1mm} & \hspace{-1mm}2.404113806117\hspace{-1mm} & \hspace{-1mm}-9.7634244078\hspace{-1mm} & \hspace*{-1.5mm}34.99888129\hspace*{-1mm} & \hspace{-1mm}\hspace{-1mm} -116.04259 \\
\hline
\multicolumn{3}{r} \emph{Eq}~\eqref{I6hexp}: & 2.404113806319 & -9.7634244476  &\hspace*{-1.0mm} \hspace*{-1.5mm}34.99888166\hspace*{-1mm} &\hspace{-1mm}\hspace{-1mm} -116.04205 \\ \hline
\end{tabular}
\end{center}
\end{scriptsize}
\label{table5}
\end{table}

\begin{table}
\caption{\footnotesize
{\color{black}Results UV \emph{3-loop} diagram of Fig~\ref{3ls-self}(i) with massive internal lines, using 36 threads on Intel(R) Xeon(R) E5-2687W v3 3.10\,GHz. DE is applied with mesh size $h=0.1265988$ and number of evaluations $N_{eval}=104$ (cf., Eq~\eqref{DEsum}). For extrapolation, $\varepsilon = \varepsilon_{\ell} = 1.15^{-\ell}, \ell=17,18,\ldots$}}
\begin{scriptsize}
\begin{center}
\begin{tabular}{ccccl}\hline
\multicolumn{2}{l}{{\sc Integral Fig~\ref{3ls-self}}(i)}& \multicolumn{3}{c}{{\sc Extrapolation}} \\
{\hspace*{-5mm}$\ell$}\hspace*{-5mm} & {\sc Result} ~${C}_{-1}$ & \hspace{-1mm}{\sc Result} ~${C}_{0}$\hspace{-1mm} & \hspace*{-0mm}{\sc Result} ~${C}_1$ & \hspace*{-1.5mm}{\sc Result} ~${C}_2$\\ \hline
\hspace*{-1mm}$17$\hspace*{-1mm}&   & \hspace{-1mm}\hspace{-1mm} & \hspace*{-1.5mm}\hspace*{-1mm} \\
\hspace*{-1mm}$18$\hspace*{-1mm}& \hspace{-1mm}0.89291935327\hspace{-1mm} & \hspace{-1mm}-1.506522015\hspace{-1mm} & \hspace*{-1.5mm}\hspace*{-1mm} & \\ 
\hspace*{-1mm}$19$\hspace*{-1mm}& \hspace{-1mm}0.91852761612\hspace{-1mm} & \hspace{-1mm}-2.187886833\hspace{-1mm} & \hspace*{-1.5mm}0.45102458\hspace*{-1mm} & \\
\hspace*{-1mm}$20$\hspace*{-1mm}& \hspace{-1mm}0.92289114325\hspace{-1mm} & \hspace{-1mm}-2.375404012\hspace{-1mm} & \hspace*{-1.5mm}7.17894748\hspace*{-1mm} & \hspace*{-1.5mm}-12.57809\hspace*{-1mm} \\
\hspace*{-1mm}$21$\hspace*{-1mm}& \hspace{-1mm}0.92353815728\hspace{-1mm} & \hspace{-1mm}-2.415386426\hspace{-1mm} & \hspace*{-1.5mm}8.09798189\hspace*{-1mm} & \hspace*{-1.5mm}-21.89098\hspace*{-1mm} \\
\hspace*{-1mm}$22$\hspace*{-1mm}& \hspace{-1mm}0.92362147849\hspace{-1mm} & \hspace{-1mm}-2.422338750\hspace{-1mm} & \hspace*{-1.5mm}8.32777907\hspace*{-1mm} & \hspace*{-1.5mm}-25.65197\hspace*{-1mm} \\
\hspace*{-1mm}$23$\hspace*{-1mm}& \hspace{-1mm}0.92363082823\hspace{-1mm} & \hspace{-1mm}-2.423351621\hspace{-1mm} & \hspace*{-1.5mm}8.37298426\hspace*{-1mm} & \hspace*{-1.5mm}-26.71586\hspace*{-1mm} \\
\hspace*{-1mm}$24$\hspace*{-1mm}& \hspace{-1mm}0.92363174411\hspace{-1mm} & \hspace{-1mm}-2.423477056\hspace{-1mm} & \hspace*{-1.5mm}8.38025256\hspace*{-1mm} & \hspace*{-1.5mm}-26.94683\hspace*{-1mm} \\
\hspace*{-1mm}$25$\hspace*{-1mm}& \hspace{-1mm}0.92363182249\hspace{-1mm} & \hspace{-1mm}-2.423490372\hspace{-1mm} & \hspace*{-1.5mm}8.38122798\hspace*{-1mm} & \hspace*{-1.5mm}-26.98707\hspace*{-1mm} \\
\hspace*{-1mm}$26$\hspace*{-1mm}& \hspace{-1mm}0.92363182726\hspace{-1mm} & \hspace{-1mm}-2.423491361\hspace{-1mm} & \hspace*{-1.5mm}8.38131792\hspace*{-1mm} & \hspace*{-1.5mm}-26.99177\hspace*{-1mm} \\
\hline
\hspace*{-1mm} {Eq}~\eqref{Ioexp}:\hspace*{-1mm} & \hspace*{-1mm}0.92363182652\hspace*{-1mm}&\hspace{-1mm} -2.423491634\hspace{-1mm}&\hspace*{-1.5mm}8.38134971\hspace*{-1mm} & \hspace*{-1.5mm}-26.99362\hspace*{-1mm} \\ \hline
\end{tabular}
\end{center}
\end{scriptsize}
\label{table5i}
\end{table}
Numerical results achieved with DE on Intel(R) Xeon(R) E5-2687W v3 3.10\,GHz are shown in Tables~\ref{table5i} and ~\ref{table5j}. Using IEEE 754-2008 binary128, extensive computation times are incurred (28 hours per iteration for Table~\ref{table5i} and 10 hours for Table~\ref{table5j}), as a trade-off for high accuracy. Similar or slightly less accuracy but far shorter computation times (between 670 and 2540 seconds per iteration) are reported in~\cite{cpp16}, using {\sc ParInt} in \emph{long double} precision on 4 nodes and 16 MPI processes per node of the \emph{thor} cluster.

\section{4-loop self-energy integrals with massless internal lines}\label{4-loop-self}

In this section we calculate the integral in Eq~\eqref{LloopIJ} for
$L=4$ and $n=4-2\varepsilon,$
\begin{equation}
I = (-1)^N \,{\Gamma\left(N-8+4\varepsilon \right)}
\int_{0}^{1}\prod_{r=1}^{N}dx_{r}\,
\delta\,(1-\sum x_{r})\,\frac{1}{U^{2-\varepsilon}\,(V-i\varrho)^{N-8+4\varepsilon}}.
\label{fourLOOP}
\end{equation}
 
\begin{table}
\caption{\footnotesize
{\color{black}Results UV \emph{3-loop} diagram of Fig~\ref{3ls-self}(j) with massive internal lines, using 36 threads on Intel(R) Xeon(R) E5-2687W v3 3.10\,GHz. DE is applied with mesh size $h=0.1253191$ and number of evaluations $N_{eval}=94$ (cf., Eq~\eqref{DEsum}). For extrapolation, $\varepsilon = \varepsilon_{\ell} = 1.15^{-\ell}, \ell=10,11,\ldots$}}
\begin{scriptsize}
\begin{center}
\begin{tabular}{ccccl}\hline
\multicolumn{2}{l}{{\sc Integral Fig~\ref{3ls-self}}(j)}& \multicolumn{3}{c}{{\sc Extrapolation}} \\
{\hspace*{-5mm}$\ell$}\hspace*{-5mm} & {\sc Result} ~${C}_{-1}$ & \hspace{-1mm}{\sc Result} ~${C}_{0}$\hspace{-1mm} & \hspace*{-0mm}{\sc Result} ~${C}_1$ & \hspace*{-1.5mm}{\sc Result} ~${C}_2$\hspace*{-1mm}\\ \hline
\hspace*{-1mm}$10$\hspace*{-1mm}&   & \hspace{-1mm}\hspace{-1mm} & \hspace*{-1.5mm}\hspace*{-1mm} \\
\hspace*{-1mm}$11$\hspace*{-1mm}& \hspace{-1mm}0.79879040550\hspace{-1mm} & \hspace{-1mm}-0.6424371659\hspace{-1mm} & \hspace*{-1.5mm}\hspace*{-1mm} \\ 
\hspace*{-1mm}$12$\hspace*{-1mm}& \hspace{-1mm}0.87785768590\hspace{-1mm} & \hspace{-1mm}-1.3301603488\hspace{-1mm} & \hspace*{-1.5mm}1.48816624\hspace*{-1mm} & \hspace*{-1.5mm}\hspace*{-1mm}\\
\hspace*{-1mm}$13$\hspace*{-1mm}& \hspace{-1mm}0.90817425166\hspace{-1mm} & \hspace{-1mm}-1.7560535047\hspace{-1mm} & \hspace*{-1.5mm}3.46958790\hspace*{-1mm} & \hspace*{-1.5mm}-3.0528730\hspace*{-1mm}\\
\hspace*{-1mm}$14$\hspace*{-1mm}& \hspace{-1mm}0.91891701861\hspace{-1mm} & \hspace{-1mm}-1.9730679987\hspace{-1mm} & \hspace*{-1.5mm}5.10026802\hspace*{-1mm} & \hspace*{-1.5mm}-8.4546800\hspace*{-1mm}\\
\hspace*{-1mm}$15$\hspace*{-1mm}& \hspace{-1mm}0.92233670919\hspace{-1mm} & \hspace{-1mm}-2.0663458474\hspace{-1mm} & \hspace*{-1.5mm}6.10815188\hspace*{-1mm} & \hspace*{-1.5mm}-13.847111\hspace*{-1mm}\\
\hspace*{-1mm}$16$\hspace*{-1mm}& \hspace{-1mm}0.92331256399\hspace{-1mm} & \hspace{-1mm}-2.1009045295\hspace{-1mm} & \hspace*{-1.5mm}6.61235817\hspace*{-1mm} & \hspace*{-1.5mm}-17.726240\hspace*{-1mm}\\
\hspace*{-1mm}$17$\hspace*{-1mm}& \hspace{-1mm}0.92356140680\hspace{-1mm} & \hspace{-1mm}-2.1120455643\hspace{-1mm} & \hspace*{-1.5mm}6.82339433\hspace*{-1mm} & \hspace*{-1.5mm}-19.918551\hspace*{-1mm}\\
\hspace*{-1mm}$18$\hspace*{-1mm}& \hspace{-1mm}0.92361796674\hspace{-1mm} &\hspace{-1mm}-2.1151864872\hspace{-1mm} & \hspace*{-1.5mm}6.89861152\hspace*{-1mm} & \hspace*{-1.5mm}-20.933028\hspace*{-1mm}\\
\hspace*{-1mm}$19$\hspace*{-1mm}& \hspace{-1mm}0.92362939967\hspace{-1mm} &\hspace{-1mm}-2.1159628755\hspace{-1mm} & \hspace*{-1.5mm}6.92167297\hspace*{-1mm} & \hspace*{-1.5mm}-21.326253\hspace*{-1mm}\\
\hspace*{-1mm}$20$\hspace*{-1mm}& \hspace{-1mm}0.92363145072\hspace{-1mm} &\hspace{-1mm}-2.1161313491\hspace{-1mm} & \hspace*{-1.5mm}6.92779243\hspace*{-1mm} & \hspace*{-1.5mm}-21.455677\hspace*{-1mm}\\
\hspace*{-1mm}$21$\hspace*{-1mm}& \hspace{-1mm}0.92363177660\hspace{-1mm} &\hspace{-1mm}-2.1161634494\hspace{-1mm} & \hspace*{-1.5mm}6.92920277\hspace*{-1mm} & \hspace*{-1.5mm}-21.492152\hspace*{-1mm}\\
\hspace*{-1mm}$22$\hspace*{-1mm}& \hspace{-1mm}0.92363182245\hspace{-1mm} &\hspace{-1mm}-2.1161688298\hspace{-1mm} & \hspace*{-1.5mm}6.92948625\hspace*{-1mm} & \hspace*{-1.5mm}-21.501020\hspace*{-1mm}\\
\hspace*{-1mm}$23$\hspace*{-1mm}& \hspace{-1mm}0.92363182806\hspace{-1mm} &\hspace{-1mm}-2.1161696094\hspace{-1mm} & \hspace*{-1.5mm}6.92953519\hspace*{-1mm} & \hspace*{-1.5mm}-21.502856\hspace*{-1mm}\\
\hspace*{-1mm}$24$\hspace*{-1mm}& \hspace{-1mm}0.92363182875\hspace{-1mm} &\hspace{-1mm}-2.1161697224\hspace{-1mm} & \hspace*{-1.5mm}6.92954359\hspace*{-1mm} & \hspace*{-1.5mm}-21.503232\hspace*{-1mm}\\
\hline
\hspace*{-1mm} {Eq}~\eqref{Ipexp}:\hspace*{-1mm} & \hspace*{-1mm}0.92363182652 \hspace*{-1mm}&\hspace{-1mm} -2.1161697185\hspace{-1mm}&\hspace*{-1.5mm}6.92954468\hspace*{-1mm} & \hspace*{-1.5mm}-21.503278\hspace*{-1mm}\\ \hline
\end{tabular}
\end{center}
\end{scriptsize}
\label{table5j}
\end{table}

%
%
%
%
\noindent
UV divergence occurs when $U$ vanishes at the boundaries. 
The $\Gamma$-function in Eq~\eqref{fourLOOP} contributes to UV divergence when $N\le 8$.
As show in in figures, the entering momentum is $p,$ and we
denote $s=p^2$.

We address the integrals adhering to Eq~\eqref{fourLOOP} 
for the diagrams of Fig~\ref{4ls-massless-diagrams},
which are denoted by $I_a^{S4},\, I_b^{S4},\,I_c^{S4},\, I_d^{S4}.$
We only consider the massless case, i.e., $m_r=0$.
In the numerical evaluation, we set the value $s=1$.

\subsection{4-loop finite integrals}

\begin{figure}
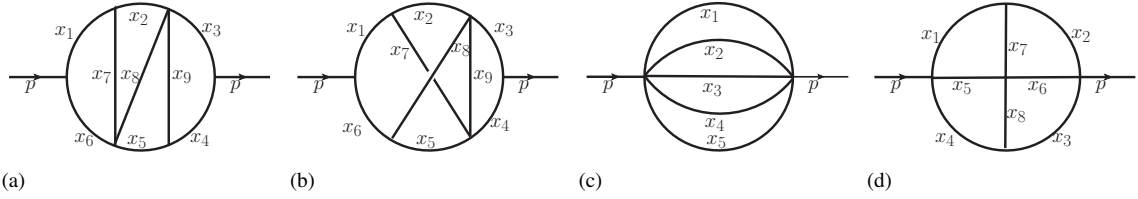

\begin{center}
\begin{subfigure}[ ]
\centering
\includegraphics[width=0.22\linewidth]{./figures/4ls-fig8a-20161029.epsi}
\end{subfigure}
\begin{subfigure}[ ]
\centering
\includegraphics[width=0.22\linewidth]{./figures/4ls-fig8b-20161029.epsi}
\end{subfigure}
\begin{subfigure}[ ]
\centering
\includegraphics[width=0.22\linewidth]{./figures/4ls-fig8c-20161029.epsi}
\end{subfigure}
\begin{subfigure}[ ]
\centering
\includegraphics[width=0.22\linewidth]{./figures/4ls-fig8d-20161029.epsi}
\end{subfigure}
\caption{4-loop self-energy diagrams with massless internal lines, 
cf., Baikov and Chetyrkin~\cite{baikov10}: 
(a) $N=9$, (b) $N=9$, (c) $N=5$, (d) $N=8$}
\label{4ls-massless-diagrams}
\end{center}
\end{figure}

Let us consider the finite integrals  $I_a^{S4},\, I_b^{S4}$ 
corresponding to the diagrams 
of Fig~\ref{4ls-massless-diagrams}(a) and (b)
 (named $M_{44}$ and $M_{45}$ in Baikov and Chetyrkin~\cite{baikov10}).
The $U, W$ functions are given in Eqs~\eqref{SEforLa}-\eqref{SEforLb}.

\begin{equation}
\mathrm{(a)}\quad
\left\{
\begin{array}{l}
U=
x_7 x_8 x_9  x_{123456}
   +x_7 x_8  x_{1256}   x_{34} 
   +x_7 x_9  x_{126}   x_{345} 
   +x_8 x_9  x_{16}   x_{2345} 
   +x_7 x_5  x_{126}   x_{34}  \\
 \quad  +x_8  x_{16}   x_{25}   x_{34} 
   +x_9 x_2  x_{16}   x_{345} 
   +x_2 x_5  x_{16}   x_{34}  
\\
W/s=
x_7 x_8 x_9  x_{123}   x_{456} 
   +x_7 x_8 \,(  x_{12}  x_3  x_{456} + x_{123}  x_4  x_{56}  )
   +x_7 x_9 \,(  x_{12}  x_3  x_{456} + x_{123}   x_{45}  x_6 ) \\
  \quad +x_8 x_9 \,( x_1  x_{23}   x_{456} + x_{123}   x_{45}  x_6 )
   +x_7 x_5 \,(  x_{12}  x_3  x_{46} + x_{123}  x_4 x_6 ) \\
  \quad +x_8 \,(  x_{12}   x_{34}  x_5 x_6+x_1 x_2  x_{34}   x_{56} + x_{16}  x_3 x_4  x_{25}  )
   +x_9 x_2 \,(  x_{13}   x_{45}  x_6+x_1 x_3  x_{456}  ) \\
  \quad +x_2 x_5 \,( x_1 x_3  x_{46} + x_{13}  x_4 x_6 ) 
\end{array}
\right.
\label{SEforLa}
\end{equation}

\begin{equation}
\mathrm{(b)}\quad
\left\{
\begin{array}{l}
U=
x_7 x_8 x_9  x_{123456}
   +x_7 x_8 \,(  x_{12}   x_{34} + x_{34}   x_{56}  )
   +x_7 x_9  x_{126}   x_{345} 
   +x_8 x_9  x_{156}   x_{234} \\
  \quad +x_7 x_5  x_{126}   x_{34} 
   +x_8 x_2  x_{156}   x_{34} 
   +x_9 \,(  x_{16}  x_2  x_{345} + x_{126}   x_{34}  x_5 )
   +x_2 x_5  x_{16}   x_{34}  
\\
W/s=
x_7 x_8 x_9  x_{123}   x_{456} 
   +x_7 x_8 \,(  x_{12}  x_4  x_{356} + x_{124}  x_3  x_{56}  )
   +x_7 x_9 \,(  x_{12}  x_3  x_{456} + x_{123}   x_{45}  x_6 ) \\
  \quad +x_8 x_9 \,( x_1  x_{23}   x_{456} + x_{123}  x_4  x_{56}  )
   +x_7 x_5 \,(  x_{12}  x_3  x_{46} + x_{123}  x_4 x_6 )
   +x_8 x_2 \,(  x_{13}  x_4  x_{56} +x_1 x_3  x_{456}  ) \\
  \quad +x_9 \,( x_1 x_2  x_{36}   x_{45} + x_{12}   x_{36}  x_4 x_5+ x_{14}   x_{25}  x_3 x_6 )
   +x_2 x_5 \,( x_1 x_3  x_{46} + x_{13}  x_4 x_6 ) 
\end{array}
\right.
\label{SEforLb}
\end{equation}

Since the corresponding integrals are finite, Eq~\eqref{fourLOOP} is evaluated
with $ \varrho = \varepsilon = 0$. 
These diagrams have $N=9$ internal lines, leading to an 8-dimensional integral in the numerical evaluation.
Table~\ref{tab:4ls-eps0} lists the results for the integrals, 
obtained with {\sc ParInt} executed on \emph{thor} in double precision, using the cubature rule 
of degree 9 in 8 dimensions (see Section~\ref{parint}), which evaluates the function at 1105 points per subregion.

The analytic values given in~\cite{baikov10} are:
\begin{equation}\label{BC-M44}
I_a^{S4}=\frac{441\,\zeta_7}{8} = 55.5852539156784\,, 
\end{equation}
\begin{equation}\label{BC-M45}
I_b^{S4}= 36\,\zeta_3^2 = 52.017868743610 \,.
\end{equation}
$I_a^{S4}$ and $I_b^{S4}$ were evaluated numerically by Smirnov and Tentyukov using FIESTA~\cite{smirnov10}.
The finite terms are given as $55.58537 \pm 0.00031$ and $52.0181 \pm 0.0003,$ 
respectively, with $1.5$ M samples.

\begin{table}
\caption{\footnotesize {\sc ParInt} accuracy and times (on 4 nodes/64 procs. \emph{thor} cluster) for the loop integrals of the
diagrams of Fig~\ref{4ls-massless-diagrams}(a) and (b) ($M_{44}$ and $M_{45}$ in Baikov and Chetyrkin~\cite{baikov10})
with $\varrho = \varepsilon = 0,$ using various numbers of function evaluations. 
}
\begin{footnotesize}
\begin{center}
\begin{tabular}{ccccc}\hline
Diagram & \# {\sc Fcn.} \hspace*{-1mm} & {\sc Integral} \hspace{-1mm} & \hspace{-1mm}{\sc Rel. err. }\hspace{-1mm} & \hspace{-2mm}{\sc Time} \\
 & {\sc Evals.} \hspace*{-1mm} & {\sc Result} \hspace{-1mm} & \hspace{-1mm}{{\sc Est.} $E_r$ }\hspace{-1mm} & \hspace{-2mm}{\sc T[s]}\hspace{-1mm} \\ 
 ~~\\
\hline
 ~~\\
Fig~\ref{4ls-massless-diagrams} (a)
    & 100B & 55.594725 & 9.87e-04 &  185.1 \\ 
    & 200B & 55.586822 & 3.53e-04 &  370.0 \\ 
    & 300B & 55.585150 & 1.80e-04 &  554.3 \\ 
\hline
    & {Eq}~\eqref{BC-M44}:  & 55.585254 & & \\ 
\hline
~~\\
Fig~\ref{4ls-massless-diagrams} (b)
    & 100B & 52.026428 & 9.84e-04 & 239.5 \\ 
    & 200B & 52.019118 & 3.63e-04 & 479.9 \\ 
    & 275B & 52.017714 & 2.20e-04 & 658.8 \\ 
\hline
    & {Eq}~\eqref{BC-M45}:  & 52.017869 & &  \\ 
\hline
\end{tabular}
\end{center}
\end{footnotesize}
\label{tab:4ls-eps0}
\end{table}

\subsection{4-loop UV-divergent integrals}\label{uv-4ls}
This section handles the integrals associated with the massless diagrams of 
Fig~\ref{4ls-massless-diagrams}(c) and (d) (the 4-loop \emph{sunrise-sunset} and
\emph{Shimadzu} diagrams 
named $M_{01}$ and $M_{36},$ respectively, in~\cite{baikov10}), 
which have a UV singularity from the $\Gamma$-function factor in Eq~\eqref{fourLOOP}.
We put  $\varrho = 0$ in
the integrals for Fig~\ref{4ls-massless-diagrams} (c) and (d), and consider the 
expansions in $\varepsilon.$

For the integral $I_c^{S4}$ in Fig~\ref{4ls-massless-diagrams}, 
the $U,\, W$ functions are
\begin{equation}
\mathrm{(c)}\quad
\left\{
\begin{array}{l}
U=x_1 x_2 x_3 x_4 + x_1 x_2 x_3 x_5 + x_1 x_2 x_4 x_5 + x_1 x_3 x_4 x_5 + x_2 x_3 x_4 x_5
\\
W/s=x_1 x_2 x_3 x_4 x_5
\end{array}
\right.\,.
\label{SEforLc}
\end{equation}

\noindent
The numerical results are compared with the expansion in Baikov and Chetyrkin~\cite{baikov10},
\begin{align}
& 
  n(\varepsilon)^4 I_c^{S4}
 = -\frac{1}{576}\,\frac{1}{\varepsilon}-\frac{13}{768}-\frac{9823}{82944}\,\varepsilon
  +\,\ldots \nonumber \\
& = -0.001736111111\,\frac{1}{\varepsilon}-0.016927083333-0.118429301698\,\varepsilon
  +\,\ldots \label{BandC-m01}
\end{align}
Note that $I = I_c^{S4}$ of Eq~\eqref{fourLOOP} is multiplied with $n(\varepsilon)^4,$ 
where $n(\varepsilon)$ is defined in Eq~\eqref{factorn}.

Based on integrations with {\sc ParInt}, using a maximum of 10B function evaluations 
and an absolute error tolerance of $10^{-12}$ (on the computation of the integral
$I_{c}^{S4}/\Gamma(-3+4\varepsilon)$), the results in Table~\ref{tab:M01} are produced using linear extrapolation. 
{\sc ParInt} returns a 0 error flag for the integrals constituting the input sequence to the
extrapolation, indicating that a successful termination
is assumed according to Eqs~\eqref{accuracy} or~\eqref{acc} for the requested accuracy.
\begin{table}
\caption{\footnotesize
{\color{black}Results UV \emph{4-loop sunrise-sunset} integral}, Fig~\ref{4ls-massless-diagrams} (c) (on 4 nodes/64 procs. \emph{thor} cluster),
err. tol. $t_a = 10^{-12},$ $T[s]$ = Time (elapsed user time in seconds);
$\varepsilon = \varepsilon_\ell = 2^{-\ell},~ \ell = 8,9,\ldots,$ ~~$E_a = $ integration estim. abs. error}
\begin{scriptsize}
\begin{center}
\begin{tabular}{cccccc}\hline
& \multicolumn{2}{c}{{\sc Integral $I_{c}^{S4}$}}
& \multicolumn{3}{c}{{\sc Extrapolation}} \\
 \raisebox{1mm}{\hspace*{-5mm}$\ell$}\hspace*{-5mm} & \hspace{-1mm}{ $E_a$ }\hspace{-1mm} & \hspace{-1mm}{\sc T[s]}
\hspace{-1mm} & {\sc Result} ~${C}_{-1}$ & \hspace{-1mm}{\sc Result} ~${C}_{0}$\hspace{-1mm} & \hspace*{-0mm}{\sc Result} ~${C}_1$ \\
\hline
\hspace*{-1mm}$8$\hspace*{-1mm}& \hspace{-1mm}3.9e-14\hspace{-1mm} & \hspace{-1mm}30.2\hspace{-1mm} & \hspace{-1mm}\hspace{-1mm} & \hspace{-1mm}\hspace{-1mm} & \hspace*{-1.5mm}\hspace*{-1mm} \\
\hspace*{-1mm}$9$\hspace*{-1mm}& \hspace{-1mm}3.8e-14\hspace{-1mm} & \hspace{-1mm}34.3\hspace{-1mm} & \hspace{-1mm} -0.001735179254977\hspace{-1mm} & \hspace{-1mm}\hspace{-1mm} & \hspace*{-1.5mm}\hspace*{-1mm} \\
\hspace*{-1mm}$10$\hspace*{-1mm}& \hspace{-1mm}3.6e-14\hspace{-1mm} & \hspace{-1mm}34.1\hspace{-1mm} & \hspace{-1mm}-0.001736115881839\hspace{-1mm} & \hspace{-1mm}-0.016918557081\hspace{-1mm} & \hspace*{-1.5mm}-0.012276556 \hspace*{-1mm} \\
\hspace*{-1mm}$11$\hspace*{-1mm}& \hspace{-1mm}4.1e-14\hspace{-1mm} & \hspace{-1mm}50.8\hspace{-1mm} & \hspace{-1mm}-0.001736111099910\hspace{-1mm} & \hspace{-1mm}-0.016927126297\hspace{-1mm} & \hspace*{-1.5mm}-0.011837812 \hspace*{-1mm} \\
\hspace*{-1mm}$12$\hspace*{-1mm}& \hspace{-1mm}4.2e-14\hspace{-1mm} & \hspace{-1mm}58.5\hspace{-1mm} & \hspace{-1mm}-0.001736111111130\hspace{-1mm} & \hspace{-1mm}-0.016927083216\hspace{-1mm} & \hspace*{-1.5mm}-0.011842959 \hspace*{-1mm} \\
\hspace*{-1mm}$13$\hspace*{-1mm}& \hspace{-1mm}1.5e-14\hspace{-1mm} & \hspace{-1mm}48.1\hspace{-1mm} & \hspace{-1mm}-0.001736111111109\hspace{-1mm} & \hspace{-1mm}-0.016927083381\hspace{-1mm} & \hspace*{-1.5mm}-0.011842916 \hspace*{-1mm} \\
\hline
\multicolumn{3}{r} \emph{Eq}~\eqref{BandC-m01}: & -0.001736111111111 & -0.016927083333 &\hspace*{-1.0mm} \hspace*{-1.5mm}-0.011842930\hspace*{-1mm} \\
\hline
\end{tabular}
\end{center}
\end{scriptsize}
\label{tab:M01}
\end{table}

For  $I_d^{S4}$ of the $N=8$ diagram shown in Fig~\ref{4ls-massless-diagrams} (d)
 (\emph{Shimadzu}, named $M_{36}$ in~\cite{baikov10,smirnov10}), the
$U, W$ functions are:

\begin{equation}
\mathrm{(d)}\quad
\left\{
\begin{array}{l}
U=
x_7 x_8 \,( x_{12} x_{3456} + x_{34}   x_{56} )
  + x_{78}  x_{1234} x_5 x_6  \\
  \quad +x_7 \,( x_5 x_3  x_{124} +x_6 x_4  x_{123} +x_3 x_4  x_{12}  )
  +x_8 \,( x_5 x_2  x_{134} +x_6 x_1  x_{234} +x_1 x_2  x_{34}  ) \\
  \quad +x_5 x_6  x_{14}   x_{23}  + x_5 x_2 x_3  x_{14}  + x_6 x_1 x_4  x_{23} 
  + x_1 x_2 x_3 x_4   
\\
W/s=
x_7 x_8  x_{12}   x_{34}   x_{56} 
  + x_{78}  x_{12}   x_{34}  x_5 x_6 
  +x_7   x_{12} x_3 x_4   x_{56} 
  +x_8   x_1 x_2  x_{34}  x_{56} \\
 \quad  +x_5 x_6 \,( x_1 x_2  x_{34}  +  x_{12}  x_3 x_4 )
  + x_{56}  x_1 x_2 x_3 x_4   
\end{array}
\right.
\label{SEforLd}
\end{equation}
 
\noindent
The expansion given in~\cite{baikov10} is
\begin{align}
& 
  n(\varepsilon)^4 I_{d}^{S4} =
\frac{5\zeta_{5}}{\varepsilon}-5\zeta_{5}-7\zeta_{3}^2 +\frac{25}{2}\zeta_{6}
+ \,(35\zeta_{5}+7\zeta_{3}^2-\frac{25}{2}\zeta_{6}-21\zeta_{3}\zeta_{4}+\frac{127}{2}\zeta_{7})\,\varepsilon +\, \ldots,\nonumber \\
& = \frac{5.184638776}{\varepsilon}- 2.582436090 +70.39915145\,\varepsilon
  +\,\ldots \label{BandC-m36}
\end{align}
This is $I = I_{d}^{S4}$ of Eq~\eqref{fourLOOP} multiplied with $n(\varepsilon)^4,$
where $n(\varepsilon)$ is defined in Eq~\eqref{factorn}.
The numerical result by FIESTA is shown 
in~\cite{smirnov10} 
and it is $5.184645 \pm 0.000042$.

The results by the DE formula~\eqref{DEsum} with $N_{eval}=49$ and mesh size $h=0.125$ in all dimensions,
and linear extrapolation, are shown in Table~\ref{tableM36}.
The starting $\varepsilon$ is $2^{-10}.$
The time required for each iteration using 64 threads on KEKSC System A, SR16000 Model M1 (POWER7(R) processor) is below 20 minutes.

\begin{table}
\caption{\footnotesize{ Results UV \emph{4-loop Shimadzu} integral, $I_{d}^{S4}$, Fig~\ref{4ls-massless-diagrams} (d), (on KEKSC 64 threads);
$\varepsilon = \varepsilon_\ell = 2^{-\ell},~ \ell = 10,11,\ldots,$.}}
\begin{scriptsize}
\begin{center}
\begin{tabular}{cccc}\hline
 \multicolumn{1}{c}{{\sc Integral $I_d^{S4}$}}
& \multicolumn{3}{c}{{\sc Extrapolation}} \\
 \raisebox{1mm}{\hspace*{-5mm}$\ell$}\hspace*{-5mm} & {\sc Result} ~${C}_{-1}$ & \hspace{-1mm}{\sc Result} ~${C}_{0}$\hspace{-1mm} & \hspace*{-0mm}{\sc Result} ~${C}_1$ \\
\hline
\hspace*{-1mm}$10$\hspace*{-1mm}& \hspace{-1mm}\hspace{-1mm} & \hspace{-1mm}\hspace{-1mm} & \hspace*{-1.5mm}\hspace*{-1mm} \\
\hspace*{-1mm}$11$\hspace*{-1mm}& \hspace{-1mm}5.18460577\hspace{-1mm} & \hspace{-1mm}-2.47956688\hspace{-1mm} & \hspace*{-1.5mm}\hspace*{-1mm} \\
\hspace*{-1mm}$12$\hspace*{-1mm}& \hspace{-1mm}5.18463921\hspace{-1mm} & \hspace{-1mm}-2.58230627\hspace{-1mm} & \hspace*{-1.5mm} 70.1367604\hspace*{-1mm} \\
\hspace*{-1mm}$13$\hspace*{-1mm}& \hspace{-1mm}5.18463922\hspace{-1mm} & \hspace{-1mm}-2.58243393\hspace{-1mm} & \hspace*{-1.5mm} 70.3982056\hspace*{-1mm} \\
\hspace*{-1mm}$14$\hspace*{-1mm}& \hspace{-1mm}5.18463923\hspace{-1mm} & \hspace{-1mm}-2.58243413\hspace{-1mm} & \hspace*{-1.5mm} 70.3991764\hspace*{-1mm} \\
\hline
\multicolumn{1}{r} {Eq}\hspace*{-0.6mm}\eqref{BandC-m36}: &5.18463878  & -2.58243609 &70.3991515 \\
\hline
\end{tabular}
\end{center}
\end{scriptsize}
\label{tableM36}
\end{table}


\section{Conclusions} \label{conc}
In this paper we describe a fully numerical method for Feynman loop integrals, based on numerical multi-dimensional integration and linear or non-linear extrapolation. 
We use three categories of numerical integration methods, iterated integration with {\sc Dqage (Dqags)} from {\sc Quadpack}, multivariate adaptive integration with {\sc ParInt}, and the DE formula.
For the numerical extrapolation, we employ nonlinear extrapolation with geometric sequences of the 
extrapolation parameter, and linear extrapolation with Bulirsch or geometric sequences. 
The main advantage of the method is its general applicability to multi-loop integrals with arbitrary physical masses and external momenta, without resorting to special problem formulations. 
Using dimensional regularization, both {\it UV-divergent} and {\it finite} terms are estimated. 
We have shown that the technique works well for sets of diagrams with up to four loops and up to four external lines, with or without UV-divergence, 
and the numerical results
reveal excellent agreement with expansions 
in the literature~\cite{laporta01,baikov10,smirnov10}. 

We also demonstrate the effectiveness of variable transformations for some integrals.
Regardless of whether or not variables are transformed, the formulation of the numerical method 
is not affected and it is only necessary to replace the integrand and Jacobian. 
The way to transform variables for a loop integral is not unique, and a general technique to find 
the most effective transformation is not currently known.
However, the effectiveness of transformations can be assessed by examining the 
behavior of the integrations and the ensuing convergence of 
the extrapolation.
The experience gained with the transformations in this paper will yield guidelines
to construct a more general procedure, which will be studied in future work.  

The computation time of the numerical multivariate integration increases 
with the number of internal lines, i.e., with the dimension of integration.
For example, though we understand the importance of scattering processes
with more external legs than five or six in two-loop order,
this is beyond the scope of the current paper, and we plan on
addressing these and related types of problems in future work.

Furthermore, while the examples shown here are limited to scalar loop integrals, 
they can easily be extended to more general cases with physical masses and external momenta,
by including the associated numerator in the integrand (see~\cite{eddacat03}). 
The present work includes integrals with massive and massless internal particles. For massive particles, a mass value of one is assigned in order to compare results with the literature~\cite{laporta01}.
Massless cases are compared, e.g., with results in~\cite{baikov10}.
In view of the numerical nature of the methods there is in principle no limitation for the general mass values,
even though challenges may arise with respect to computing time vs. computational precision.
We are helped in dealing with this trade-off by the evolution in computer architecture and 
the computational techniques.

\section*{Acknowledgments}
We acknowledge the support from the National Science Foundation under Award Number 1126438,
and the Center for High Performance Computing and Big Data at Western Michigan University.
This work is further supported by Grant-in-Aid for Scientific
Research (15H03668) of JSPS, and the Large Scale Simulation Program Nos. 15/16-06 and 16/17-21 of KEK.
\appendix 
\label{app:KKmethod} 
\section{Analytic method for Fig~\ref{2-loop-UV-diagrams}(c)}
\noindent
This Appendix derives the analytic result
for the 2-loop \emph{half-boiled egg} diagram of Fig~\ref{2-loop-UV-diagrams}(c).

By the variable transformation
\begin{center}
$x_1= \tau\,(1-\xi)$,~
$x_2= \tau\,\xi$,~
$x_3= (1-\tau)\,\tau'\,(1-\xi')$,~
$x_4= (1-\tau)\,\tau'\,\xi'$,~
$x_5= (1-\tau)\,(1-\tau'),$
\end{center} 
the functions $U, V$ are given by
\begin{equation}
U=\tau \,F,\quad F = 1-\tau+\tau\,\xi\,(1-\xi) \nonumber
\end{equation}
\begin{equation}
W/s=\tau \,G, \quad G=(1-\tau)\,(1-\tau')\,((1-\tau)\,\tau'+\tau\,\xi\,(1-\xi)) \nonumber
\end{equation}
and
\begin{equation}
J_c^{S2}=\int_0^1 d\tau ~\tau(1-\tau)^2 
\int_0^1 d\xi \int_0^1 d\tau' \,\tau' 
\frac{1}{U^{1-3\varepsilon}\,(UV)^{1+2\varepsilon}}
\label{Jc}
\end{equation}
where we assume $m_3=m_4$ to perform the $\xi'$-integration.

The integral is divergent at $\tau=0$ to produce $\displaystyle{1/\varepsilon}$
singularity, and the separation of the singularity is done as follows.
Let us denote $H=M^2 F- s\,G,$ and let us use the suffix 0 for the function
defined at $\tau=0$, i.e., $F_0=F\,(\tau=0)=1$ and $H_0=H\,(\tau=0)
=[\tau'm_3^2+(1-\tau')\,m_5^2]-s\,(1-\tau')\,\tau'$.

The integral is separated into two terms as
\begin{equation}
J_c^{S2}=I_A+I_B \nonumber
\end{equation}
where the first term $I_A$ has a UV singularity. The integrands of
$I_A$ and $I_B$ are given according to Eq~\eqref{Jc} and
\begin{equation}
\frac{1}{U^{1-3\varepsilon}(UV)^{1+2\varepsilon}}
=\frac{1}{\tau^{2-\varepsilon}}
\frac{1}{F^{1-3\varepsilon}H^{1+2\varepsilon}}
=\frac{1}{\tau^{2-\varepsilon}}\left[
\frac{1}{F_0^{1-3\varepsilon}H_0^{1+2\varepsilon}}
+\left( \frac{1}{F^{1-3\varepsilon}H^{1+2\varepsilon}}
-\frac{1}{F_0^{1-3\varepsilon}H_0^{1+2\varepsilon}} \right)
\right] \, . \nonumber
\end{equation}

The divergent term $I_A$ is trivial in the $\xi$-integral and is calculated as 
\[
I_A=
 \int_0^1 d\tau  \,\frac{(1-\tau)^2}{\tau^{1-\varepsilon}}
\int_0^1  d\tau' \frac{\tau'}{H_0^{1+2\varepsilon}}
\]
\begin{equation}
=\left(\frac{1}{\varepsilon}-\frac{3}{2}+\frac{7}{4}\varepsilon-\frac{15}{8}\varepsilon^2\right)
\times \left( I_A^{(0)}+I_A^{(1)}\varepsilon+I_A^{(2)}\varepsilon^2+I_A^{(3)}\varepsilon^3\right) 
\label{expIA}
\end{equation}
and the non-divergent term $I_B$ is
\[
I_B=
\int_0^1d\tau
\int_0^1 d\xi \int_0^1  d\tau'  \,\frac{(1-\tau)^2 \tau'}{\tau^{1-\varepsilon}}
\left(\frac{1}{F^{1-3\varepsilon}H^{1+2\varepsilon}}-\frac{1}{H_0^{1+2\varepsilon}}\right)
\]
\begin{equation}
=I_B^{(1)}+I_B^{(2)}\varepsilon+I_B^{(3)}\varepsilon^2 
\label{expIB}
\end{equation}

The terms $ I_A^{(0)}, I_A^{(1)}, I_A^{(2)}, I_A^{(3)}$ and
$I_B^{(1)}, I_B^{(2)}, I_B^{(3)}$ in Eqs~\eqref{expIA} and~\eqref{expIB} can be obtained by
expanding the integrands in powers of $\varepsilon$.
We evaluated these terms numerically by {\sc Dqage} and {\sc Dqags} from {\sc Quadpack} and the DE formula, for the values $s=M^2=1,$ yielding 
the expression in Eq~(\ref{Ihexp}).


%

\section*{References}

\bibliographystyle{iopart-num}
\bibliography{./bib2,./wi,./bibjk}

\end{document}